\documentclass[a4paper,12pt]{article}
\pdfoutput=1
\usepackage[svgnames]{xcolor}
\usepackage[colorlinks=true,urlcolor = purple,linkcolor=teal,citecolor=magenta,bookmarks=false]{hyperref}
\usepackage{amsfonts,amsmath,amssymb}
\usepackage{graphicx}
\usepackage{dsfont}
\usepackage{graphicx}
\usepackage{amssymb}
\usepackage{amsmath}
\usepackage{amsthm}
\usepackage{mathtools}
\usepackage{empheq}
\usepackage{bm}
\usepackage{ulem}
\usepackage{stmaryrd}
\usepackage[nomessages]{fp}
\usepackage[font=footnotesize,labelfont=bf]{caption}

\newtheorem{lemma}{Lemma}
\newtheorem{theorem}{Theorem}
\newtheorem{corrollary}{Corrallary}
\newtheorem{proposition}{Proposition}
\newcommand{\bc}{\begin{center}}
\newcommand{\ec}{\end{center}}
\def\ba#1{\begin{array}{#1}\displaystyle}
\newcommand{\ea}{\end{array}}

\newcommand{\beq}{\begin{equation}}
\newcommand{\eeq}{\end{equation}}
\newcommand{\beqa}{\begin{eqnarray}}
\newcommand{\eeqa}{\end{eqnarray}}

\newcommand{\bi}{\begin{itemize}}
\newcommand{\ei}{\end{itemize}}

\newcommand{\Tr}{{\rm Tr}}

\newcommand{\cqfd}{  {$\square$} }

\begin{document}


\vspace{1cm}

\begin{center}
\textbf{{\Large  Solution to the Quantum Symmetric Simple Exclusion Process~: the Continuous Case}}
\end{center}
\begin{center}
{Denis Bernard${}^{\clubsuit}$ and Tony Jin${}^{\spadesuit}$}
\end{center}

\vspace{0.5cm}
\noindent
\small{$^\clubsuit$ Laboratoire de Physique de l'\'Ecole Normale Sup\'erieure, CNRS, ENS $\&$ PSL University, Sorbonne Universit\'e, Universit\'e de Paris, 75005 Paris, France.}\\
\small{${}^{\spadesuit}$ DQMP, University of Geneva, 24 Quai Ernest-Ansermet, CH-1211 Geneva, Switzerland.}
\vspace{0.5cm}

\centerline{\today}

\vspace{1.5cm}

\noindent {\bf Abstract}\\
The Quantum Symmetric Simple Exclusion Process (Q-SSEP) is a model for quantum stochastic dynamics of fermions hopping along the edges of a graph with Brownian noisy amplitudes and driven out-of-equilibrium by injection-extraction processes at a few vertices. We present a solution for the invariant probability measure of the one dimensional Q-SSEP in the infinite size limit by constructing the steady correlation functions of the system density matrix and quantum expectation values. These correlation functions code for a rich structure of fluctuating quantum correlations and coherences. Although our construction does not rely on the standard techniques from the theory of integrable systems, it is based on a remarkable interplay between the permutation groups and polynomials. We incidentally point out a possible combinatorial interpretation of the Q-SSEP correlation functions via a surprising connexion with geometric combinatorics and the associahedron polytopes.
\vspace{1cm}

{\setlength{\parskip}{0pt plus 1pt} \tableofcontents}

\section{Introduction}

Non-equilibrium phenomena, classical or quantum, are ubiquitous in Nature, but their understanding is more difficult, and thus yet less profound, than the equilibrium ones. Last decade has witnessed important conceptual progresses in this direction for classical systems, starting from the exact analysis of simple models~\cite{Kipnis99,Liggett99,Spohn91}, such as the Symmetric Simple Exclusion Process (SSEP) or its assymetric partner (ASEP) ~\cite{SSEP,Derrida_Review,Mallick_Review}, via the understanding of fluctuation relations~\cite{Fluctu95,Jar97,Crooks99} and their interplay with time reversal~\cite{Maes99,Maes_bis}. These progresses culminated in the formulation of the macroscopic fluctuation theory (MFT) which is an effective theory describing transports and their fluctuations in diffusive classical systems~\cite{MFT}.  

The questions whether macroscopic fluctuation theory may be extended to the quantum realm and which form this extension will take are still open and timely. If such theory can be formulated, it should aim at describing not only diffusive transports and their fluctuations but also quantum coherent phenomena, in particular quantum interferences, quantum correlations and entanglement spreading and their fluctuations, in out-of-equilibrium quantum many body systems. A substantial amount of information has been gained on this question for integrable quantum many-body systems via the formulation of the generalized hydrodynamics \cite{GHD1,GHD2}. But transports in these systems are mainly ballistic. Other pieces of information on a possible form of such theory for diffusive systems has recently been gained by studying model systems based say on random quantum circuits for which a membrane picture \cite{Membrane1,Membrane2,Membrane3,Membrane4,Membrane5} for entanglement production in many-body systems is starting to emerge.

Another route has been taken in a series of works \cite{BBJ1,BJ2019,JKB2020} consisting in analysing model systems which provide quantum extensions of the classical exclusion processes SSEP or ASEP.  One exemple of such models is the quantum symmetric simple exclusion processes, named Q-SSEP \cite{BJ2019}. These models are formulated as noisy quantum many body systems whose mean dynamics reduce to those of the classical exclusion processes, as a consequence of decoherence phenomena. This quasi-classical reduction only applies to the average dynamics since decoherence is at play only in the mean dynamics. Fluctuations are beyond the quasi-classical regime and survive decoherence. In particular, it has been shown that fluctuations of off-diagonal quantum correlations and coherences possess a rich structure in Q-SSEP. 

The aim of the following is to give an explicit construction of the invariant steady probability measure of Q-SSEP (under the hypothesis that a locality conjecture, which we checked in few instances, is valid). That is, it aims at constructing the probability measure on quantum many-body states or density matrices relevant to Q-SSEP which is invariant under the stochastic dynamics of Q-SSEP. This measure encodes for the fluctuations of all quantum expectation values and in particular of the quantum coherences. We hope that, in the same way as the solution of the classical SSEP played a role in the formulation of the classical MFT,  this construction will open the route towards a formulation of the quantum extension of the macroscopic fluctuation theory to quantum many body systems.

In the following Section \ref{sec:summary}, we start this article by describing the definition of Q-SSEP and its main expected characteristics. We formulate the problem of constructing the invariant measure of Q-SSEP in a way that is largely independent of the context of Q-SSEP (although solving Q-SSEP is of course our motivation for tackling this problem). A summary of the main results we obtained as well as open questions are also given in Section \ref{sec:summary}. All proofs and technical details are presented in the remaining Sections \ref{sec:preliminaries}, \ref{sec:proof-solution}, \ref{sec:regular}, \ref{sec:trans-deformation}, \ref{sec:general-deformation} and \ref{sec:appendix}.

\medskip

{\bf Acknowledgements:} We thank Michel Bauer for many discussions on this topics, for his help with generating functions, and for pointing out the connexion with associahedra. We also thank Marko Medenjak for discussions. We acknowledge support from the Agence nationale de la recherche under contract ANR-20-CE47-0014-01.

\section{Summary} \label{sec:summary}

	\subsection{Context}

The quantum SSEP is a model for stochastic quantum many-body dynamics of fermions hopping on the edges of a graph but with Brownian amplitudes and injection/extraction processes at a few vertices modelling interaction with external reservoirs. Here we shall be interested in the one dimensional case (1D) defined on a line interval with injection/extraction processes at the two ends of the interval. See \cite{BBJ1,BJ2019}.

\begin{figure} 
	\centering
	\includegraphics[scale=0.7]{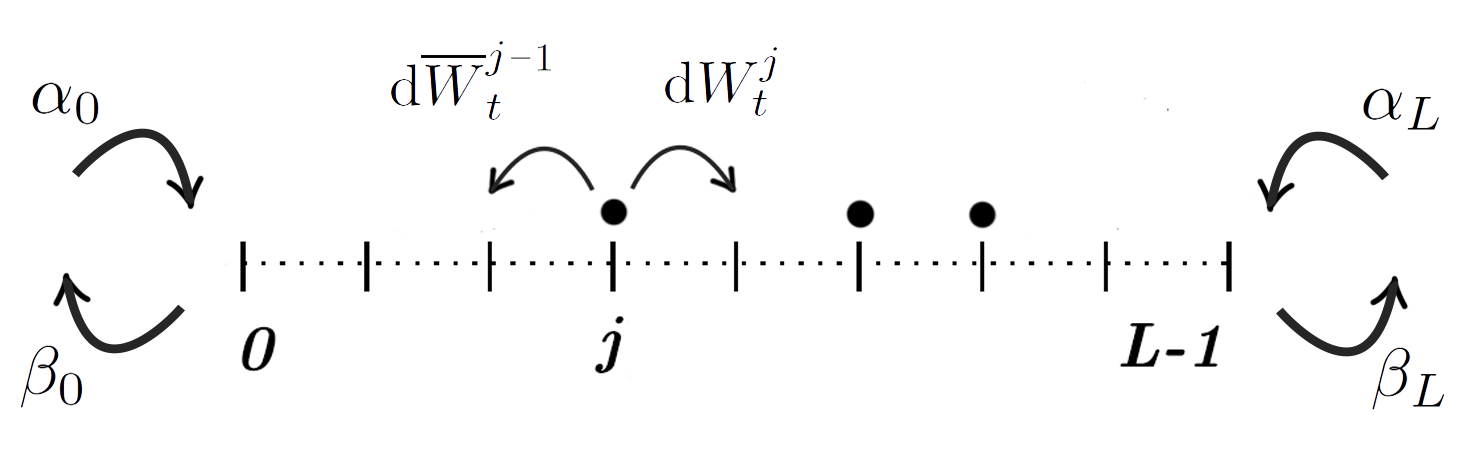}
	\caption{The open quantum SSEP. Particles are injected and extracted at the left and right boundaries at rates $\alpha_0$, $\beta_0$ and $\alpha_L$, $\beta_L$ respectively. In the bulk the fermions undergo stochastic hopping between nearest neighbours.}
	\label{fig:hopping}
\end{figure}

The bulk dynamics is unitary but stochastic. It induces a unitary evolution of the system density matrix $\rho_t$ onto $e^{-idH_t}\,\rho_t\,e^{idH_t}$ with Hamiltonian increments
\beq \label{eq:defHXXsto}
dH_{t}=\sqrt{D}\,\sum_{j=0}^{L-1}\big(c_{j+1}^{\dagger}c_{j}\,dW_{t}^{j}+c_{j}^{\dagger}c_{j+1}\,d\overline W_{t}^{j}\big),
\eeq
for a chain of length $L$, where $c_{j}$ and $c_{j}^{\dagger}$ are canonical fermionic operators, one pair for each site of the chain, with $c_{j}c_{k}^{\dagger}+c_{k}^{\dagger}c_{j}=\delta_{j;k}$, and $W_{t}^{j}$ and  $\overline W_{t}^{j}$ are pairs of complex conjugated Brownian motions, one pair for each edge along the chain, with quadratic variations $dW_{t}^{j}d\overline{W}_{t}^{k}=\delta^{j;k}\,dt$. The parameter $D$ has the dimension of a diffusion constant. See Figure \ref{fig:hopping}.

The boundary dynamics is deterministic but dissipative. Assuming the interaction between the chain and the reservoirs to be Markovian, it is modelled by Lindblad terms. The resulting equations of motion for the density matrix read
\beq \label{eq:Q-flow}
 d\rho_{t}=-i[dH_{t},\rho_{t}] -\frac{1}{2}[dH_{t},[dH_{t},\rho_{t}]]+\mathcal{L}_\mathrm{bdry}(\rho_{t})dt,
 \eeq
with $dH_t$  as above and $\mathcal{L}_\mathrm{bdry}$ the boundary Lindbladian. The two first terms result from expanding the unitary increment $\rho_t \to e^{-idH_t}\,\rho_t\,e^{idH_t}$ to second order (because the Brownian increments scale as $\sqrt{dt}$). The third term codes for the dissipative boundary dynamics
with $\mathcal{L}_\mathrm{bdry}=\alpha_{0}{\cal L}_{0}^{+}+\beta_{0}{\cal L}_{0}^{-}+\alpha_{L}{\cal L}_{L-1}^{+}+\beta_{L}{\cal L}_{L-1}^{-}$ and, for $j=0,\, L-1$,
\begin{align*} \label{eq:L-bdry}
{\cal L}_{j}^{+}(\rho) & =c_{j}^{\dagger}\rho c_{j}-\frac{1}{2}(c_{j}c_{j}^{+}\rho+\rho c_{j}c_{j}^{\dagger}),\\
{\cal L}_{j}^{-}(\rho) & =c_{j}\rho c_{j}^{\dagger}-\frac{1}{2}(c_{j}^{\dagger}c_{j}\rho+\rho c_{j}^{\dagger}c_{j}),
\end{align*}
where the parameters $\alpha_j$ (resp. $\beta_j$) are the injection (resp. extraction) rates.
Equation \eqref{eq:Q-flow} is a (classical) stochastic differential equation for (quantum) density matrices.

This model describes stochastic non-equilibrium physics in quantum many-body systems. At large time, the system reaches a non-equilibrium steady state with a non-trivial mean density profiles $n_j^*:=\lim_{t\to\infty}\mathbb{E}[\Tr(\hat n_j\rho_t)]$, with $\hat n_j:=c_j^\dag c_j$ the local number operator, 
\[ n_j^*= \frac{n_a(L+b-j) + n_b(j+a)}{L+a+b},\]
where $n_{a}:=\frac{\alpha_{0}}{\alpha_{0}+\beta_{0}}$, $n_{b}:=\frac{\alpha_{L}}{\alpha_{L}+\beta_{L}}$ with
$a:=\frac{1}{\alpha_{0}+\beta_{0}}$, $b:=\frac{1}{\alpha_{L}+\beta_{L}}$. In the large size limit, $L\to\infty$ at $x=i/L$ fixed, this profile, $n^*(x)= n_a + x(n_b-n_a)$, interpolates linearly the two boundary mean occupations $n_a$ and $n_b$ \cite{Profile,BJ2019}. 

The fact that the above mean density profile coincides with that in the classical SSEP \cite{SSEP,Derrida_Review,Mallick_Review} reflects that decoherence is at work in the mean dynamics. However, as shown in \cite{BJ2019}, beyond this quasi-classical mean behavior, quantum-ness persists at large time in the sub-leading (in $L$) fluctuations, and a rich structure of fluctuating quantum correlations and coherences is encoded in the steady probability measure of \eqref{eq:Q-flow}.

We shall be interested in the invariant probability measure of \eqref{eq:Q-flow}, called the Q-SSEP invariant measure, in the large size limit, and in the steady correlations of the quantum coherences.

Quantum coherences are defined as the fermion two-point functions $G_{ji}:=\Tr(c^\dag_i c_j \rho_t)$. They specify the system density matrix, since the dynamical equation \eqref{eq:Q-flow} are quadratic. They are random as is the density matrix $\rho_t$. We shall therefore be interested in their multiple point connected correlation functions in the Q-SSEP steady probability measure.  For instance, at large size, $L\to\infty$ with $x=i/L$, $y=j/L$ fixed, their second moments behave, for $0\leq x<y\leq 1$,  as \cite{BJ2019}
\begin{equation*} \label{eq:N=2-QC}
 \mathbb{E}[G_{ij} G_{ji}]^c= \frac{1}{L}(\Delta n)^2\, x(1-y) +O(L^{-2}),
\end{equation*}
with $\Delta n := n_b-n_a$ the difference between the boundary densities, while the other two-point functions are $\mathbb{E}[G_{ii}^2]^c =  \frac{1}{L}(\Delta n)^2\, x(1-x) +O(L^{-2})$ and $\mathbb{E}[G_{ii}G_{jj}]^c =- \frac{1}{L^2}(\Delta n)^2\, x(1-y) +O(L^{-3})$. 

\begin{figure} 
	\centering
	\includegraphics[scale=0.6]{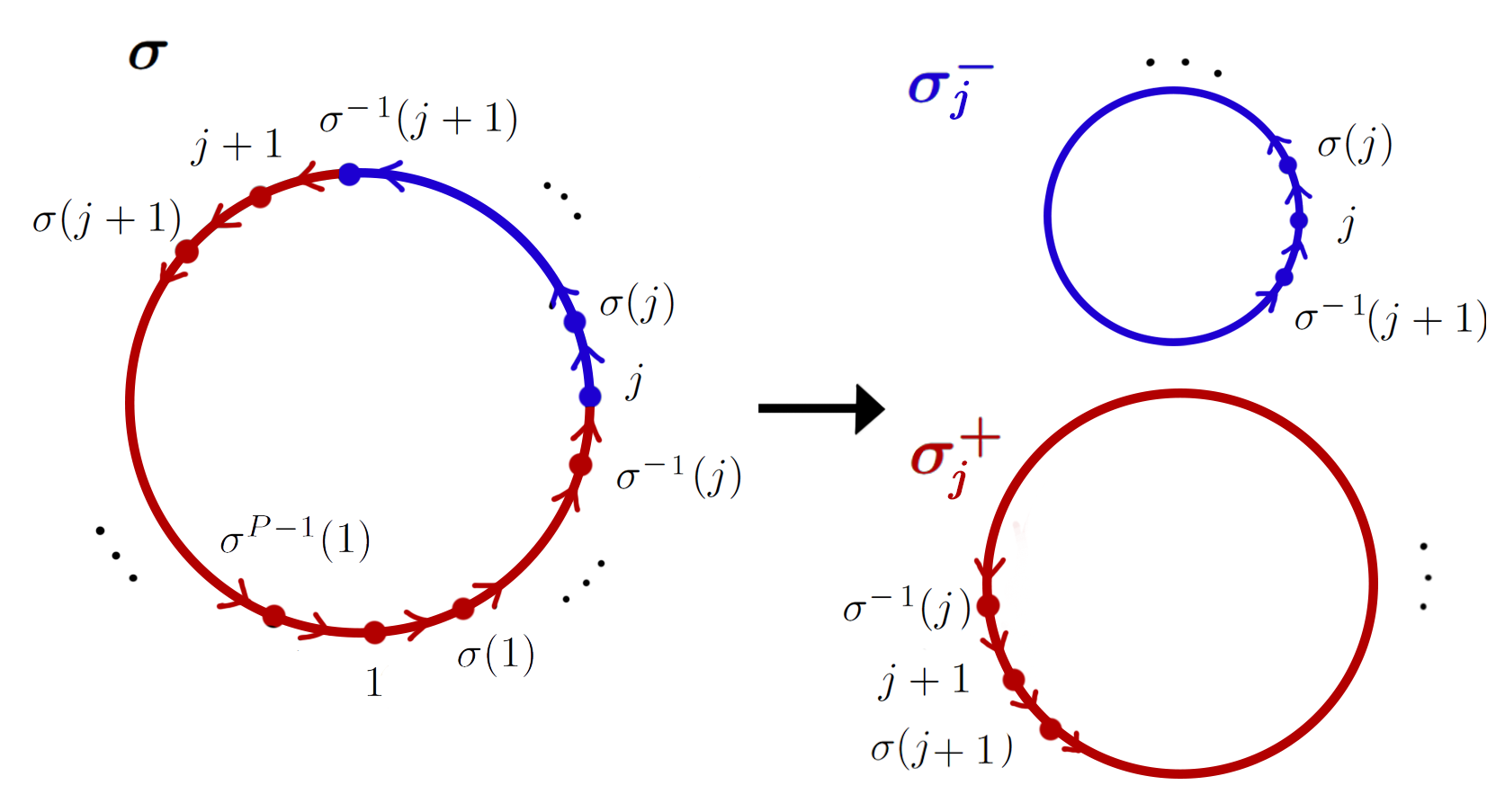}
	\caption{ A cyclic permutation $\sigma$ is represented by a loop. Equation (\ref{eq:gluing}) involves $\sigma^-_j$ and $\sigma^+_j$ which are obtained by breaking the loop $\sigma$ at points $j$ and $j+1$ and forming smaller loops with the two remaining strands.}
	\label{fig:sigma-loop}
\end{figure}

It has been shown in \cite{BJ2019} that, in the large size limit, the leading contributions among the multi-point expectation values $\mathbb{E}[G_{i_1j_1}\cdots G_{i_Pj_P}]^c$ come from the correlation functions of cyclic products $G_{i_1i_P}\cdots G_{i_3i_2}G_{i_2i_1}$. In this limit, these correlation functions scale proportionally to $1/L^{P-1}$, with $P$ the number of insertion points,
\[ 
\mathbb{E}[G_{i_1i_P}\cdots G_{i_3i_2}G_{i_2i_1}]^c =\frac{1}{L^{P-1}}\, g_P(x_1,\cdots,x_P) + O(\frac{1}{L^{P}}),
\]
with $x_k=i_k/L$. The expectation values $\mathbb{E}[G_{i_1j_1}\cdots G_{i_Pj_P}]$ are non-vanishing only if the indices $j_k$'s are permutations of the indices $i_l$'s. To such product $G_{i_1j_1}\cdots G_{i_Pj_P}$ we may associate an oriented graph (connected or not) with a vertex for each point $i_k$ and an oriented edge from $i$ to $j$ for each occurence of $G_{ji}$ in the product. Cyclic products $G_{i_1i_P}\cdots G_{i_3i_2}G_{i_2i_1}$ correspond to single oriented loop graphs. See Figure \ref{fig:sigma-loop}. Other cumulants of $G_{ji}$'s not corresponding to single loop diagrammes are sub-leading and decrease faster at large $L$ than single loop expectation values. Single loops are thus the elementary building blocks in the large size limit.

These correlation functions depend on how the ordering of the points ${\bm x}:=(x_1,\cdots,x_P)$ along the chain matches or un-matches that along the oriented loop. Fixing an order along the chain interval, these different orderings are indexed by single cycle permutations of the permutation group of $P$ elements. The rule for this correspondence is that by turning around the oriented loop indexed by the single cycle permutation $\sigma$ one successively encounters the points labeled as $x_{1}$, $x_{\sigma(1)}$, $x_{\sigma^{2}(1)}$, $\cdots$, up to closing the loop back to $x_{\sigma^{P}(1)}=x_1$. For $0\leq x_1<\cdots<x_P \leq1$, we denote by $[\sigma]({\bm x})$ the expectation values of the loop associated to the single cycle permutation $\sigma$, 
 \beq \label{eq:Q-loop}
\mathbb{E}[G_{i_1i_{\sigma^{P-1}(1)}}\cdots G_{i_{\sigma^2(1)}i_{\sigma(1)}}G_{i_{\sigma(1)}i_1}]^c =\frac{1}{L^{P-1}}\, [\sigma]({\bm x})+ O(\frac{1}{L^{P}}),
\eeq
The aim of the following is to determine the expectation values $[\sigma]({\bm x})$ for all oriented loops with an arbitrary number of marked points.

	\subsection{Formulation of the problem} \label{sec:problem} 
	
Stationarity of the measure under the Q-SSEP flow \eqref{eq:Q-flow} imposes constraints on the correlation functions $[\sigma]({\bm x})$. The problem of determining this invariant measure, and more particularly all these correlation functions, can be formulated algebraically, without making explicit reference to the Q-SSEP -- although Q-SSEP is of course the (initial) motivation to solve this problem.

Let us pick $P$ points on the interval $[0,1]$ and fix an ordering of them along the chain interval, say $0\leq x_1<\cdots<x_P \leq1$. We then consider labeled loops by placing these $P$ points on a loop but without respecting the ordering, so that the order of the points on the clockwise oriented loops may differ from the order of the points on the chain interval. The order of the points on a given labeled loop is in a one-to-one correspondence with a single cycle permutation $\sigma$, such that a clockwise exploration of the loop successively reveals the point $x_1$, $x_{\sigma(1)}$, $x_{\sigma^2(1)}$, up to $x_{\sigma^P(1)}=x_1$. To each of these labeled loops, or alternatively to each single cycle permutation $\sigma$ of $P$ elements, we associated a function $[\sigma]({\bm x})$ of ${\bm x}:=(x_1,\cdots,x_P)$ which we call the loop expectation value or the loop correlation function. These functions satisfy a series of conditions which are the conditions for the stationarity of the Q-SSEP measure, see \cite{BJ2019}. There are three types of conditions (bulk relations, boundary conditions and exchange relations)~:

\noindent
(i) \underline{\it Bulk relations and locality} : For all $x_j$ distincts, the function $[\sigma]({\bm x})$ is harmonic: 
\begin{equation} \label{eq:harmonic}
 \sum_j \Delta_{x_j}[\sigma]({\bm x})=0 ,
 \end{equation}
where $\Delta_{x_j}$ is the Laplacian, i.e. $\Delta_{x}\,f({\bm x}):=\nabla_{x_j}^2f({\bm x})$. We shall look for solutions which enforce this condition locally. 
We put this hypothesis as a working conjecture:

\noindent
{\bf Conjecture.} {\it  
We conjecture that :\\
(c1) The Q-SSEP invariant measure is unique;\\
(c2) This measure satisfies a locality property, meaning that $\Delta_{x_j}[\sigma]({\bm x})=0$ for all $j$, as long as the points in ${\bm x}$ are distincts.\\
We shall prove that, assuming the conjecture {\it (c2)}, the Q-SSEP invariant measure is unique.
}

This locality conjecture is supported by the explicit computation of the steady 2, 3 and 4 point functions, at finite size $L$, done in our previous paper \cite{BJ2019}.

\noindent
(ii) \underline{\it Boundary conditions} : There are two boundary conditions at the two ends of the chain interval, for $P>1$,
\beq \label{eq:bdry}
[\sigma]({\bm x})\vert_{x_1=0}=0,\quad [\sigma]({\bm x})\vert_{x_P=1}=0.
\eeq

\noindent
(iii) \underline{\it Echange relations} : These are relations imposing conditions on $[\sigma]({\bm x})$ at the boundary $x_j=x_{j+1}$ and involve the pair of loops $\sigma$ and $\tau_j\circ\sigma:=\tau_j\sigma\tau_j^{-1}$ with $\tau_j$ the transposition exchanging $j$ and $j+1$. The first is a continuity equation,
\beq \label{eq:continu}
[\sigma]({\bm x})\vert_{x_j=x_{j+1}}= [\tau_j\circ\sigma]({\bm x})\vert_{x_j=x_{j+1}}.
\eeq
The second is a Neumann like boundary condition,
\beq \label{eq:gluing}
\big(\nabla_{x_j}-\nabla_{x_{j+1}}\big)\Big([\sigma]({\bm x})+ [\tau_j\circ\sigma]({\bm x})\Big)\vert_{x_j=x_{j+1}} = 2\, \nabla_{x_j}[\sigma_j^-]({\bm x})\, \nabla_{x_{j+1}}[\sigma_j^+]({\bm x}),
\eeq
where $\sigma_j^\pm$ are the two loops arising from the decomposition of the product $\tau_j\,\sigma$ into its two connected cycles, with $\sigma_j^-$ containing $j$ and $\sigma_j^+$ containing $j+1$.
Note that both sides of the above equation are independent of $x_j$ and $x_{j+1}$, since the $[\sigma]({\bm x})$'s are polynomials of degree at most one in each variable.

Hence, by the bulk harmonic condition \eqref{eq:harmonic} imposed locally,  to any labeled loop, or alternatively to any single cycle permutation $\sigma$, we associate a polynomial $[\sigma]({\bm x})$ of degree at most one is each of its variable.
By the boundary conditions \eqref{eq:bdry}, these polynomials may be factorized as $[\sigma]({\bm x})=x_1\sigma^{(1)}({\bm x})$, with $\sigma^{(1)}({\bm x}):=\nabla_{x_1}[\sigma]({\bm x})$ independent of $x_1$, and $\sigma({\bm x})= \sigma^{(P)}({\bm x})(1-x_P)$, with $\sigma^{(P)}({\bm x}):=-\nabla_{x_P}[\sigma]({\bm x})$ independent of $x_P$.
The exchange relations \eqref{eq:gluing} relate polynomials associated to different labeled loops. Since the transpositions $\tau_j$ generate the group of permutations of $P$ elements, these relations allow to explore the conjugacy class of all single cycle permutations. Together with the boundary conditions, they determine uniquely the solution, as we shall prove below.

The case $P=1$ is peculiar. From Q-SSEP computations, we know that the expectation value of a loop with a single point linearly interpolates between the two boundary densities, that is, $[(1)](x)= n_a + (n_b-n_a)x$ with $n_a$ (resp. $n_b$)  the densities at the left (resp. right) boundary of the chain interval. Actually, the only information we shall need in the following is $\nabla_x[(1)]=(\Delta n)$ with $\Delta n:=n_b-n_a$.

It is easy to verify that loop expectation values with $P$ points are proportional to $(\Delta n)^P$, that is~: $[\sigma]({\bm x})\propto (\Delta n)^P$ with $P$ the number of points in the label loop $\sigma$. From the reflexion symmetry exchanging the chain boundaries, we have: $[\sigma](x_1,\cdots,x_P)=(-)^P\,[\sigma^{rev}](1-x_P,\cdots,1-x_1)$ with $\sigma^{rev}(k):=\sigma(P+1-k)$.

In the following, to simplify notation, we shall set $\Delta n=1$ (unless explicitly specified).

	\subsection{Summary of the results}
	
The aim of the sequel is to present the solution to the constraints (\ref{eq:harmonic},\ref{eq:bdry},\ref{eq:continu},\ref{eq:gluing}) assuming the locality conjecture. Since the gluing condition \eqref{eq:gluing} relates correlation functions with $P$ points to correlation functions with less points, the construction of the solution will be recursive on the number of points.

We shall prove that, if the locality conjecture holds, then the solution of the stationarity conditions  (\ref{eq:harmonic},\ref{eq:bdry},\ref{eq:continu},\ref{eq:gluing}) is unique and, for any labeled loop $\sigma$ with $P$ marked points, can be written as
\beq \label{eq:solution-00}
[\sigma]({\bm x}) = \sum_{j=1}^{P-2} x_1\cdots x_j\, [\sigma]^o_{j+1}({\bm x}) + x_1\cdots x_{P-1}(1-x_P)\,[\sigma]^o_{P},
\eeq
with ${\bm x}:=(x_1,\cdots,x_P)$ with $0\leq x_1<\cdots<x_P\leq 1$, where the polynomial coefficients $[\sigma]^o_{j+1}({\bm x})$ are called hole coefficients and are given by
\beq \label{eq:sigma0-00}
[\sigma]^o_{j+1}({\bm x})=(\nabla_{x_{j+1}}\cdots\nabla_{x_1})\sum_{k=1}^{j}\big[ (\tau_{k+1}\cdots\tau_j\circ\sigma)^-_k\big]({\bm x})\,\big[ (\tau_{k+1}\cdots\tau_j\circ\sigma)^+_k\big]({\bm x}) ,
\eeq
for $j=1,\cdots, P-1$. This formula relates the hole coefficients of a loop correlation function with $P$ points to multiple derivatives of smaller loop correlation functions with less than $P$ points, via a quadratic relation. Most of the following constructions or arguments will be based on an interplay between hole coefficients, multiple derivatives and the permutation group.
See Theorem \ref{th:solution} and its proof in Section \ref{sec:proof-solution} for details.

Equation \eqref{eq:sigma0-00} is recursive by construction. It allows to compute loop correlation functions for loops with $P$ points knowing all loop correlation functions for loop with up to $P-1$ points. As such one may say that it solves the problem of determining the Q-SSEP invariant measure (at least the multi-point correlation functions within this invariant measure). The following results aim at making this recursive procedure more concrete and at deciphering (part of) the structure underlying it. These results will be formulated in terms of generating functions coding for (and summing over) loop expectation values with an arbitrary number of points. 

First we shall notice that the formulas (\ref{eq:solution-00},\ref{eq:sigma0-00}) becomes rather explicit when dealing with the loops $\omega_P$, which we call the regular loops, for which the order along the chain and the loop coincide. These are associated to the cyclic permutation $\omega_P:=(12\cdots P)$ of the $P$ first integers with $P$ the number of points. We introduce the generating function $\mathcal{C}_k(z)$ of multiple derivatives of the regular loop expectation values $[\omega_P]$ defined as the formal power series in $z$,
\[
\mathcal{C}_k(z) := \sum_{N\geq 0} C_{N+k+1;k}\,z^{N},
\]
with $C_{N+k+1;k}:=\nabla_{x_{N+1}}\cdots\nabla_{x_1}[\omega_{N+k+1}]({\bm x})$. Since $[\omega_{N+k+1}]({\bm x})$ are polynomials of degree at most one in each of their variables, by the locality conjecture, all coefficients $C_{N+k+1;k}$, and hence also the generating functions $\mathcal{C}_k(z)$, depend only on the remaining $k$ variables $x_{N+k+1},\cdots,x_{N+2}$ on which the derivatives are not acting. We call them the floating variables and renamed them as $y_l:=x_{N+k+1-l}$ for $l=1,\cdots,k$. The formal power series $\mathcal{C}_k(z)$ sum over the number of points in the regular loops fixing the number of floating variables. For the regular loops, the formulas (\ref{eq:solution-00},\ref{eq:sigma0-00}) can be recasted into the following recurrence relation:
\begin{equation} \label{eq:Ck-series} 
\mathcal{C}_{k+1}(z)= \Big[ \big(\mathfrak{c}(z) + \frac{y_k}{z}\big)\mathcal{C}_{k}(z)\Big]_+ ,
\end{equation}
 where  $\big[\cdots\big]_+$ means the part with positive degrees of the Laurent series and $\mathfrak{c}(z):=\big(\sqrt{1+4z}-1\big)/2z=1-z+2z^2-5z^3+\cdots$ is the generating function of alternating Catalan numbers. See the Proposition \ref{prop:Ck} in Section \ref{sec:regular} for details. The regular loop expectation values are recovered from the generating functions $\mathcal{C}_k(z)$ by evaluating it at $z=0$ via the reconstruction formula
 \beq \label{eq:omega-Ck}
  [\omega_{k+1}]({\bm x}) = x_1\,\mathcal{C}_k(0)(y_0=x_{k+1},y_1=x_{k},\cdots,y_{k-1}=x_2),
\eeq 
with the initial condition $\mathcal{C}_{k=0}(z)=\mathfrak{c}(z)$.
Furthermore, we shall show that, as a consequence of \eqref{eq:Ck-series} and \eqref{eq:omega-Ck}, there exists yet another  formal power series $\overline{D}_\omega(y_0,y_1,\cdots)$ of an infinite number of variables such that, for each $P\geq 2$,
\begin{equation} 
 [\omega_P]({\bm x})=x_1\, \overline{D}_\omega(x_{P},x_{P-1},\cdots,x_2,0,0,\cdots).
 \end{equation}
 This formal power series is recursively constructed using \eqref{eq:Ck-series}.
 See Theorem \ref{the:D-omega} in Section \ref{sec:regular} for details.
 
 Second we shall view any labeled loop as a deformation of a regular loop. Indeed, because labeled loops are in one-to-one correspondence with single cycle permutations, they form a closed orbit under the adjoint action of the permutation group. Hence any labeled loops with $P$ points is of the form $\mu_n\circ\omega_P:=\mu_n\,\omega_P\,\mu_n^{-1}$ with $\mu_n$ some permutation. Any permutation is characterized by its support which is the minimal length interval outside which it acts trivially as the identity map. A permutation with support of length $|\mu|$ is canonically associated to an element $\mu$ of the permutation group $\mathbb{S}_{|\mu|}$ of $|\mu|$ elements, which we call the profile of the deformation.  Any permutation $\mu_n$ can be viewed as a translation by $n-1$ steps of its canonically associated profile. That is: the permutation $\mu_n$ possesses the same profile as the permutation $\mu$ but starting at point $n$ instead at point $1$: $\mu_n(j+n-1)=\mu(j)+n-1$ for $j=1,\cdots,|\mu|$. We then introduce generating functions for the expectation values of deformations of the regular loop of a given profile $\mu$ defined as the formal power series in $z$, for $q\geq 1$,
\[
 \mathcal{D}_k^{(\mu;q)}(z) :=\sum_{N\geq 0} C_{N+k+1;k}^{\mu_{q+N}}\, z^N ,
 \]
 with $C_{N+k+1;k}^{\mu_{q+N}}:=\nabla_{x_{N+1}}\cdots\nabla_{x_1}[\mu_{q+N}\circ\omega_{N+k+1}]({\bm x})$. Here the integer  label $q$ codes for the distance between the location of the insertion of the deformation $\mu_{q+N}$ and the position of the last derivative $\nabla_{x_{N+1}}$ in $C_{N+k+1;k}^{\mu_{q+N}}$, so that these generating functions sum over the number of points in the loops and the location of the insertion of the deformations. As above, the functions $ \mathcal{D}_k^{(\mu;q)}(z)$ depend on $k$ floating variables $(y_0,\cdots,y_{k-1})$. We shall prove in Proposition \ref{prop:DS-general} in Section \ref{sec:general-deformation} that the formulas (\ref{eq:solution-00},\ref{eq:sigma0-00}) imply that the following recurrence relations for these generating functions
\beq \label{eq:Dgene-recur-intro}
\mathcal{D}_{k+1}^{(\mu;q)}(z)= \Big[ \big(\mathfrak{c}(z) + \frac{y_k}{z}\big)\mathcal{D}_{k}^{(\mu;q-1)}(z)\Big]_+,
\eeq
 for $q\geq 2$. These relations are structurally of the same form as \eqref{eq:omega-Ck} but they strongly depart from them by their initial conditions $\mathcal{D}_{k}^{(\mu;q)}(z)$ at $q=1$. The later is obtained by a procedure which consists in traversing the deformation starting from the right end of the chain interval that we describe in Proposition \ref{prop:DS-general-through} in Section \ref{sec:general-deformation}. The number of steps involve in this procedure increases with the size of the support of the deformation (see Proposition \ref{prop:DS-general-through}). The expectation values of deformations of the regular loops are reconstructed from these generating functions via
\beq \label{eq:deform-Dk}
  [\mu_q\circ\omega_{k+1}]({\bm x})= x_1\, \mathcal{D}_k^{(\mu;q)}(0)(y_0=x_{k+1},\cdots,y_{k-1}=x_2).
\eeq
Similarly as for the regular loop, as a consequence of \eqref{eq:Dgene-recur-intro} and \eqref{eq:deform-Dk}, there exists formal power series $\overline{D}^{\mu;q}(y_0,y_1,\cdots)$ of an infinite number of variables such that, for each $P\geq 2$,
\begin{equation} 
 [\mu_q\circ\omega_{P}]({\bm x})=x_1\, \overline{D}^{\mu;q}(x_{P},x_{P-1},\cdots,x_2,0,0,\cdots).
 \end{equation}
That is: the expectation values of deformations with profile $\mu$ and $P$ points are obtained by evaluating the generating function $\overline{D}^{\mu;q}$ but with $y_l=x_{P-l}$, for $l=0,\cdots,P-2$, and all other variables $y_{P-1},y_{P},\cdots$ set to zero.  See Theorem \ref{th:defom-Dseries} in Section \ref{sec:general-deformation} for details. 

The particular case of deformations implemented by a simple transposition is described with some details in Section \ref{sec:trans-deformation}.

\subsection{Remarks and open questions}

The results we have obtained are modulo the locality conjecture, although we checked it in a few cases. For instance, the formulas (\ref{eq:solution-00},\ref{eq:sigma0-00}) are implied by the conditions (\ref{eq:harmonic},\ref{eq:bdry},\ref{eq:continu},\ref{eq:gluing}) if we assume that the expectation values $[\sigma]({\bm x})$ are harmonic in each of their variables, but we do not provide a full proof that they imply all conditions encoded in (\ref{eq:gluing}). A proof of the later statement will be welcome.

As it will be explicit in the following, in particular in the exemples, the loop expectation values $[\sigma]({\bm x})$ are polynomials with integer coefficients. We do not have an a priori explanation why this should be so, but formulas (\ref{eq:solution-00},\ref{eq:sigma0-00}) give an a posteriori explanation. Also, when specialized to all its variables being equal, $x_j=-t$ for all $j$, these polynomials are remarkably generating functions counting objects of certain degree in discrete and computational geometry \cite{discret-geom}. The coefficients of the loop expectation values with $n+2$ marked points specialized at $x_j=-t$ for all $j$ are the number of faces of given dimension in the associahedron polytope of dimension $n$, as we shall notice in \eqref{eq:count} at the end of Section \ref{sec:regular}. We do not have a good understanding of this surprising fact. Remarkably, an interplay between the geometry of associahedron polytopes and scattering theory has recently been revealed \cite{arkani_et_al}. This remark raises the question whether the polynomials $[\sigma]({\bm x})$ have a combinatorial interpretation.

The tools and structures we used to prove the formulas (\ref{eq:solution-00},\ref{eq:sigma0-00}), or the properties of the generating functions we introduced, are calling for a deeper algebraic structure. These proofs rely on one hand on manipulating two simple operations on $[\sigma]({\bm x})$, the evaluation at $0$ and the derivative with respect to one of the variables, and on the other hand on finding relations between hole coefficients and multiple derivatives. This is for instance illustrated by the formula \eqref{eq:move!!} which is the first step towards proving (\ref{eq:solution-00},\ref{eq:sigma0-00}). On polynomials of degree at most one in each of its variables, as $[\sigma]({\bm x})$ are, these operations can be formulated as annihilating a particle or a hole using a language alluded to at the end of Section \ref{sec:forward}. In this language, formula \eqref{eq:move!!} yields information on the behavior of $[\sigma]({\bm x})$ under the move of a hole from one position to the next. An algebraic framework encompassing this information will be welcome. The computations we present below resemble the computations of expectation values of algebraic operators or of matrix product states but, paradoxically, without knowing explicitly these operators. If such algebraic framework was available, it will most probably be more economical (as, for instance, it is simpler to write expectation values of creation/annihilation operators than computing them explicitly using Wick's theory, if an analogy is needed). It is tempting to suggest that a proof of the locality conjecture will be more direct using such algebraic framework, if it exists (as it is with the matrix product state representation of the classical SSEP invariant measure, if an analogy is needed).

It may be surprising, but remarkable, that explicit expressions (\ref{eq:solution-00},\ref{eq:sigma0-00}) for $[\sigma]({\bm x})$ are available although we did not use any known structure from the theory of integrable systems \cite{integrable}, classical or quantum, besides the permutation group. In a way, at the present stage, the model seems to be solvable without being integrable in the usual sense. Is the potential algebraic structure we discussed above linked to some realisation of the Yang-Baxter equation or of quantum groups~? Are the quadratic relations we found, as in \eqref{eq:sigma0-00} or in (\ref{eq:Ck-series},\ref{eq:Dgene-recur-intro}), related in some disguised form to bilinear Hirota's equations or Sato's tau functions ? Do cluster algebras, via their connexions with exchange relations and associahedron polytopes \cite{fomin,cluster-asso}, play a role in the algebraic structure underlying Q-SSEP  ?  We do not know yet how to answer these questions.

In parallel, the generating functions as well as the operations on them which we introduced, such as in (\ref{eq:Ck-series},\ref{eq:Dgene-recur-intro}), are reminiscent of vertex operators or Sato's tau functions. Since single cycle permutations form an orbit under the adjoint action of the permutation group, and since these generating functions exist for any element in this orbit, we are lead to conjecture the existence of vertex like operators representing the action of permutations of any size on these generating functions (which depend on an infinite number of variables). We hope to report soon on this problem. Permutations may be viewed as discrete diffeomorphisms \cite{Brenier}, and so do the deformations of the regular loop we considered which can be viewed as discrete diffeomorphisms from the chain interval to the loops. Adopting this point of view, these vertex operators, if they can be constructed, are going to implement discrete non-commutative diffeomorphisms. 

The scaling of the correlation functions \eqref{eq:Q-loop} with the system size is such that it ensures the existence of a large deviation principle for the quantum correlation and coherence fluctuations \cite{BJ2019}. It will be welcome to have a variational principle for determining the large deviation rate function, as in the classical SSEP \cite{Derrida_Review,Mallick_Review,BD2004}. This may well be a possible route towards the quantum extension of the macroscopic fluctuation theory \cite{MFT} aiming at describing quantum coherence and entanglement fluctuations in out-of-equilibrium quantum many body systems.

Of course, the first applications of the results presented here may deal with the computation of various entanglement entropy fluctuations or entanglement productions and spreadings in Q-SSEP systems. It will be interesting to understand whether there exists a controllable coarse-grained hydrodynamics description of entanglement spreading and how the later is connected to the recently emerging membrane picture \cite{Membrane1,Membrane2,Membrane3,Membrane4,Membrane5} for entanglement production in many-body systems.

\section{Preliminaries} \label{sec:preliminaries}

Since solving the problem formulated in Section \ref{sec:problem} requires dealing with permutations and polynomials, we first need to recall a few basics facts and introduce simple notations.

	\subsection{Permutations, cycles and loops}
	
	Let $\mathbb{S}_P$ be the permutation group of $P$ elements labeled $1,2,\cdots,P$. Any permutation can be decomposed into independent product of cycles. The number of cycles (counted with multiplicities) specifies the conjugacy classes in $\mathbb{S}_P$. The number of single cycle permutations in $\mathbb{S}_P$ is $(P-1)!$. 

Let $\tau_j$, for $j=1,\cdots, P-1$, be the transpositions of $j$ and $j+1$. They generate the permutation group. Transpositions are involutions: $\tau_j^2=1$. They satisfy the fundamental relation (known in another context as the Yang-Baxter equation),
\[ \tau_j\tau_{j+1}\tau_j = \tau_{j+1}\tau_j\tau_{j+1} .\]
The triple product $\tau_j\tau_{j+1}\tau_j$ is the transposition of $j$ and $j+2$.

For $\sigma_1$ and $\sigma_2$ any two permutations, the adjoint action of $\sigma_2$ on $\sigma_1$ is defined by $\sigma_2\circ\sigma_1:= \sigma_2\sigma_1\sigma_2^{-1}$. In particular, the adjoint action  of the transposition $\tau_j$ on $\sigma$ is:
\[ \tau_j\circ\sigma := \tau_j\sigma\tau_j.\]
Adjoint actions act on conjugacy classes (by definition) and hence preserve the number of cycles entering in the decomposition of the permutation into product of cycles. In particular, if $\sigma$ is a single cycle permutation so is $\tau_j\circ\sigma$. 

A contrario,  if $\sigma$ is a single cycle permutation then the product $\tau_j\sigma$ is not a single cycle permutation but it decomposes into two cycles, one including $j$, the other one $j+1$. Let us denote these cycles $\sigma_j^\pm$, so that $\tau_j\sigma=\sigma_j^-\,\sigma^+_j$. They are:
\begin{eqnarray*}
\sigma_j^-&:& j\to\sigma(j)\to\cdots\to \sigma^{-1}(j+1)\to j,\\
\sigma_j^+&:& j+1\to\sigma(j+1)\to\cdots\to\sigma^{-1}(j)\to j+1.
\end{eqnarray*}
A similar statement applies to the reversed product $\sigma\tau_j$.

Oriented labeled loops with $P$ points marked along the loop are in one-to-one correspondence with single cycle permutations, defined by following the ordering of the points along the loop. For $\sigma$ a single cycle permutation, the corresponding labeled loop is $1\to\sigma(1)\to\sigma^2(1)\to\cdots\to\sigma^{P-1}(1)\to 1$. As a consequence, if $\sigma$ is a single cycle permutation, then $\tau_j\sigma$ is associated to two disconnected loops with less marked points obtained by cutting the original loop $\sigma$ at the point $j$ and $j+1$.

To simplify notation, when needed, we shall denote (as usual) a single cycle permutation $\sigma$, or equivalently an oriented labeled loop, by specifying its ordered sequence of points $(i_1 i_2 \cdots i_P)$, with $i_k:=\sigma^{k-1}(1)$.

More information, and notations, shall be needed to construct the generating functions of expectation values of arbitrary loop. We prefer to postpone the introduction of these notations to Section \ref{sec:general-deformation}.
	
	\subsection{Polynomials} \label{sec:poly}

By construction, solving the problem of Section \ref{sec:problem} requires dealing with polynomials of degree at most one in each of its variables. 
We need to introduce an appropriate triangular decomposition of such polynomials, which amounts to introduce an order among monomials. We call it the hole decomposition. 

Let $Q({\bm x})$, with ${\bm x}:=(x_1,\cdots,x_P)$, be a polynomial in $P$ variables of degree at most one in each of its variables. It can be uniquely written as 
\begin{equation} \label{eq:hole1}
Q({\bm x})= \sum_{j=0}^{P} x_1\cdots x_j\, Q^o_{j+1}({\bm x}) ,
\end{equation}
with $Q^o_{j+1}({\bm x})$ independent of $x_1,\cdots,x_j,x_{j+1}$, and where by convention the first term is $Q^o_{1}({\bm x})$, independent of $x_1$, and the last term involve a coefficient $Q^o_{P+1}$ independent of all variables. 
The polynomial coefficients $Q^o_{j+1}({\bm x})$ are called the hole coefficients of $Q$. They are defined as
\[ 
Q^o_{j+1}({\bm x}) := \nabla_{x_j}\cdots\nabla_{x_1}Q({\bm x})\vert_{x_{j+1}=0},
\]
for $j=1,\cdots,P-1$. By convention, $Q^o_{1}({\bm x})=Q({\bm x})\vert_{x_1=0}$ and $Q^o_{P+1} := \nabla_{x_P}\cdots\nabla_{x_1}Q({\bm x})$.
	
Indeed, since $Q({\bm x})-Q({\bm x})\vert_{x_1=0}$ vanishes at $x_1=0$, we can write $Q({\bm x})-Q({\bm x})\vert_{x_1=0}=x_1 Q^r_1({\bm x})$, with $Q^r_1({\bm x})$ independent of $x_1$ and of degree at most one in all other variables. We can then expand $Q^r_1({\bm x})$ in power series of $x_2$ (up to degree one) and write $Q^r_1({\bm x})=Q^o_2({\bm x}) + x_2Q^r_2({\bm x})$ so that $Q({\bm x})-Q({\bm x})\vert_{x_1=0} = x_1\, Q^o_2({\bm x}) + x_1x_2\,Q^r_2({\bm x})$, with $Q^o_2({\bm x}):=\nabla_{x_1}Q({\bm x})\vert_{x_2=0}$ and  $Q^r_2({\bm x})$  independent of $x_1$ and $x_2$ and of degree at most one in all other variables. We can then expand $Q^r_2({\bm x})$ in power series of $x_3$, etc. Hence, by iteration, we can write,
\[ Q({\bm x})-Q({\bm x})\vert_{x_1=0} = \sum_{j=1}^k x_1\cdots x_j\, Q^o_{j+1}({\bm x}) + x_1\cdots x_{k+1}\,Q^r_{k+1}({\bm x}),\]
with both $Q^o_{j+1}({\bm x}):=\nabla_{x_j}\cdots\nabla_{x_1} Q({\bm x})\vert_{x_{j+1}=0}$ 
and  $Q^r_{j+1}({\bm x})$  independent of $x_1,\cdots,x_{j+1}$ and of degree one in all other variables. This can be pushed up to $j=P-1$. 

The construction below will be grounded on an interplay between the hole coefficients and multiple derivative of the polynomials. Thus, let us define strings of derivatives acting on polynomials,
\begin{equation}
 \mathfrak{D}_kQ({\bm x}):= \nabla_{x_k}\cdots\nabla_{x_1} Q({\bm x}) ,
 \end{equation}
which depends on the $P-k$ variables $x_{k+1},\cdots,x_P$. Clearly, we have
\[ \mathfrak{D}_kQ({\bm x}) = Q_{k+1}^o({\bm x})  + \sum_{j=k+1}^P x_{k+1}\cdots x_j\, Q_{j+1}^o({\bm x})  ,\]
or equivalently,
\begin{equation} \label{eq:D-hole}
\mathfrak{D}_kQ({\bm x})= Q_{k+1}^o({\bm x})  + x_{k+1} \mathfrak{D}_{k+1}Q({\bm x}) .
\end{equation}
	
{\bf Remark.} One may use an alternative language to describe the monomials entering into polynomials of degree at most one by declaring that, if the variable $x_j$ is raised to the power $0$ in that monomial then there is a hole at position $j$ and if $x_j$ is raised to the power $1$ then there is a particle at position $j$. Alternatively, we can speak about spins by declaring that the spin is down at position $j$ is there a hole at that position and up if there is particle. The hole decomposition consists then in introducing a basis of states in the spin system in which the spins from position $1$ to $j$ are up, with $j$ running along the chain.

In the following we set $\nabla_j:=\nabla_{x_j}$.

\section{Solution to Q-SSEP in the continuum} \label{sec:proof-solution}

The construction is going to be recursive since, for any labeled loops with $P$ points, the data entering the right hand side of the gluing condition \eqref{eq:gluing} involve loops with a smaller number of points. Thus we assume to have solved for the stationarity conditions (\ref{eq:harmonic},\ref{eq:bdry},\ref{eq:continu},\ref{eq:gluing})  for all labeled loop expectation values up to $P-1$ points and look for the $P$ point expectation values.

\subsection{Local moves} \label{sec:moves}

The first step consists at looking at the consequences of the exchange relations (\ref{eq:continu},\ref{eq:gluing}).

For any given labeled loop $\sigma$ and $[\sigma]({\bm x})$ its associated polynomial, let us single out its dependency on the pair of neighbour variables $x_j$ and $x_{j+1}$. Since $[\sigma]({\bm x})$ is a polynomial of degree at most one in each of its variables, it can be decomposed as,
\[
[\sigma]({\bm x}) = A_j(\sigma) + B_j(\sigma) x_j + C_j(\sigma) x_{j+1} + D_j(\sigma) x_jx_{j+1},
\]
where all coefficients $A_j, B_j, C_j, D_j$ are independent of $x_j$ and $x_{j+1}$ but depend on all other variables in ${\bm x}$ and on $\sigma$.

\begin{lemma}
The exchange relations (\ref{eq:continu},\ref{eq:gluing}) are equivalent to:
\begin{subequations} \label{eq:moves}
\begin{align}
A_j(\tau_j\circ\sigma) &= A_j(\sigma), \label{eq:moves-a}\\
B_j(\tau_j\circ\sigma) &= C_j(\sigma) + \nabla_{j}\nabla_{{j+1}}[\sigma_j^-\cdot\sigma_j^+], \label{eq:moves-b}\\
C_j(\tau_j\circ\sigma) &= B_j(\sigma) - \nabla_{j}\nabla_{{j+1}}[\sigma_j^-\cdot\sigma_j^+], \label{eq:moves-c}\\
D_j(\tau_j\circ\sigma) &= D_j(\sigma). \label{eq:moves-d}
\end{align}
\end{subequations}
The second and third relations (\ref{eq:moves-b},\ref{eq:moves-c}) are actually equivalent (because $\nabla_{j}\nabla_{{j+1}}[\sigma_j^-\cdot\sigma_j^+]$ is invariant upon replacing $\sigma$ by $\tau_j\circ\sigma$). 
Here and in the following, we set $[\sigma_j^-\cdot\sigma_j^+]({\bm x}):= [\sigma_j^-]({\bm x})[\sigma_j^+]({\bm x})$.
\end{lemma}

{\it Proof.}
The continuity condition at $x_j=x_{j+1}$ is equivalent to
\begin{eqnarray*} 
A_j(\tau_j\circ\sigma) &=& A_j(\sigma),\\
B_j(\tau_j\circ\sigma) + C_j(\tau_j\circ\sigma) &=& B_j(\sigma) + C_j(\sigma),\\
D_j(\tau_j\circ\sigma) &=& D_j(\sigma). 
\end{eqnarray*}
The Neumann like gluing condition is equivalent to
\[
\big(B_j(\tau_j\circ\sigma) - C_j(\tau_j\circ\sigma)\big) 
+ \big(B_j(\sigma) - C_j(\sigma)\big) = 2\,\nabla_{j}\nabla_{{j+1}}[\sigma_j^-\cdot\sigma_j^+].
\]
Solving the above equations for $B_j$ and $C_j$ yields eqs.\eqref{eq:moves}. \cqfd

These relations induce moves inside the conjugacy class of single cycle permutations.

\begin{lemma} 
The moves \eqref{eq:moves} define an action of the permutation group into the set of polynomials of degree one in each $P$ variables, provided the polynomials $[\sigma_j^\pm]$ in less variables satisfy the exchange relations \eqref{eq:moves}.
\end{lemma}

{\it Proof.} We have to check the fundamental relation of the permutation group
\[ \tau_j\tau_{j+1}\tau_j = \tau_{j+1}\tau_j\tau_{j+1} .\]
To verify it, we pick a triplet of points $j$, $j+1$, $j+2$, and decompose the polynomials $[\sigma]({\bm x})$ accordingly:
\begin{eqnarray*}
[\sigma]({\bm x}) &=& a(\sigma)
+ b_0(\sigma) x_j + b_1(\sigma)x_{j+1} + b_2(\sigma) x_{j+2} \\
&~& + c_{01}(\sigma) x_jx_{j+1} + c_{02}(\sigma) x_jx_{j+2} + c_{12}(\sigma) x_{j+1}x_{j+2}
+ d(\sigma) x_jx_{j+1}x_{j+2},
\end{eqnarray*}
with all coefficients $a$, $b_0$, etc., dependent on all variables except the triplet $x_j, x_{j+1}, x_{j+2}$. To compute the actions of $\tau_j\tau_{j+1}\tau_j$ and $\tau_{j+1}\tau_j\tau_{j+1}$ on polynomials we have to apply successively the moves  \eqref{eq:moves} for the pairs $(j,j+1)$ and $(j+1,j+2)$, in the appropriate order. Since $a$ and $d$ are both invariant under $\tau_j$ and $\tau_{j+1}$, the action of the group generated by $\tau_j$ and $\tau_{j+1}$ on them is clearly compatible. Next, for the pair $(j,j+1)$ we have 
\begin{eqnarray*}
b_0(\tau_j\circ\sigma)+c_{02}(\tau_j\circ\sigma) x_{j+2} &=& b_1(\sigma)+c_{12}(\sigma) x_{j+2} 
+\nabla_{j}\nabla_{{j+1}}\big[\sigma_j^-\cdot\sigma_j^+\big],\\
b_1(\tau_j\circ\sigma)+c_{12}(\tau_j\circ\sigma) x_{j+2} &=& b_0(\sigma)+c_{02}(\sigma) x_{j+2} 
-\nabla_{j}\nabla_{{j+1}}\big[\sigma_j^-\cdot\sigma_j^+\big],
\end{eqnarray*}
while $a$, $b_2$, $c_{01}$, $d$  are $\tau_j$-invariant. For the pair $(j+1,j+2)$ we similarly have 
\begin{eqnarray*}
b_1(\tau_{j+1}\circ\sigma)+c_{01}(\tau_j\circ\sigma) x_{j} &=& b_2(\sigma)+c_{02}(\sigma) x_{j} 
+\nabla_{j+1}\nabla_{j+2}\big[\sigma_{j+1}^-\cdot\sigma_{j+1}^+\big],\\
b_2(\tau_{j+1}\circ\sigma)+c_{02}(\tau_j\circ\sigma) x_{j} &=& b_1(\sigma)+c_{12}(\sigma) x_{j} 
-\nabla_{{j+1}}\nabla_{{j+2}}\big[\sigma_{j+1}^-\cdot\sigma_{j+1}^+\big],
\end{eqnarray*}
while $a$, $b_0$, $c_{12}$, $d$  are $\tau_{j+1}$-invariant.  Using these relations, we can evaluate all coefficients $b_k(\tau_j\tau_{j+1}\tau_j\circ\sigma)$ and $c_{kl}(\tau_j\tau_{j+1}\tau_j\circ\sigma)$, as well as all coefficients $b_k(\tau_{j+1}\tau_{j}\tau_{j+1}\circ\sigma)$ and $c_{kl}(\tau_{j+1}\tau_{j}\tau_{j+1}\circ\sigma)$. Since $\tau_{j+1}\tau_{j}\tau_{j+1}=\tau_j\tau_{j+1}\tau_j$ is the permutation $\tau_{j;j+2}$ exchanging $j$ and $j+2$, these computations yield formulas of the form
\begin{eqnarray*}
b_0(\tau_j\tau_{j+1}\tau_j\circ\sigma) &=& b_2(\sigma) + \mathrm{"extra\ terms\ (a)"}, \\
b_0(\tau_{j+1}\tau_{j}\tau_{j+1}\circ\sigma) &=& b_2(\sigma) + \mathrm{"extra\ terms\ (b)"}, 
\end{eqnarray*}
and similar formulas for $b_1$, $b_2$ and for $c_{01}$, $c_{02}$, $c_{12}$. The compatibility conditions with the fundamental relation in the permutation group are then that the "extra terms" of type (a) and (b) should coincide. These conditions reduce to the following one (equivalent to a set of two relations because it is polynomial of degree one in $y$)
\begin{eqnarray} \label{eq:compatible}
&&\nabla_{j}\nabla_{{j+1}}\big[(\tau_{j;j+2}\circ\sigma)_j^-\cdot(\tau_{j;j+2}\circ\sigma)_j^+\big]\vert_{x_{j+2}=y}\nonumber\\
&+& \nabla_{{j+1}}\nabla_{{j+2}}\big[(\tau_{j;j+2}\circ\sigma)_{j+1}^-\cdot(\tau_{j;j+2}\circ\sigma)_{j+1}^+\big]\vert_{x_j=y}\\
&=& \nabla_{j}\nabla_{{j+1}}\big[\sigma_j^-\cdot \sigma_j^+\big]\vert_{x_{j+2}=y}
+ \nabla_{{j+1}}\nabla_{{j+2}}\big[ \sigma_{j+1}^-\cdot \sigma_{j+1}^+\big]\vert_{x_j=y},\nonumber
\end{eqnarray}
with $y$ a book-keeping variable (with $x_{j-1}<y<x_{j+3}$). 
To finish the proof of the compatibility of the moves \eqref{eq:moves}, we have to show that this relation is satisfied provided the exchange relations for cycles with less points are fulfilled. We remark that $(\tau_{j;j+2}\circ\sigma)_j^\pm$ is identical to $\sigma_{j+1}^\pm$ up to permuting $j$ and $j+2$, since
\[ \tau_{j;j+2}\circ(\tau_{j+1}\sigma) = \tau_j(\tau_{j;j+2}\circ\sigma) ,\]
because $\tau_{j;j+2}\tau_j\tau_{j;j+2}=\tau_{j+1}$. Thus we can compare $\nabla_{j}\nabla_{{j+1}}\big[(\tau_{j;j+2}\circ\sigma)_j^-\cdot(\tau_{j;j+2}\circ\sigma)_j^+\big]\vert_{x_{j+2}=y}$ and $\nabla_{{j+1}}\nabla_{{j+2}}\big[ \sigma_{j+1}^-\cdot \sigma_{j+1}^+\big]\vert_{x_j=y}$ using the exchange relations \eqref{eq:moves} (applied to $\sigma_{j+1}^\pm$ depending to which component $j$ belongs to) and write
\begin{eqnarray*}
&& \hskip +1.5 truecm \nabla_{j}\nabla_{{j+1}}\big[(\tau_{j;j+2}\circ\sigma)_j^-\cdot(\tau_{j;j+2}\circ\sigma)_j^+\big]\vert_{x_{j+2}=y} \\
&=& \nabla_{{j+1}}\nabla_{{j+2}}\big[ \sigma_{j+1}^-\cdot \sigma_{j+1}^+\big]\vert_{x_j=y} 
 + \nabla_{j}\nabla_{{j+1}}\nabla_{{j+2}}[\sigma_{j;j+1}^{(0)}\cdot\sigma_{j;j+1}^{(1)}\cdot\sigma_{j;j+1}^{(2)}] ,
\end{eqnarray*}
where the three cycles $\sigma_{j;j+1}^{(\alpha)}$, containing $j+\alpha$ respectively, $\alpha = 0,1,2$, are those entering the decomposition of the product $\tau_{j+1}\tau_j\sigma$. Similarly, using again \eqref{eq:moves}, we have
\begin{eqnarray*}
&& \hskip +1.5 truecm \nabla_{{j+1}}\nabla_{{j+2}}\big[(\tau_{j;j+2}\circ\sigma)_{j+1}^-\cdot(\tau_{j;j+2}\circ\sigma)_{j+1}^+\big]\vert_{x_j=y} \\
&=& \nabla_{j}\nabla_{{j+1}}\big[\sigma_j^-\cdot \sigma_j^+\big]\vert_{x_{j+2}=y}
 - \nabla_{j}\nabla_{{j+1}}\nabla_{{j+2}}[\sigma_{j;j+1}^{(0)}\cdot\sigma_{j;j+1}^{(1)}\cdot\sigma_{j;j+1}^{(2)}] .
\end{eqnarray*}
Adding the two last equations yields \eqref{eq:compatible} and hence proves the compatibility of the moves \eqref{eq:moves}, provided that the lower point correlation functions do satisfy the exchange relations (\ref{eq:continu},\ref{eq:gluing}).
\cqfd

{\bf Remark.} If $[\sigma]({\bm x})$ is a family of polynomials satisfying the moves \eqref{eq:moves}, then adding to them polynomials $Q_\sigma({\bm x})$ invariant by simultaneous transpositions of the variables $x_j$ and $x_{j+1}$ and of their positions inside the loop provides another solution. 

This follows from the fact that such polynomials $Q^\sigma({\bm x})$ are solutions of the equations \eqref{eq:moves} without the non-linear terms in the right hand side. Indeed, because they are invariant under the simultaneous transformations $\sigma \to \tau_j\circ \sigma$, $x_j\to x_{j+1}$, $ x_{j+1}\to x_j$, such polynomials can be decomposed as
\[ Q^\sigma({\bm x}) = a_j(\sigma) + b_j(\sigma) x_j + c_j(\sigma) x_{j+1} + d_j(\sigma) x_jx_{j+1} ,
\]
with $a_j(\tau_j\circ\sigma) = a_j(\sigma)$, $b_j(\tau_j\circ\sigma) = c_j(\sigma)$, $c_j(\tau_j\circ\sigma) = b_j(\sigma)$, and $d_j(\tau_j\circ\sigma) = d_j(\sigma)$.
\medskip

\subsection{Forward propagation} \label{sec:forward}

The second step consists at examining the consequences of the boundary conditions \eqref{eq:bdry} and their compatibility with the exchange relations (\ref{eq:continu},\ref{eq:gluing}). Indeed, on one hand, although they allow to move within the single cycle conjugacy classes, the exchange relations (\ref{eq:continu},\ref{eq:gluing}) do not fully specify the polynomials $[\sigma]({\bm x})$, as we noted above. On the other hand, the moves \eqref{eq:moves} that the exchange relations (\ref{eq:continu},\ref{eq:gluing}) they imply do not preserve the boundary condition unless the polynomials $[\sigma]({\bm x})$ satisfy peculiar conditions. The determination of the polynomials $[\sigma]({\bm x})$ will come from using the interplay between the boundary conditions \eqref{eq:bdry}  and the moves \eqref{eq:moves}.

The moves \eqref{eq:moves}, and in particular the second relation \eqref{eq:moves-b}, $B_j(\sigma) = C_j(\tau_j\circ\sigma) + \nabla_{j}\nabla_{{j+1}}\big[\sigma_j^-\cdot\sigma_j^+\big]$, which we rewrite as,
\beq \label{eq:move!!}
\nabla_{j}[\sigma]({\bm x})\vert_{x_{j+1}=0} = \nabla_{{j+1}}[\tau_j\circ\sigma]({\bm x})\vert_{x_j=0} + \nabla_{j}\nabla_{{j+1}}\big[\sigma_j^-\cdot\sigma_j^+\big]({\bm x}),
\eeq
allow to move within the conjugacy class of single loop but also along the chain, as it exchanges derivatives and evaluations at $0$ at points $x_j$ and $x_{j+1}$. The strategy to determine the polynomials $[\sigma]({\bm x})$ will consist in moving recursively the evaluation at $0$ to the point $x_1$ where it vanishes thanks to the boundary condition.

Given a single loop $\sigma$, let us decompose $[\sigma]({\bm x})$ on the pair of the two first variables $x_1$ and $x_2$ as $[\sigma]({\bm x}) = A_1(\sigma) + B_1(\sigma) x_1 + C_1(\sigma) x_{2} + D_1(\sigma) x_1x_{2}$,
with $A_1,\ B_1,\ C_1,\ D_1$ independent of $x_1$ and $x_2$. The boundary $[\sigma]({\bm x})\vert_{x_1=0}=0$ then imposes that $A_1=0$ and $C_1=0$, so that
\[
[\sigma]({\bm x}) = x_1\big( B_1(\sigma) + D_1(\sigma) x_{2}\big).
\]
The move \eqref{eq:move!!} then tell us that
\[ 
B_1(\sigma)({\bm x}) = \nabla_{1}[\sigma]({\bm x}) \vert_{x_2=0} = \nabla_{2}\nabla_{1}\big[\sigma_1^-\cdot\sigma_1^+\big]({\bm x}).
\]
Note that $B_1(\sigma)=B_1(\tau_1\circ\sigma)$.
Thus, $\nabla_{1}\sigma({\bm x}) \vert_{x_2=0}$ is determined since, by the induction hypothesis, we assumed having solved for the single loop correlation functions with $P'<P$ marked points and $\sigma_1^\pm$ are single loops with less than $P$ points. We can iterate this procedure to get

\begin{lemma} 
By iterating the move \eqref{eq:move!!}, for any single loop $\sigma$ and for $j=1,\cdots, P-1$, we have,
\begin{eqnarray} \label{eq:propag}
&& ~~~~~~~~ (\nabla_{1}\cdots\nabla_{j})[\sigma]({\bm x})\vert_{x_{j+1}=0} \\
&=& (\nabla_{{1}}\cdots\nabla_{j+1})\sum_{k=1}^{j}\big[ (\tau_{k+1}\cdots\tau_j\circ\sigma)^-_k\cdot (\tau_{k+1}\cdots\tau_j\circ\sigma)^+_k\big]({\bm x}), \nonumber 
\end{eqnarray}
where the last term in the sum is $[\sigma_j^-\cdot\sigma_j^+\big]({\bm x})$ by convention. 
Notice that, for all $k>1$, $(\tau_{k+1}\cdots\tau_j\circ\sigma)^\pm_k$ are loops with less than $P$ points.
\end{lemma}

{\it Proof.} 
We shall prove \eqref{eq:propag} by induction. For $j=1$, we already proved that
\[ \nabla_{1}[\sigma]({\bm x})\vert_{x_{2}=0} = \nabla_{{2}}\nabla_{1}\big[\sigma_1^-\cdot\sigma_1^+\big]({\bm x}).\]
To understand how the induction is at work, let us now do it for $j=2$. We look at the move \eqref{eq:move!!} for $j=2$, 
\[
\nabla_{2}[\sigma]({\bm x})\vert_{x_{3}=0} = \nabla_{{3}}[\tau_2\circ\sigma]({\bm x})\vert_{x_2=0}
 + (\nabla_{{3}}\nabla_{2})\big[\sigma_2^-\cdot\sigma_2^+\big]({\bm x}).
\]
Taking the derivative w.r.t. $x_1$ yields,
\[ 
\nabla_{2}\nabla_{1}[\sigma]({\bm x})\vert_{x_{3}=0} = \nabla_{{3}}\nabla_{1} [\tau_2\circ\sigma]({\bm x})\vert_{x_2=0} + (\nabla_{{3}}\nabla_{2}\nabla_{1})\big[\sigma_2^-\cdot\sigma_2^+\big]({\bm x}).
\]
Now, $\nabla_{1} [\tau_2\circ\sigma]({\bm x})\vert_{x_2=0}$ has been previously identified as $(\nabla_{{2}}\nabla_{1})\big[(\tau_2\circ\sigma)_1^-\cdot(\tau_2\circ\sigma)_1^+\big]({\bm x})$. Hence we get
\[ 
\nabla_{2}\nabla_{1} [\sigma]({\bm x})\vert_{x_{3}=0} = (\nabla_{{3}}\nabla_{{2}}\nabla_{1})\big([(\tau_2\circ\sigma)_1^-\cdot(\tau_2\circ\sigma)_1^+]({\bm x}) + [\sigma_2^-\cdot\sigma_2^+]({\bm x})\big),
\]
as in \eqref{eq:propag} for $j=2$. For higher $j$, we assume \eqref{eq:propag} to be valid up $j-1$ for all single loops. We then start with \eqref{eq:move!!}, 
\[ 
\nabla_{j}[\sigma]({\bm x})\vert_{x_{j+1}=0} = \nabla_{j+1}[\tau_j\circ\sigma]({\bm x})\vert_{x_j=0} + \nabla_{j}\nabla_{{j+1}}\big[\sigma_j^-\cdot\sigma_j^+\big]({\bm x}),
\]
and act on it with the successive derivatives $\nabla_{{j-1}}\cdots\nabla_{1}$ to get
\begin{eqnarray*}
&&\nabla_{j}\nabla_{{j-1}}\cdots\nabla_{1}[\sigma]({\bm x})\vert_{x_{j+1}=0} \\
&=&  \nabla_{{j+1}}\nabla_{{j-1}}\cdots\nabla_{1}[\tau_j\circ\sigma]({\bm x})\vert_{x_j=0} 
 + \nabla_{{j+1}}\nabla_{j}\nabla_{{j-1}}\cdots\nabla_{1}\big[\sigma_j^-\cdot\sigma_j^+\big]({\bm x}),
\end{eqnarray*}
Now, by the induction hypothesis $\nabla_{{j-1}}\cdots\nabla_{1}[\tau_j\circ\sigma]({\bm x})\vert_{x_j=0}$ is given by \eqref{eq:propag} but for $j-1$. Inserting this formula in the above equation yields \eqref{eq:propag}. \cqfd

Notice that $(\nabla_{j}\cdots\nabla_{1})[\sigma]({\bm x})\vert_{x_{j+1}=0} $ is independent of $x_1,\cdots,x_{j+1}$ but depends on the other variables $x_{j+2},\cdots,x_P$ (both sides of \eqref{eq:propag} are indeed independent of $x_1,\cdots, x_{j+1}$).

{\bf Remark.} Since $[\sigma]({\bm x})$ are polynomials of degree at most one in each of its variables, evaluating at the origin and taking derivatives are two basic and fundamental operations. Knowing their actions on $[\sigma]$ determine it completely. Eq.\eqref{eq:move!!} can be interpreted as a move exchanging derivative at point $x_j$ and evaluation at $x_{j+1}=0$ with evaluation at $x_j=0$ and derivative at $x_{j+1}$. In terms of the particle/hole language we eluded to in Section \ref{sec:poly}, equation \eqref{eq:move!!} codes for the moves exchanging particle and hole at nearby positions $j$ and $j+1$. The strategy for proving \eqref{eq:propag} consists in moving the holes to position $1$ where it is annihilated.

	\subsection{Synthesis}
	
The end of the proof consists now in using the formula \eqref{eq:propag} to reconstruct the correlation functions $[\sigma]({\bm x})$. 
In the l.h.s. of \eqref{eq:propag} we recognize the hole coefficients of the polynomials $[\sigma]({\bm x})$.

Recall that, according to the hole decomposition as we defined it in Section \ref{sec:poly},  the polynomials $[\sigma]({\bm x})$ can be written as
\[ 
[\sigma]({\bm x}) =  [\sigma]^o_1({\bm x})  + \sum_{j=1}^{P-1} x_1\cdots x_j\, [\sigma]^o_{j+1}({\bm x}) + x_1\cdots x_P\,[\sigma]^o_{P+1},
\]
where $[\sigma]_{k+1}^o$ are the hole coefficients of $[\sigma]$ with $[\sigma]^o_1({\bm x}):= [\sigma]({\bm x})\vert_{x_{1}=0}$ and 
\[ [\sigma]^o_{j+1}({\bm x}):=\nabla_{j}\cdots\nabla_{1}[\sigma]({\bm x})\vert_{x_{j+1}=0} ,\]
for $j=1,\cdots, P-1$, and $[\sigma]^o_{P+1}$ a constant independent on all variables. 
These polynomial coefficients $[\sigma]_{j+1}^o$, which we called the hole coefficients of $[\sigma]({\bm x})$, are independent of the first $j+1$ variables $x_1,\cdots,x_{j+1}$. 

The boundary condition \eqref{eq:bdry} further constraints this hole decomposition. We have $[\sigma]^o_1({\bm x})=0$ since $[\sigma]({\bm x})\vert_{x_{1}=0}=0$ by \eqref{eq:bdry}. Similarly, we have $[\sigma]^o_{P+1}=-[\sigma]^o_{P}$ by the boundary condition $[\sigma]({\bm x})\vert_{x_P=1}=0$. This boundary condition also imposes that $[\sigma]^o_{j+1}({\bm x})\vert_{x_P=1}=0$, for $j=1,\cdots,P-2$.

Collecting the information about this presentation of $[\sigma]({\bm x})$, about the boundary conditions and about the formula \eqref{eq:propag} for the hole coefficients of $[\sigma]({\bm x})$, we get

\begin{theorem} \label{th:solution}
Assuming the locality conjecture of Section \ref{sec:problem} to be true, 
for any single loop $\sigma$ with $P$ marked points, the solution of the stationarity conditions  (\ref{eq:harmonic},\ref{eq:bdry},\ref{eq:continu},\ref{eq:gluing}), is given by
\beq \label{eq:solution}
[\sigma]({\bm x}) = \sum_{j=1}^{P-2} x_1\cdots x_j\, [\sigma]^o_{j+1}({\bm x}) + x_1\cdots x_{P-1}(1-x_P)\,[\sigma]^o_{P},
\eeq
with $[\sigma]^o_j({\bm x}):=\nabla_{j}\cdots\nabla_{1}[\sigma]({\bm x})\vert_{x_{j+1}=0}$ given by, for $j=1,\cdots, P-1$,
\beq \label{eq:sigma0}
[\sigma]^o_{j+1}({\bm x})=(\nabla_{{j+1}}\cdots\nabla_{1})\sum_{k=1}^{j}\big[ (\tau_{k+1}\cdots\tau_j\circ\sigma)^-_k\cdot (\tau_{k+1}\cdots\tau_j\circ\sigma)^+_k\big]({\bm x}).
\eeq
The last term in the sum is $[\sigma_j^-\cdot\sigma_j^+\big]({\bm x})$ by convention.
\end{theorem}

{\it Proof.} This theorem is direct consequence of the analysis of the previous Sections \ref{sec:moves} and \ref{sec:forward}. Of course it relies on assuming the locality conjecture. 
\cqfd

\medskip

A first (immediate) consequence of \eqref{eq:solution} is that the top hole coefficients are actually independent of $\sigma$.

\begin{corrollary} 
The top hole coefficients are independent of $\sigma$, i.e. $[\sigma]_{P+1}^o=-[\sigma]_{P}^o:=C_{P}$ for all $\sigma$, but not on $P$, the number of points. They satisfy the recurrence relation
\begin{equation} \label{eq:Ctop}
C_{P} = -\sum_{n=1}^{P-1} C_{n}C_{P-n} .
\end{equation}  
They are equal to the alternating Catalan numbers (up to the multiplicative factor $(\Delta n)^P$ which we set to $1$ by convention).
\end{corrollary} 

{\it Proof.} This is proved by induction on $P$. Recall that $[\sigma]_{P+1}^o=\nabla_P\cdots\nabla_1[\sigma]$. By the boundary conditions \eqref{eq:bdry}, we have $[\sigma]_{P+1}^o=-[\sigma]_{P}^o$. The later $[\sigma]_{P}^o$ is given by formula \eqref{eq:sigma0} with $j=P-1$, which thus involves the derivatives with respect to all $P$ points. The r.h.s. of \eqref{eq:sigma0} then involves the derivatives of loop correlation functions w.r.t. all their points. These derivatives are thus independent of the loop (but not of the number of points) by the induction hypothesis. Hence, $[\sigma]_{P+1}^o$ is independent of the labeled loop $\sigma$ and satisfy the recurrence relation \eqref{eq:Ctop}.
\cqfd

The above recurrence relation is solved by introducing a generating function:
\begin{equation} \label{eq:C-catalan}
\mathfrak{c}(z):=\sum_{N\geq 0} z^N\, C_{N+1}=C_{1} +z\,C_{2}+\cdots 
\end{equation}
Eq.\eqref{eq:Ctop} then reads  $z\mathfrak{c}(z)^2+\mathfrak{c}(z)-C_{1}=0$. Recall that $C_{1} = \nabla_1[(1)]=\Delta n$ and that we choose to set $\Delta n=1$, so that \eqref{eq:Ctop} becomes $z\mathfrak{c}(z)^2+\mathfrak{c}(z)-1=0$. Hence,
\[\mathfrak{c}(z):=\frac{\sqrt{1+4z}-1}{2z}=1-z+2z^2-5z^3+14z^4+\cdots\]
is the solution of $z\mathfrak{c}^2+\mathfrak{c}-1=0$ that is non-singular at $0$, which is known to be the generating function for (alternating sign) Catalan numbers.

{\bf Remark.} What has been (really) proved in the above theorem is that \eqref{eq:solution} is the only possible solution assuming the locality property. The conditions (i) and (ii), and in particular the boundary conditions, are satisfied by construction.  Formula \eqref{eq:propag} on which equation \eqref{eq:solution} is based is extracted by taking derivatives of the exchange relations (iii), and it is therefore a priori not equivalent to it, although it may be a posteriori (and it is conjecturally equivalent to it). We have checked directly on various exemples that indeed \eqref{eq:solution} is a solution of all exchange relations (\ref{eq:continu},\ref{eq:gluing}), thus checking the locality conjecture in few cases. See below Section \ref{sec:ex1} and Appendix \ref{sec:exchange-Tn}.

	\subsection{Exemples} \label{sec:ex1}
	
Let us compute the first few loop correlation functions. Since these correlation functions are invariant by reversing the loop orientation, there are $(P-1)!/2$ inequivalent loop correlation functions with $P$ marked points.

By construction, for $P=1$, we have $\nabla_1[(1)]=(\Delta n)$ with $\Delta n =1$ by convention.

For $P=2$, the only single loop corresponds to the permutation $\tau_{1}=(12)$ of the two points. We have $[(12)](x_1,x_2)]= (\Delta n)^2\, x_1(1-x_2)$.

For $P=3$, there is only one independent loop configuration. It is the cyclic permutation of $(123)$. As in  \eqref{eq:solution}, we decompose $[(123)]({\bm x})$ as (with $c_j^o:=[\sigma]_{j+1}^o$ for $\sigma=(123)$)
\[ [(123)](x_1,x_2,x_3)=x_1c_1^o + x_1x_2(1-x_3)c_2^o, \]
with
\begin{eqnarray*}
c_1^o&=& \nabla_1\nabla_2[(1)][(23)]=(\Delta n)^3 (1-x_3),\\
c_2^o &=& \nabla_1\nabla_2\nabla_3\big( [(3)][(12)]+[(1)][(23)]\big) = (\Delta n)^3 (-2).
\end{eqnarray*}
Thus we get
\begin{equation}
 [(123)](x_1,x_2,x_3)= (\Delta n)^3\, x_1(1-2x_2)(1-x_3) ,
 \end{equation}
as expected (see \cite{BJ2019}).

For $P=4$, there are three independent loop configurations corresponding to the three permutations $(1234)$, $(1324)$ and $(1342)$. Let us compute them using \eqref{eq:solution}. For $(1234)$, we write (with $c_j^o:=[\sigma]_{j+1}^o$ for $\sigma=(1234)$)
\[ [(1234)]({\bm x}) = x_1c_1^o + x_1x_2c_2^o + x_1x_2x_3(1-x_4)c_3^o,\]
with
\begin{eqnarray*}
c_1^o&=& \nabla_1\nabla_2[(1)][(234)]=(\Delta n)^4 (1-2x_3)(1-x_4),\\
c_2^o &=& \nabla_1\nabla_2\nabla_3\big( [(2)][(134)]+[(24)][(13)]\big) = (\Delta n)^4 (-3)(1-x_4),\\
c_3^o&=& \nabla_1\nabla_2\nabla_3\nabla_4\big( [(134)][(2)]+[(24)][(13)]+[(3)][(124)] \big)=(\Delta n)^4 (+5).
\end{eqnarray*}
Thus we get
\begin{equation}
 [(1234)](x_1,x_2,x_3,x_4)= (\Delta n)^4\, x_1(1-3x_2-2x_3+5x_2x_3)(1-x_4) .
 \end{equation}
For $(1324)$, we write (with $c_j^o:=[\sigma]_{j+1}^o$ for $\sigma=(1324)$)
\[ [(1324)]({\bm x}) = x_1c_1^o + x_1x_2c_2^o + x_1x_2x_3(1-x_4)c_3^o,\]
with
\begin{eqnarray*}
c_1^o&=& \nabla_1\nabla_2[(14)][(23)]=(\Delta n)^4 (1-x_3)(1-x_4),\\
c_2^o &=& \nabla_1\nabla_2\nabla_3\big( [(3)][(124)]+[(2)][(134)]\big) = (\Delta n)^4 (-4)(1-x_4),\\
c_3^o&=& \nabla_1\nabla_2\nabla_3\nabla_4\big([(143)][(2)]+[(2)][(143)]+[(32)][(14)] \big)=(\Delta n)^4 (+5).
\end{eqnarray*}
Thus we get
\begin{equation}
 [(1324)](x_1,x_2,x_3,x_4)= (\Delta n)^4\, x_1(1-4x_2-x_3+5x_2x_3)(1-x_4) .
 \end{equation}
For $(1342)$, we write (with $c_j^o:=[\sigma]_{j+1}^o$ for $\sigma=(1342)$)
\[ [(1342)]({\bm x}) = x_1c_1^o + x_1x_2c_2^o + x_1x_2x_3(1-x_4)c_3^o,\]
with
\begin{eqnarray*}
c_1^o&=& \nabla_1\nabla_2[(1)][(234)]=(\Delta n)^4 (1-2x_3)(1-x_4),\\
c_2^o &=& \nabla_1\nabla_2\nabla_3\big( [(21)][(34)]+[(2)][(134)]\big) = (\Delta n)^4 (-3)(1-x_4),\\
c_3^o&=& \nabla_1\nabla_2\nabla_3\nabla_4\big( [(14)][(23)]+[(142)][(3)]+[(3)][(142)] \big)=(\Delta n)^4 (+5).
\end{eqnarray*}
Thus we get
\begin{equation}
 [(1342)](x_1,x_2,x_3,x_4)= (\Delta n)^4\, x_1(1-3x_2-2x_3+5x_2x_3)(1-x_4) .
 \end{equation}

For $P=5$, there are $4!/2=12$ independent loop configurations. We compute the correlation function for the circular cyclic permutation $(12345)$ using \eqref{eq:solution} -- see below -- and checked the exchange relation \eqref{eq:moves} for it. We have
\begin{equation} \label{eq:Dregular5}
[(12345)]({\bm x})=(\Delta n)^4\, x_1 (1-4 x_2-3 x_3-2 x_4+9 x_2 x_3+7 x_2 x_4+5 x_3 x_4-14 x_2 x_3 x_4)( 1-x_5)
\end{equation}
It is easy to checked that it satisfies the reversing symmetry ${\bm x} \to {\bm x}^{rev}=1-{\bm x}$.
Eqs.\eqref{eq:moves} for the pair $(1;2)$ is satisfied by construction, eqs.\eqref{eq:moves} for the pairs $(4;5)$ and $(3;4)$ are respectively equivalent to those for the pairs $(1;2)$ and $(2;3)$, by the reversing symmetry. Thus we only have to check \eqref{eq:moves} for the pair $(2;3)$ for $P=5$. Using \eqref{eq:solution}, we get
\begin{equation} \label{eq:deformedT32}
[(13245)]({\bm x})= (\Delta n)^4\, x_1(1 -6x_2-x_3-2x_4+9x_2x_3+10 x_2x_4+2x_3x_4-14x_2x_3x_4)(1-x_5).
\end{equation}
It is then easy to verify that \eqref{eq:moves} are indeed fulfilled.

{\bf Remark.}
Quite generally, if the exchange relations \eqref{eq:moves} hold, then
\begin{equation*}
 [\tau_1\circ \sigma]({\bm x}) = [\sigma]({\bm x}), 
 \end{equation*}
for any loop $\sigma$. Indeed, using the boundary condition at $x_1=0$ for both $[\tau_1\circ \sigma]$ and $[\sigma]$, and the moves \eqref{eq:moves}, we have $A_1(\tau_1\circ \sigma)=A_1(\sigma)=0$, $C_1(\tau_1\circ \sigma)=C_1(\sigma)=0$, as well as $B_1(\tau_1\circ \sigma)= \nabla_{1}\nabla_{2}[\sigma_1^-\cdot\sigma_1^+]=B_1(\sigma)$ and $D_1(\tau_1\circ \sigma)=D_1(\sigma)$. Hence $[\tau_1\circ \sigma]=[\sigma]$.

\section{The regular loop} \label{sec:regular}

A special role is played by the circular cyclic permutation $(123 \cdots P)$, which we denote as $\omega_P$. For this permutation, the orders along the chain and along the loop coincide. 

	\subsection{Recurrence relations}

We are interested in the regular loop expectation values $[\omega_P]({\bm x})$.
For $0\leq k\leq P$, let us define multiple derivatives as in Section \ref{sec:poly}
\begin{equation}
D_{k|P-k}({\bm x}_{\geq k+1}):= \nabla_{k}\cdots\nabla_{1}[\omega_P]({\bm x}),
 \end{equation}
which depend on the $P-k$ variables ${\bm x}_{\geq k+1}:=(x_{k+1},\cdots,x_P)$. In particular, $D_{0|P}:=[\omega_P]$ and $D_{P|0}=C_{P}$ is independent on all variables and equal to the alternating Catalan numbers as defined in \eqref{eq:C-catalan}. We have

\begin{lemma}
The hole coefficients of the regular loops $[\omega_P]$ with $P$ points are given by
\begin{equation} \label{eq:hole-regular}
[\omega_P]^o_{j+1} = \sum_{n=1}^{j}  D_{n|0} D_{j+1-n|P-j-1}({\bm x}_{\geq j+2})
 \end{equation}
or alternatively, the regular loop correlation functions are :
\begin{eqnarray}  \label{eq:cP1}
[\omega_P]({\bm x}) &=& \sum_{j=1}^{P-2} x_1\cdots x_{j} \sum_{n=1}^{j}  D_{n|0} D_{j+1-n|P-j-1}({\bm x}_{\geq j+2})\\
&& + x_1\cdots x_{P-1}(1-x_P)\sum_{n=1}^{P-1}D_{n|0}D_{P-n|0}. \nonumber
\end{eqnarray}
This yields recurrence relations for the multiple derivatives $D_{k|P-k}$ and the hole coefficients of $[\omega_P]({\bm x})$.
\end{lemma}

{\it Proof.}
The proof relies on the fact that cutting the loop $\omega_P=(12\cdots P)$ according to the formula (\ref{eq:solution},\ref{eq:sigma0}) produces smaller loops which also preserve the ordering of the points. Let us do it progressively. First, we have:
\begin{eqnarray*}
(\omega_P)_j^- &:& (j) ,\\
(\omega_P)_j^+ &:& (12 \cdots, j-1,j+1, \cdots P) .
\end{eqnarray*}
Let us now look at the permutation $\tau_{k+1}\cdots\tau_j\circ\omega_P$ entering \eqref{eq:sigma0}. By adjoint action the product $\tau_{k+1}\cdots\tau_j$ amounts to slide the point $k+1$ along the loop so that $\tau_{k+1}\cdots\tau_j\circ\omega_P$ is the following cycle:
\[ 
(12\cdots k,k+2,k+3, \cdots ,j+1, k+1, j+2 \cdots P) .
\]
By breaking it at the pair $(k;k+1)$ according to formula \eqref{eq:sigma0}, we get the following smaller loops
\begin{eqnarray*}
(\tau_{k+1}\cdots\tau_j\circ\omega_P)_k^- &:& (k, k+2, k+3,\cdots, j+1) ,\\
(\tau_{k+1}\cdots\tau_j\circ\omega_P)_k^+ &:& (12 \cdots, k-1, k+1, j+2 ,\cdots P) .
\end{eqnarray*}
In both of these loops, the order along the loop coincides with that along the chain (even if the labelling is not regular).
Hence, the formula \eqref{eq:sigma0} only involves correlation functions of well ordered loops, of different sizes, for which the order along the loop and along the chain coincide. Applying formula (\ref{eq:solution},\ref{eq:sigma0}) gives \eqref{eq:cP1}.
\cqfd

Another way of writing equation \eqref{eq:cP1} is (by renaming the indices)
\begin{eqnarray} \label{eq:cP2}
[\omega_P]({\bm x}) &=& \sum_{n=1}^{P-2} C_{n} \sum_{k=1}^{P-n-1} x_1\cdots x_{n+k-1} D_{k|P-n-k}({\bm x}_{\geq k+n+1})\\
&& - x_1\cdots x_{P-1}(1-x_P)\,C_{P}. \nonumber
\end{eqnarray}
where we used that $D_{n|0}=C_{n}$ and $\sum_{n=1}^{P-1}C_{n}C_{P-n}=-C_{P}$ (which can be again checked from \eqref{eq:cP1}).
Notice that the terms factorizing $C_{n}$ at $n$ fixed all involve loops with $P-n$ points and their multiple derivatives.

	\subsection{Generating functions}

The previous recursive construction of the regular loop correlation functions may be compactly written using generating functions carrying the same information as $[\omega_P]$ and generalizing the observation that the top hole coefficients $C_{P}$ are alternating Catalan numbers. 

Because the previous construction involves strings of derivatives w.r.t. $x_1,\cdots,x_P$ and thus involves functions which depend on the remaining variables attached to the right end of the chain interval, let us first rename the variables and set 
\begin{equation}
x_k=: y_{P-k},\quad \mathrm{for}\ k=1,\cdots P.
\end{equation}
That is: $x_P=y_0$, $x_{P-1}=y_1$, and so on.

Let us also rename the multiple derivatives and the hole coefficients of the regular loop expectation values as $C_{P;k} := D_{P-k|k}({\bm x}_{\geq P-k+1})$ and ${S}_{P;k} :=  [\omega_P]^o_{P-k}({\bm x}_{\geq P-k+1})$, and define generating functions $\mathcal{C}_k(z)$ and $\mathcal{O}_k(z)$ for them by
\begin{subequations} \label{eq:def-DSomega}
\begin{align}
\mathcal{C}_k(z) &= \sum_{N\geq 0} C_{N+k+1;k}\,z^{N},
\quad \mathrm{with}\ C_{N+k+1;k} := D_{N+1|k}, \label{eq:def-DSomega-a}\\
 \mathcal{O}_k(z)&= \sum_{N\geq 0} {S}_{N+k+2;k}\,z^{N},
 \quad \mathrm{with}\ {S}_{N+k+2;k} :=  [\omega_{N+k+2}]^o_{N+2} , \label{eq:def-DSomega-b}
\end{align}
\end{subequations} 
for $k\geq 0$. These functions depend on the $k$ variables $(y_{k-1},\cdots,y_1,y_0)$, i.e. on the last $k$ positions. Notice that $C_{N;0}=C_{N}$, the alternating Catalan numbers. We have

\begin{proposition} \label{prop:Ck}
The relations \eqref{eq:cP1} for the regular loop expectation values are equivalent to
\begin{subequations} \label{eq:D0-gene}
\begin{align}
\mathcal{C}_{k+1}(z) &= \mathcal{O}_{k}(z) +  y_{k}z^{-1}\big( \mathcal{C}_{k}(z)-\mathcal{C}_{k}(0)\big) \label{eq:D0-gene-a},\\
\mathcal{O}_k(z) &= \mathfrak{c}(z)\mathcal{C}_k(z), \label{eq:D0-gene-b}
\end{align}
\end{subequations} 
for $k\geq0$ with the initial condition $\mathcal{C}_0(z)=\mathfrak{c}(z)$. These equations are equivalent to \eqref{eq:Ck-series}.
\end{proposition}

{\it Proof.}
Using $y_{k}=x_{P-k}$, the general relation \eqref{eq:D-hole} between multiple derivatives and hole coefficients (for $[\omega_P]$ the regular loop correlation function) translates into, 
\[ C_{P,k} = {S}_{P;k-1} + y_{k-1} C_{P;k-1} ,\]
for $k\geq 1$.
Similarly, the recurrence relation \eqref{eq:hole-regular} for the hole coefficients of the regular loop correlation function translates into
\[ 
{S}_{P;k} = \sum_{n=1}^{P-k-1} C_{n} C_{P-n;k} ,
\]
for $P-2\geq k\geq 0$. Using $C_{k+1;k}=\mathcal{C}_{k}(0)$, these equations are easily shown to be equivalent to \eqref{eq:D0-gene}.
\cqfd

Equations \eqref{eq:D0-gene} provide recurrence relations to determine the regular loop expectation values, and their derivatives and hole coefficients. It starts with $\mathcal{C}_0(z)=\mathfrak{c}(z)$. Eq.\eqref{eq:D0-gene-a} gives $\mathcal{O}_0(z)=\mathfrak{c}(z)^2=z^{-1}(1-\mathfrak{c}(z))$, then eq.\eqref{eq:D0-gene-a} yields $\mathcal{C}_{1}(z)=\mathfrak{c}(z)^2(1-y_0)$, similarly $\mathcal{C}_2(z)=\big(\mathfrak{c}(z)^3+y_1z^{-1}(\mathfrak{c}(z)^2-1)\big)(1-y_0)$, etc. Notice that, by induction, $(1-y_0)$ always factorizes in $\mathcal{C}_k(z)$, as it should.

Furthermore, there is a remarkable stabilization phenomena which ensures that there exists a single formal power series of an infinite number of variables coding for the regular loop expectation values for each $P\geq 2$.

\begin{theorem} \label{the:D-omega}
There exists a single formal power series $\overline{D}_\omega(y_0,y_1,\cdots)$ of an infinite number of variables such that, for each $P\geq 2$,
\begin{equation} \label{eq:Dbar-rond}
 [\omega_P]({\bm x})=x_1\, \overline{D}_\omega(x_{P},x_{P-1},\cdots,x_2,0,0,\cdots).
 \end{equation}
 That is, $[\omega_P]$ is obtained by putting all variables $y_{j}$ with $j\geq P$ to zero and otherwise by setting $y_k=x_{P-k}$, for $k=0,\cdots, P-1$.
\end{theorem}

{\it Proof.} The relation \eqref{eq:D0-gene-b} implies that $S_{k+2;k}=C_{k+1;k}$, with the convention $\Delta n=1$ (a relation which also follows from \eqref{eq:hole-regular} without appealing to generating functions).
Observe then that $C_{k+1;k}=\nabla_1[\omega_{k+1}]$ and $[\omega_{k+1}]=x_1\nabla_1[\omega_{k+1}]$ so that $[\omega_{k+1}]=x_1C_{k+1;k}$. Observe also that $S_{k+2;k}$ is the first hole coefficient of $[\omega_{k+2}]$ so that $S_{k+2;k}=\nabla_1[\omega_{k+2}]\vert_{x_2=0}$, that is $S_{k+2;k}=C_{k+2;k+1}\vert_{y_k=0}$. Thus, we have the claimed stabilisation phenomena,  
\[ C_{k+1;k}=C_{k+2;k+1}\vert_{y_k=0}. \]
Since $C_{k+1;k}=\mathcal{C}_k(0)$ we can alternatively write
\begin{eqnarray*}
[\omega_{k+1}]&=& x_1\,\mathcal{C}_k(0)(y_0,\cdots,y_{k-1}) \\
&=& x_1\,\mathcal{C}_{k+1}(0)(y_0,\cdots,y_{k-1},y_k=0).
\end{eqnarray*}By iterating at infinitum, we infer that there exists a function  $\overline{D}_\omega$ depending an infinite number of variables $y_0,y_1,\cdots$ such that \eqref{eq:Dbar-rond} holds.
\cqfd

The recurrence relations \eqref{eq:D0-gene} can be translated into equations for $\overline{D}_\omega$, adding recursively one $y$-variable at a time, starting from  $\overline{D}_\omega(0,0,\cdots)=1$. 

As an illustration, keeping only the $5$ first variables, one finds
\begin{eqnarray*} 
\overline{D}_\omega(y_0,y_1,y_2,y_3,y_4,0,\cdots)& & =\big(42\,y_{1}\,y_{2}\,y_{3}\,y_{4}-28\,y_{2}\,y_{3}\,y_{4}-23\,y_{1}\,y_{3}\,y_{4}+14\,y_{3}\,y_{4} \\ 
& & -19\,y_{1}\,y_{2}\,y_{4
 }+12\,y_{2}\,y_{4} +9\,y_{1}\,y_{4}-5\,y_{4}-14\,y_{1}\,y_{2}\,y_{3}\\ 
 & & +9\,y_{2}\,y_{3} +7\,y_{1}\,y_{3}-4\,y_{3}+5\,y_{1}\,y_{2}-3\,y_{2}-2\,y_{1}+1\big)(1-y_0).
\end{eqnarray*}
One may also check that setting $y_4=0$ reproduces \eqref{eq:Dregular5} as the stabilisation phenomena requires.

{\bf Remark.} 
Surprisingly \cite{michel}, the expectation values $[\omega_{n+1}]({\bm x})$ connect to objects in discrete and computational geometry, and in particular to the so-called associahedron, also called Stasheff polytopes \cite{discret-geom}. Let us specialize $[\omega_{n+1}]({\bm x})$ to $x_j=-t$, for all $j$. It can then be written as
\[ [\omega_{n+1}]({\bm x})\vert_{x_j=-t} = - t\, \Phi_n(t)\, (1+t),\]
with $\Phi(t)$ a polynomial of degree $n-1$ and positive integer coefficients. For instance,
\begin{eqnarray}
\Phi_2(t) &=& 1 + 2t, \nonumber\\
\Phi_3(t) &=& 1 + 5t+5t^2, \label{eq:count}\\
\Phi_4(t) &=& 1 + 9t+21t^2+14t^3, \nonumber\\
\Phi_5(t) &=& 1 + 14t+56t^2+84t^3+42t^4. \nonumber
\end{eqnarray}
These polynomials count the $(n-k)$ dimensional faces in the associahedron of order $n$ \cite{discret-geom}. This coincidence actually extends to any loop expectation values as they are all equal once specialized to $x_j=-t$, for all $j$, thanks to the continuity condition \eqref{eq:continu}.
We do not have a good understanding of this connection except that both structures are related to the moduli spaces of configurations of particles on the line. Anyway, it is remarkable that the loop expectation values are polynomials with integer coefficients (of alternating signs).

\section{Simple deformations of the regular loop} \label{sec:trans-deformation}

We now look at the action of the transposition $\tau_n$, exchanging $n$ and $n+1$, on the circular cyclic permutation $\omega_P$ and its correlation functions. To first deal with this peculiar deformation before the cases of general deformations will allow to reveal the structures we expect to be valid in general.
We have $\tau_n\circ\omega_P=(12\cdots ,n-1,n+1,n,n+2,\cdots,P)$.

	\subsection{Transpositions on the regular loop} \label{sec:omega-Tn}

We look at the expectation values $[\tau_n\circ\omega_P]({\bm x})$. 
As above, we introduce a notation for its multiple derivatives and hole coefficients. 
Generalising the definition of previous Section, we set
\begin{equation} \label{eq:def-Dn}
D^{(n)}_{k|P-k}({\bm x}_{\geq k+1}):= \nabla_{k}\cdots\nabla_{1}[\tau_n\circ\omega_P]({\bm x}) .
\end{equation}
In particular, $D^{(n)}_{0|P}:=[\tau_n\circ\omega_P]$ and $D^{(n)}_{P|0}=C_{P}$ independently of $n$ and equal to the alternating Catalan numbers defined in \eqref{eq:C-catalan}.
Recall also the definition of the hole coefficients,
\begin{equation} \label{eq:def-Sn}
[\tau_n\circ\omega_P]_{k+1}^o({\bm x}_{\geq k+2}) := \nabla_{k}\cdots\nabla_{1}[\tau_n\circ\omega_P]({\bm x})\vert_{x_{k+1}=0},
\end{equation}
as well as the relation \eqref{eq:D-hole} between the multiple derivatives and the hole coefficients.
We have
	
\begin{lemma}
The hole coefficients of $[\tau_n\circ\omega_P]$ satisfy the following recurrence relations, with $j\geq n+1$ in the first line, and $1\leq j \leq n-2$ in the last line :
\begin{subequations} \label{eq:Dnsys}
\begin{align}
[\tau_n\circ\omega_P]_{j+1}^o &= \sum_{k=1}^{n}D^{(n-k+1)}_{j+1-k|0}D_{k|P-j-1}({\bm x}_{\geq j+2})  +\sum_{k=n+1}^j D_{j+1-k|0} D^{(n)}_{k|P-j-1}({\bm x}_{\geq j+2}),  \label{eq:Dnsys-a}\\
[\tau_n\circ\omega_P]_{n+1}^o &=  \sum_{k=1}^{n-1}D_{n-k|0}D_{k+1|P-n-1}({\bm x}_{\geq n+2}) + D_{1|0}D_{n|P-n-1}({\bm x}_{\geq n+2}) \label{eq:Dnsys-b} ,\\
[\tau_n\circ\omega_P]_{n}^o &= \sum_{k=1}^{n-1}D_{n-k|1}(x_{n+1})D_{k|P-n-1}({\bm x}_{\geq n+2})  \label{eq:Dnsys-c}, \\
[\tau_n\circ\omega_P]_{j+1}^o &= \sum_{k=1}^{j}D_{j+1-k|0}D^{(n+k-j-1)}_{k|P-j-1}({\bm x}_{\geq j+2})  \label{eq:Dnsys-d}.
\end{align}
\end{subequations}
with ${\bm x}_{\geq j}:=(x_{j},\cdots,x_P)$. Recall that $D^{(m)}_{j|0}=C_{j}$ for all $m,j$.
\end{lemma}

{\it Proof.}
Thanks to \eqref{eq:sigma0}, we have to look at all  sub-loops $(\tau_{k+1}\cdots\tau_j\circ(\tau_n\circ\omega_P))_k^\pm$, for $j=1,\cdots, P-1$ and $k=1,\cdots,j$. Let us introduce a notation:
\[ 
\omega^n_{k;j|P}:=\tau_{k+1}\cdots\tau_j\circ(\tau_n\circ\omega_P)
\]
Let us prove it by inspection, on a case by case basis (this is not the simplest proof but it allows to learn how to deal with \eqref{eq:sigma0}, and this will be useful when dealing with the general case). We shall use repeatedly the fact that $\tau_{k+1}\cdots\tau_j$ is the cyclic permutation $(k+1,k+2,\cdots,j+1)$. \\
- For $j\geq n+3$, we have $(\tau_n\circ\omega_P)_j^-=(j)$ and $(\tau_n\circ\omega_P)_j^+=\tau_n\circ(1\cdots,j-1,j+1,\cdots P)$. Then, by induction, one easily proves that, for $k\geq n+2$,
\begin{eqnarray*}
(\omega^n_{k;j|P})_k^- &=& (k,k+2,\cdots,j+1),\\
(\omega^n_{k;j|P})_k^+ &=& \tau_n\circ(1\cdots, k-1,k+1,j+2,\cdots P).
\end{eqnarray*}
For $j\geq n+3$ and $k=n+1$, one has $(\omega^n_{k;j|P})_{n+1}^- =(n+1,n,n+3,\cdots,j+1)$ and $(\omega^n_{k;j|P})_{n+1}^+= (1\cdots, n-1,n+2,j+2,\cdots P)$.
For $j\geq n+3$ and $k=n$, one has $(\omega^n_{k;j|P})_{n}^- = (n,n+3,\cdots,j+1)$ and $(\omega^n_{k;j|P})_{n}^+ = (1\cdots, n-1,n+2,n+1,j+2,\cdots P)$.
For $j\geq n+3$ and $k\leq n-1$, one find 
\begin{eqnarray*}
(\omega^n_{k;j|P})_{k}^- &=& (k,k+2,\cdots,n,n+2,n+1,n+3,\cdots,j+1),\\
(\omega^n_{k;j|P})_{k}^+ &=& (1\cdots, k-1,k+1,j+2,\cdots P).
\end{eqnarray*}
- For $j=n+2$, we have $(\tau_n\circ\omega_P)_{n+2}^-=(n+2)$ and $(\tau_n\circ\omega_P)_{n+2}^+=\tau_n\circ(1,\cdots,n+1,n+3,\cdots P)$. For $j=n+2$ and $k=n+1$, we have $(\omega^n_{k;j|P})_{n+1}^-=(n+1,n,n+3)$ and $(\omega^n_{k;j|P})_{n+1}^+=(1,\cdots,n-1,n+2,n+4,\cdots P)$. By induction, one proves that for $j=n+2$ and $k\leq n-1$,
\begin{eqnarray*}
(\omega^n_{k;j|P})_{k}^- &=& (k,k+2,\cdots,n,n+2,n+1,n+3),\\
(\omega^n_{k;j|P})_{k}^+ &=& (1\cdots, k-1,k+1,n+4,\cdots P).
\end{eqnarray*}
- The cases with $j=n+1$ yields similar results and we leave it to the reader.\\
- For $j=n$, we have $(\tau_n\circ\omega_P)_{n}^-=(1,\cdots,n-1,n,n+2,\cdots P)$ and $(\tau_n\circ\omega_P)_{n}^+=(n+1)$. For $j=n$ and $k=n-1$, we have $(\omega^n_{k;j|P})_{n-1}^-=(n-1)$ and $(\omega^n_{k;j|P})_{n-1}^+=(1,\cdots,n-2,n,n+1\cdots P)$. then, by induction, for $j=n$ and $k\leq n-2$
\begin{eqnarray*}
(\omega^n_{k;j|P})_{k}^- &=& (k,k+2,\cdots,n-1,n),\\
(\omega^n_{k;j|P})_{k}^+ &=& (1\cdots, k-1,k+1,n+1,n+2\cdots P).
\end{eqnarray*}
- For $j=n-1$ and $k\leq n-1$, we have
\begin{eqnarray*}
(\omega^n_{k;j|P})_{k}^- &=& (k,k+2,\cdots,n+1),\\
(\omega^n_{k;j|P})_{k}^+ &=& (1\cdots, k-1,k+1,n+2\cdots P).
\end{eqnarray*}
- Finally, the cases $j\leq n-2$ are simple because the decomposition of the loop does not touch the pair $(n,n+1)$ on which the transposition is acting. We get
\begin{eqnarray*}
(\omega^n_{k;j|P})_{k}^- &=& (k,k+2,\cdots,j+1),\\
(\omega^n_{k;j|P})_{k}^+ &=& (1\cdots, k-1,k+1,j+2\cdots P).
\end{eqnarray*}
Using these decompositions in  \eqref{eq:sigma0} and taking the derivatives $\nabla_1\cdots\nabla_{j+1}$ yields \eqref{eq:Dnsys}.
\cqfd

The two middle equations (\ref{eq:Dnsys-b},\ref{eq:Dnsys-c}) involve hole coefficients with indices in the zone where the transposition is acting and, as a consequence, they are structurally different from equations \eqref{eq:hole-regular} for the hole coefficients of the regular loop. The two other equations (\ref{eq:Dnsys-a},\ref{eq:Dnsys-d}) involve hole coefficients away from the zone where the transposition is acting and are structurally similar to \eqref{eq:hole-regular}. As a consequence

\begin{corrollary} 
We have:\\
(i) $D^{(n)}_{P-k|k}=D_{P-k|k}$ independently of $n$ for $P\geq P-k\geq n+1$.\\
(ii) The pair $[\omega_P]$ and $[\tau_n\circ\omega_P]$ satisfy the exchange relation \eqref{eq:moves}.
\end{corrollary}

{\it Proof.}
(i) The proof is by induction as above. 
Recall the relation \eqref{eq:D-hole} between multiple derivatives and hole coefficients which in the present case reads
\beq \label{eq:DS-Tn} 
D^{(n)}_{k|P-k}= [\tau_n\circ\omega_P]_{k+1}^o + x_{k+1} D^{(n)}_{k+1|P-k-1} . 
\eeq
Recall that $[\tau_n\circ\omega_P]_{P+1}^o=-[\tau_n\circ\omega_P]_{P}^o$ are independent of $n$ and that $D^{(n)}_{j|0}=C_j$, for all $n,j\geq 1$. Using this fact, \eqref{eq:Dnsys-a} can be simplified into
\[
[\tau_n\circ\omega_P]_{j+1}^o = \sum_{l=1}^{n}C_{j+1-l}D_{l|P-j-1}  +\sum_{l=n+1}^j C_{j+1-l} D^{(n)}_{l|P-j-1}.
\]
For $j=P-1$, this equation only involves factors $D^{(n)}_{k|P-j-1}=D^{(n)}_{k|0}=C_k$. Hence $[\tau_n\circ\omega_P]_{P}^o=[\omega_P]_{P}^o$ independently of $n$ and, from \eqref{eq:DS-Tn}, $D^{(n)}_{P-1|1}$ is independent of $n$. The proof now is by induction on $k$. Assume that $D^{(n)}_{P-l|l}=D_{P-l|l}$ holds for $l$ up to $k$ for $0\leq k\leq P-n-2$ for all $P\geq n+1$. Write \eqref{eq:Dnsys-a} for $j=P-k$ to get
\[
[\tau_n\circ\omega_P]_{P-k+1}^o = \sum_{l=1}^{n}C_{j+1-l}D_{l|k-1}  +\sum_{l=n+1}^{P-k} C_{j+1-l} D^{(n)}_{l|k-1}.
\]
By the induction hypothesis $D^{(n)}_{l|k-1}=D_{l|k-1}$ for $l\leq P-k$ and hence $[\tau_n\circ\omega_P]_{P-k+1}^o=[\omega_P]_{P-k+1}^o$ and thus  $D^{(n)}_{P-k|k}=D_{P-k|k}$ from \eqref{eq:DS-Tn}. The induction stops at $P-k=n+1$.\\
(ii) The proof is given in Appendix \ref{sec:exchange-Tn}. 
\cqfd

The fact that the pair $[\omega_P]$ and $[\tau_n\circ\omega_P]$ satisfies the exchange relation \eqref{eq:moves} provides an independent check of the locality conjecture.

{\bf Remark.} From \eqref{eq:Dnsys} with $n=1$ one can check that $[\tau_1\circ\omega_P]=[\omega_P]$ as expected. See the remark at the end of Section \ref{sec:ex1}.

\subsection{Generating functions for transposition deformations}

Generating functions for these deformations have to include summations over the location of the deformation. Thus let us again renamed the coefficients (\ref{eq:def-Dn},\ref{eq:def-Sn}) by setting $C^{(n)}_{P;k} := D^{(n)}_{P-k|k}({\bm x}_{\geq P-k+1})$ and $
{S}^{(n)}_{P;k}:= [\tau_n\circ\omega_P]^o_{P-k}({\bm x}_{\geq P-k+1})$.
Let us now introduce generating functions for these deformed loops defined by, for $k\geq 0$, $q\geq 0$,
\begin{subequations} \label{eq:def-DSTn}
\begin{align}
\mathcal{D}_k^{(q)}(z) &:=\sum_{N\geq 0} C_{N+k+1;k}^{(q+N)}\, z^N,
\quad \mathrm{with}\  C_{N+k+1;k}^{(q+N)}:= D_{N+1|k}^{(q+N)} , \label{eq:def-DSTn-a}\\
\mathcal{S}_k^{(q)}(z) &:=\sum_{N\geq 0} {S}_{N+k+2;k}^{(q+N+1)}\,z^N 
\quad \mathrm{with}\  {S}_{N+k+2;k}^{(q+N+1)}:= [\tau_{q+N+1}\circ\omega_{N+k+2}]^o_{N+2} . \label{eq:def-DSTn-b}
\end{align}
\end{subequations}
These generating functions include sums over the number of points. They depend on the $k$ variables $(y_{k-1},\cdots,y_1,y_0)$, i.e. on the last $k$ positions, which we call the floating variables. Recall that we set $y_k=x_{P-k}$ with $P$ the number of points. For $\mathcal{D}_k^{(q)}(z)$ the number of points is $N+k+1$, the derivatives act on the first $N+1$ points, there are $k$ floating variables. The deformation is located at $(q+N;q+N+1)$ so that it is inside the `floating variable zone' for $q\geq 2$. For $\mathcal{S}_k^{(q)}(z)$, the number of points are $N+k+2$ and there are $k$ floating variables. In both cases, $q$ parametrizes the distance between the deformation and the boundary between the two zones (independently of the total number of points). Hence this generating functions involve summing over different locations of the deformation.

We shall look at \eqref{eq:Dnsys} in reverse order.

\begin{proposition}
We have: \\
(i) The generating functions $ \mathcal{D}_k^{(q)}(z)$ and $ \mathcal{S}_k^{(q)}(z)$ are related by
\begin{equation} \label{eq:D-T}
 \mathcal{D}_{k+1}^{(q+1)}(z) =  \mathcal{S}_k^{(q)}(z) + y_kz^{-1}\big( \mathcal{D}_k^{(q)}(z)- \mathcal{D}_k^{(q)}(0)\big),
\end{equation}
for $k\geq 0$ and all $q\geq0$.\\
(ii) Eqs.\eqref{eq:Dnsys} are equivalent to 
\begin{subequations} \label{eq:Dn-gener}
\begin{align}
 \mathcal{S}_{k+2}^{(q)}(z) &= \mathfrak{c}(z)\, \mathcal{D}_{k+2}^{(q)}(z),\ \mathrm{for}\ q\geq 2, \label{eq:Dn-gener-a} \\
\mathcal{S}_{k+1}^{(1)}(z) &= \mathcal{C}_1(z)\,\mathcal{C}_{k}(z),\ \mathrm{for}\ q=1 , \label{eq:Dn-gener-b}\\
\mathcal{S}_k^{(0)}(z) &= \mathfrak{c}(z)\big(\mathcal{C}_{k}(z)-\mathcal{C}_{k}(0)\big)+\mathfrak{c}(0)\,\mathcal{C}_{k}(z),\ \mathrm{for}\ q=0  ,\label{eq:Dn-gener-c}\\
\mathcal{D}_k^{(0)}(z) &= \mathcal{C}_k(z),\ \mathrm{for}\ q= 0. \label{eq:Dn-gener-d}
\end{align}
\end{subequations}
In \eqref{eq:Dn-gener-b}, the floating variable in $\mathcal{C}_1(z)$ is $y_{k}$. Recall that $\mathfrak{c}(0)=1$ (by convention $\Delta n=1$).
\end{proposition}

{\it Proof.}
(i) Eq.\eqref{eq:D-T} is a direct consequence of the relation $D^{(n)}_{j|P-j}= [\sigma^{(n)}_P]^o_{j+1}+x_{j+1} D^{(n)}_{j+1|P-j-1}$ which is equivalent to $C^{(n)}_{P;k+1}=S^{(n)}_{P;k}+y_kC^{(n)}_{P;k}$.\\
(ii) By checking (carefully) that the generating functions reproduce the correct equations.\\ 
Eq.\eqref{eq:Dnsys-d} codes for the cases for which the deformation is inside the `floating variable zone'.
Eqs.(\ref{eq:Dnsys-b},\ref{eq:Dnsys-c}) involve quantities at the location of the deformation.\\
Eqs. \eqref{eq:Dnsys-a} implies that $D^{(n)}_{P-k|k}=D_{P-k|k}$ for $0\leq k\leq P-n-1$ which is equivalent to $D^{(n)}_{N+1|k}=D_{N+1|k}$ for $N\geq n$, that is $\mathcal{D}_k^{(0)}(z) = \mathcal{C}_k(z)$. 
\cqfd

Let us discuss a bit how to solve these equations recursively. First notice that $\mathcal{D}_k^{(q)}(0)= D_{1|k}^{(q)}=C^{(q)}_{k+1;k}$, which is the derivative w.r.t. $x_1$ of the deformed loop. Thus
\begin{equation}
 [\tau_q\circ\mathfrak{c}_{k+1}](x_1,\cdots,x_{k+1})= x_1\, \mathcal{D}_k^{(q)}(0)(y_0,\cdots,y_{k-1}),\ \mathrm{with}\ x_j=y_{k+1-j}.
 \end{equation}
Hence, to determine the correlation functions of the deformed loops $\tau_n\circ\omega_P$ we have to solve for $\mathcal{D}_k^{(q)}(0)$, for $q\geq 1$. This is done by iterating \eqref{eq:D-T} and \eqref{eq:Dn-gener-a}, using (\ref{eq:Dn-gener-b},\ref{eq:Dn-gener-c},\ref{eq:Dn-gener-d}) as initial conditions. At this stage, we assume the functions $\mathcal{C}_k(z)$, associated to the regular loop, to be known from \eqref{eq:D0-gene}.

Indeed, from \eqref{eq:Dn-gener-d}, we have $\mathcal{D}_k^{(0)}(z)=\mathcal{C}_k(z)$ and from \eqref{eq:Dn-gener-c} we know $\mathcal{S}_k^{(0)}(z)$. Applying $\eqref{eq:D-T}$, we then get to know $\mathcal{D}_k^{(1)}(z)$ via
\[ \mathcal{D}_{k+1}^{(1)}(z) =  \big(y_k+ z\mathfrak{c}(z)\big)\frac{\mathcal{C}_k(z)- \mathcal{C}_k(0)}{z} +  \mathfrak{c}(0)\mathcal{C}_k(z) .\]
Then, from \eqref{eq:Dn-gener-b} we have a formula for $\mathcal{S}_k^{(1)}(z)$, and applying again \eqref{eq:D-T}, we know $\mathcal{D}_k^{(2)}(z)$ via
\begin{eqnarray*} 
\mathcal{D}_{k+1}^{(2)}(z) &=&  y_k\Big( y_{k-1} \frac{\mathcal{C}_{k-1}(z)-\mathcal{C}_{k-1}(0)-z\mathcal{C}_{k-1}'(0)}{z^2} + 2\mathfrak{c}(z)\frac{\mathcal{C}_{k-1}(z)-\mathcal{C}_{k-1}(0)}{z} \\
&& + \mathfrak{c}(0)(\mathcal{C}_{k-1}(z)-\mathcal{C}_{k-1}(0))\Big) 
 + \mathcal{C}_1(z)\,\mathcal{C}_{k-1}(z) .
\end{eqnarray*}
We can propagate the induction by alternating recursively the two equations \eqref{eq:D-T} and \eqref{eq:Dn-gener-a}. Indeed, given $\mathcal{D}_k^{(2)}(z)$, we get $\mathcal{S}_k^{(2)}(z)=\mathfrak{c}(z)\mathcal{D}_k^{(2)}(z)$ by \eqref{eq:Dn-gener-a}, then \eqref{eq:D-T} yields $\mathcal{D}_k^{(3)}(z)$, etc...

As a check we can verify that deforming by $\tau_1=(12)$ has no effect. Indeed, from the formula above, we have $\mathcal{D}_{k+1}^{(1)}(0)= y_k \mathcal{C}'_k(0) + \mathfrak{c}(0)\mathcal{C}_k(0)$. On the other hand, we know that $ \mathcal{C}_{k+1}(z) =  y_k(\mathcal{C}_k(z)- \mathcal{C}_k(0))z^{-1} +  \mathfrak{c}(0)\mathcal{C}_k(z)$, hence $\mathcal{C}_{k+1}(0)= y_k \mathcal{C}'_k(0)+\mathfrak{c}(0)\mathcal{C}_k(0)$, and $\mathcal{D}_{k+1}^{(1)}(0)=\mathcal{C}_{k+1}(0)$ as expected.

Similarly, we note that $\mathcal{D}_{k+1}^{(2)}(0) = y_k\big(\frac{1}{2}y_{k-1} \mathcal{C}''_{k-1}(0) + 2\mathfrak{c}(0)\mathcal{C}'_{k-1}(0)\big) + \mathcal{C}_1(0)\mathcal{C}_{k-1}(0)$, that is:
\[  [\tau_2\circ\mathfrak{c}_{k+2}]({\bm x}) = y_{k+1}\Big(y_k\big(\frac{1}{2}y_{k-1} \mathcal{C}''_{k-1}(0) + 2\mathfrak{c}(0)\mathcal{C}'_{k-1}(0)\big) + \mathcal{C}_1(0)\mathcal{C}_{k-1}(0)\Big) ,\]
with $y_l=x_{k-l+2}$ and $\mathcal{C}_{k-1}$ depends on $(y_0,\cdots,y_{k-2})$ and $\mathcal{C}_1$ on $y_{k-1}$.

We can check it for $k=3$ and $q=2$ (which we computed before in \eqref{eq:deformedT32}). We have $\mathcal{C}_0(z)=\mathfrak{c}(z)$, $\mathcal{C}_1(z)=\mathfrak{c}(z)^2(1-y_0)$, $\mathcal{C}_2(z)=\big(\mathfrak{c}(z)^3+y_1z^{-1}(\mathfrak{c}(z)^2-1)\big)(1-y_0)$ with $\mathfrak{c}(z)=1-z+2z^2-5z^3+\cdots$. Hence we have
\begin{eqnarray*}
\mathcal{C}_2(0) &=& (1-2\,y_1)(1-y_0) ,\\
\mathcal{C}_2'(0) &=& (-3+5\,y_1)(1-y_0) ,\\
\mathcal{C}_2''(0) &=& 2(9-14\,y_1)(1-y_0) ,
\end{eqnarray*}
and thus
\[  [(13245)] = y_4\big(1-2y_1-y_2-6y_3+2y_1y_2+10y_1y_3+9y_2y_3-14y_1y_2y_3\big)(1-y_0),\]
which coincides with \eqref{eq:deformedT32} once the substitution $y_l=x_{5-l}$ has been done.

\section{General deformations of the regular loop} \label{sec:general-deformation}

We now look at expectation values of generic labeled loops. These loops are in the orbit of the regular loop under the action of the permutation group because they are in one-to-one correspondence with single cycle permutations which form a conjugacy class of the permutation group.  Hence any generic labeled loops is the image of the regular loop by the adjoint action of the permutation, say $\mu_n$, on the regular cyclic permutation. We call this permutation $\mu_n$ the deformation.

From the analysis of the transposition deformation done in previous Section \ref{sec:trans-deformation}, the strategy to deal with generic deformations of the regular loop is clear: \\
(a) Away from the support of the deformation towards the right boundary (i.e. on the $x=1$ side), the multiple derivatives and hole coefficients of the correlation functions are undeformed and identical to those of the regular loop, their generating functions are simply expressed in terms of the functions $\mathcal{C}_k(z)$ of the regular loop;\\
(b) Outside the support of the deformation but toward the left boundary (i.e. on the $x=0$ side), the generating functions the multiple derivatives and hole coefficients will be coupled as above and recurrence relations of the form \eqref{eq:D0-gene}, or (\ref{eq:D-T},\ref{eq:Dn-gener-a}), can be applied once the initial condition is known.\\
(c) This initial condition is obtained by propagating the multiple derivatives and hole coefficients generating functions through the deformation (starting from the undeformed side).\\
Of course, the difficulty resides in the last step which codes for the peculiarities of each deformation. As above, the construction relies on an interplay between multiple derivatives and hole coefficients and on using the formulas (\ref{eq:solution},\ref{eq:sigma0}) relating hole coefficients and multiple derivatives of smaller loops. These relations can be summarized into generating functions which encoded for both the profile of the deformation and its location along the chain interval.

	\subsection{Preliminaries} \label{sec:def-extra}

Given a permutation in $\mathbb{S}_P$ and its associated deformation of the regular loop, we define its support as the smallest interval such that the permutation acts trivially as the identity on its complement. Because it acts bijectively on its support, a permutation of support of size $|\mu|$ can be viewed as an element of $\mathbb{S}_{|\mu|}$. Given $\mu\in\mathbb{S}_{|\mu|}$ we define $\mu_n$ as the permutation having the same profile as  $\mu$ but with support $[n,|\mu|+n-1]$, so that the associated deformation of the regular $\mu_n\circ\omega_P$ is the following single cycle permutation
\[ 
1\to 2 \to \cdots \to n-1 \to \mu_n(n) \to \mu_n(n+1) \cdots \to \mu_n(n+|\mu|-1) \to n+|\mu| \to \cdots \to P ,
\]
or equivalently,
\[ 
(12\cdots,n-1,\mu_n(n),\mu_n(n+1),\cdots,\mu_n(n+|\mu|-1),n+|\mu|,\cdots,P ).
\]
The $\mu_q$'s are translation versions of each other such that $\mu_n(j)+n'-n=\mu_{n'}(j+n'-n)$. In particular, $\mu_n(j+n-1)=\mu(j)+n-1$ for $j=1,\dots, |\mu|$.

To describe below the recurrence relations between the multiple derivatives and the hole coefficients of the loop expectation values within the support of their deformation we have to introduce some notations. (The main difficulty is to introduce a good enough notation). We shall define different truncations of a deformation which we call extractions.

Let $\mu\in \mathbb{S}_{|\mu|}$ be a deformation of length $|\mu|$ and $\mu_n$ its translation. Let $m$ and $M$ in $[1,|\mu|]$ be two integers ordered according to $\mu$, ie. along the cycle associated to $\mu$, that is such that $1\leq \mu^{-1}(m)<\mu^{-1}(M)\leq |\mu|$.\\
-- We define $\mu^{(m,M)}$ the truncation of $\mu$ between $m$ and $M$ as the restriction of $\mu$ on the interval between $\mu^{-1}(m)$ (included) to $\mu^{-1}(M)-1$ (included). I.e. $\mu^{(m,M)}$ is the map (the map is read from the top line to the bottom line)
\[ 
\mu^{(m,M)}:=\left(\begin{matrix}
 \mu^{-1}(m) & \mu^{-1}(m)+1 & \cdots & \mu^{-1}(M)-1 \\
 m & \mu(\mu^{-1}(m)+1) & \cdots & \mu(\mu^{-1}(M)-1) 
\end{matrix}\right)
\]
Its support has length $|\mu^{(m,M)}|:=\mu^{-1}(M)-\mu^{-1}(m)$. The map $\mu^{(m,M)}$ are bijections but between different spaces as their image spaces are in general not the ordered integer numbers between $\mu^{-1}(m)$ and $\mu^{-1}(M)-1$. However, what matters is not the exact labeling of the points in the image space but their ordering.
Since there is a unique map between the set of images of $\mu^{(m,M)}$ and $[1,|\mu^{m)(M}|]$ preserving the order, we may identify the truncation $\mu^{(m,M)}$ with a deformation in $\mathbb{S}_{|\mu^{(m,M)}|}$.  \\
-- We define the complementary truncation $\mu^{m)(M}$ has the restriction of $\mu$ on $[1,\mu^{-1}(m)-1]\cup[\mu^{-1}(M),|\mu|]$. Ie. $\mu^{m)(M}$ is the map
\[ 
\mu^{m)(M}:=\left(\begin{matrix}
1 & \cdots & \mu^{-1}(m) -1 & \mu^{-1}(M) & \cdots &| \mu| \\
\mu(1) & \cdots &\mu(\mu^{-1}(m)-1) &  M& \cdots & \mu(|\mu|) 
\end{matrix}\right)
\]
Its length is $|\mu|+\mu^{-1}(m)-\mu^{-1}(M)=:|\mu^{m)(M}|$. Of course $|\mu|=|\mu^{m)(M}|+ |\mu^{m)(M}|$. Similarly, it can be identified as a deformation in $\mathbb{S}_{|\mu^{m)(M}|}$.\\
-- We extend this definition for non ordered pair of numbers $m$ and $M$ (choosing the obvious definition) by exchanging the role of $m$ and $M$ depending whether $\mu^{-1}(m)<\mu^{-1}(M)$ or $\mu^{-1}(M)<\mu^{-1}(m)$. We set
\begin{eqnarray*}
&\mu^{[m,M]}:= \mu^{(m,M)},\ \mu^{m][M}:= \mu^{m)(M},\quad \mathrm{if}\ \mu^{-1}(m)<\mu^{-1}(M),\\
&\mu^{[m,M]}:= \mu^{(M,m)},\ \mu^{m][M}:= \mu^{M)(m},\quad \mathrm{if}\ \mu^{-1}(M)<\mu^{-1}(m) .
\end{eqnarray*}
By construction, $\mu^{[m,M]}$ and $\mu^{m][M}$ are complementary in $\mu$.\\
-- We define $\mu^{m]}$, with length $|\mu^{m]}|:=\mu^{-1}(m) -1$, by
\[ 
\mu^{m]}:=\left(\begin{matrix}
1 & \cdots & \mu^{-1}(m) -1  \\
\mu(1) & \cdots &\mu(\mu^{-1}(m)-1)  
\end{matrix}\right)
\]
and $\mu^{[M}:$, with length $|\mu^{[M}|:= |\mu|-\mu^{-1}(M)$, by
\[ 
\mu^{[M}:=\left(\begin{matrix}
\mu^{-1}(M)+1 & \cdots &| \mu| \\
 \mu(\mu^{-1}(M)+1)& \cdots & \mu(|\mu|) 
\end{matrix}\right)
\]
The two extractions $\mu^{[M}$ and $\mu^{M]}$ are not complementary in $\mu$ since the point $M$ is missing in the union of their images.\\
-- For $1\leq\ell\leq M$, let  $\omega^{\ell}_{M}$ be the cyclic permutation $\omega^{\ell}_{M}:=(\ell,\ell+1,\cdots,M)$. In particular $\omega^{1}_{M}=\omega_M$ in the cyclic permutation of the $M$ first integers. We define the following composition
\begin{eqnarray*}
\mu_\varpi^{M]}:= \omega^{1}_{M}\,\mu^{M]} &,&\ \mu_\varpi^{[M}:= \omega^{1}_{M}\,\mu^{[M},\\
\mu_\varpi^{[m,M]}:= \omega^{m+1}_{M}\,\mu^{[m,M]}&,&\ \mu_\varpi^{m][M}:= \omega^{m+1}_{M}\,\mu^{m][M} .
\end{eqnarray*}
We used the implicit convention that $\omega^{\ell}_M$ acts trivially as the identity on all integers not in $\{\ell,\ell+1,\cdots,M\}$.

Given these extractions of the deformation, we define the number of points, greater than $M$, which are in the support of their images. Namely, we set
\begin{eqnarray*}
k^{[m,M]} &=& \# \{j\in[M+1,|\mu|]\ \mathrm{s.t.}\ j\in \mathrm{Im}\, \mu_\varpi^{[m,M]}\}\\
k^{M]} &=& \# \{j\in[M+1,|\mu|]\ \mathrm{s.t.}\ j\in \mathrm{Im}\, \mu_\varpi^{M]}\}
\end{eqnarray*}	

	\subsection{Recurrence relations}
\begin{figure}[!ht]
	\includegraphics[scale=0.5]{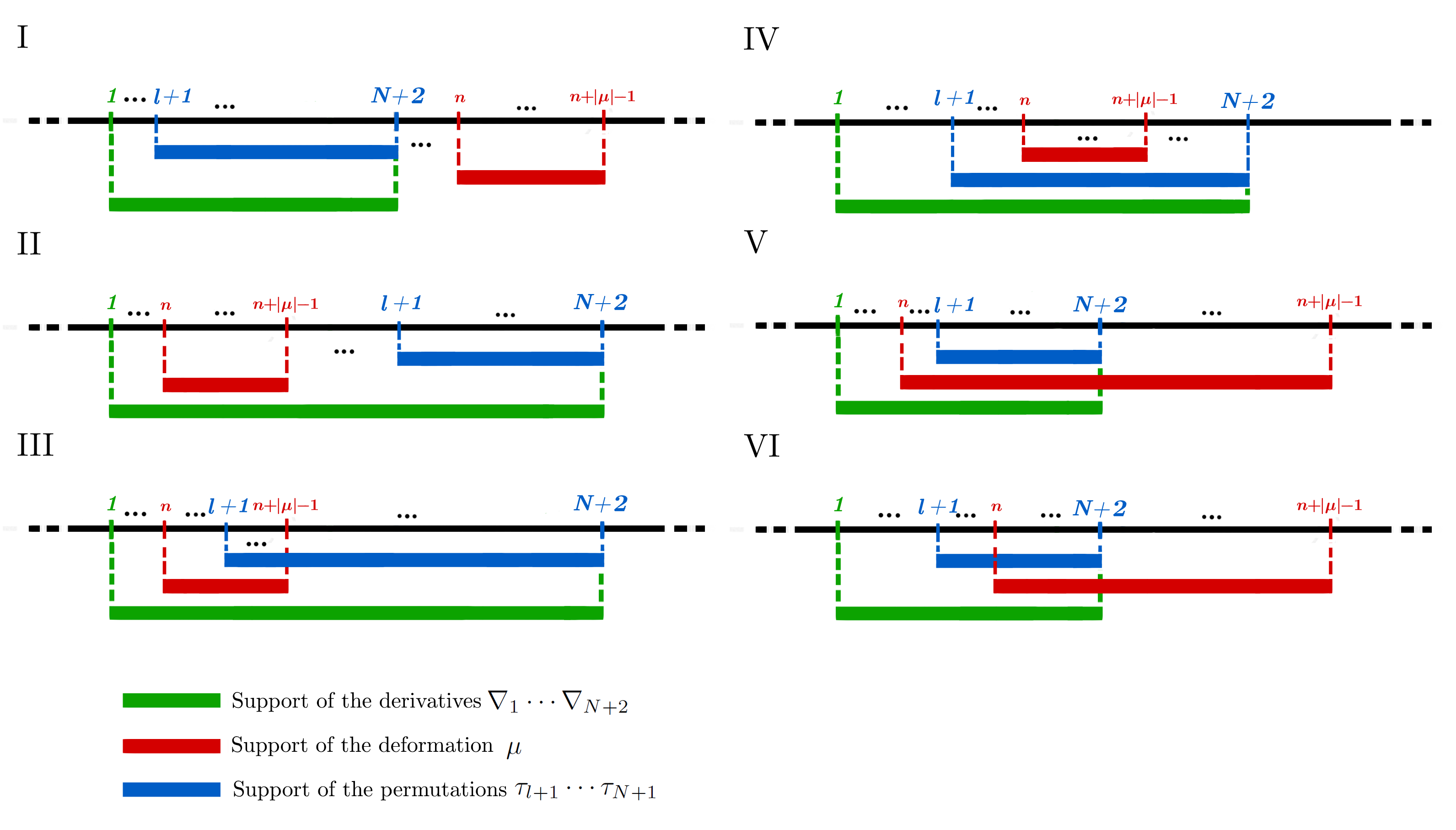}
	\caption{The different possible configurations depending on how the support of the derivatives, the deformations and the permutations overlap with each other.}
	\label{fig:differentcases}
\end{figure}

We aim at computing $[\mu_n\circ\omega_P]$ for any deformation $\mu_n$ and any number of points $P>|\mu_n|$. As before we define multiple derivatives and hole coefficients but adapting the notation such that it is directly suitable for the generating functions:
\begin{subequations} \label{eq:def-general} 
\begin{align} 
 C^{\mu_n}_{N+k+1;k} & := \nabla_1\cdots\nabla_{N+1}[\mu_n\circ\omega_{N+k+1}], \label{eq:def-general-a}\\
  S^{\mu_n}_{N+k+2;k} &:= \nabla_1\cdots\nabla_{N+1}[\mu_n\circ\omega_{N+k+2}]\vert_{x_{N+2}=0} , \label{eq:def-general-b}
\end{align} 
\end{subequations} 
 with $N+k+1$ points in total for $C^{\mu_n}_{N+k+1;k}$ and $N+k+2$ points for $S^{\mu_n}_{N+k+2;k}$. In both cases there are $k$ floating variables $(y_{k-1},\cdots,y_0)$ where we set $y_l=x_{P-l}$ as before.
We recall the relation \eqref{eq:sigma0} between the hole coefficients and the multiple derivatives
\begin{eqnarray} \label{eq:Smu}
 S^{\mu_n}_{N+k+2;k} = \nabla_{N+2}\cdots\nabla_1\sum_{l=1}^{N+1} \Big[ (\tau_{l+1}\cdots\tau_{N+1}\circ\hat\mu_n)^-_l\, (\tau_{l+1}\cdots\tau_{N+1}\circ\hat\mu_n)^+_l\Big] ,
\end{eqnarray}
with $\hat \mu_n:=\mu_n\circ\omega_{N+k+1}$.
We will proceed by induction on $N$ to find the coefficients (\ref{eq:def-general-a}), (\ref{eq:def-general-b}). To compute (\ref{eq:Smu}), one has to go through a case by case inspection of the different possible configurations, depending on the interplays between the support of the deformation $\mu$, that of the permutations $\tau_{l+1}\cdots\tau_{N+1}$, and that of the derivatives $\nabla_1\cdots\nabla_{N+2}$. The different cases are labelled from I to VI and are summed up in Figure \ref{fig:differentcases}. To make all notations clear, specific examples of the different situations are presented in the Appendix \ref{app:exemple} as well.

Away from the deformation, we have the simple relations.

\begin{lemma}
We have :\\
(i) for $N\leq n-3$,
\begin{equation} \label{eq:Dmu-sys1}
S^{\mu_n}_{N+k+2;k} =\sum_{m=1}^{N+1} C_{m} C^{\mu_{n-m}}_{N+k+2-m;k} .
\end{equation}
(ii) for $N\geq n+|\mu|-2$,
\begin{equation} \label{eq:Cmu-away}
C^{\mu_n}_{N+k+1;k}=C_{N+k+1;k} ,
\end{equation}
independently of the deformation $\mu$, where $C_n$ are the alternating Catalan numbers \eqref{eq:C-catalan}.
This implies that $S^{\mu_n}_{N+k+2;k}=S_{N+k+2;k}$ for $N+1\geq n+|\mu|-2$.
\end{lemma}

{\it Proof.}
To simplify notation let $\hat \mu_n:=\mu_n\circ\omega_{N+k+1}$ and $\Omega\mu_{n}^{l}:=\tau_{l+1}\cdots\tau_{N+1}\mu_n$. Recall that $\tau_{l+1}\cdots\tau_{N+1}$ is the following cyclic permutation
\[ \tau_{l+1}\cdots\tau_{N+1} = (l+1,l+2,\cdots,N+1,N+2).\] 
(i) We start from \eqref{eq:Smu}.\\
\noindent
- Case I:
For $N\leq n-3$, the permutation $\tau_{l+1}\cdots\tau_{N+1}$ never touches the support of $\mu_n$. As a consequence,
\begin{eqnarray*}
(\tau_{l+1}\cdots\tau_{N+1}\circ\hat\mu_n)^-_l &=& (l,l+2,\cdots,N+2),\\
(\tau_{l+1}\cdots\tau_{N+1}\circ\hat\mu_n)^+_l &=& (1,\cdots,l-1,l+1,N+3,\cdots)=\mu_{n+l-N-2}.
\end{eqnarray*}
Setting $l=N+2-m$, with $1\leq m\leq N+1$, proves \eqref{eq:Dmu-sys1}.\\
(ii) We prove it by induction on the size of the support of the deformation, assuming it to be true for all deformations of size smaller than $|\mu|$. For $N\geq n+ |\mu|-2$, the derivatives $\nabla_1\cdots\nabla_{N+2}$ act on all the deformation zone. Again we start from  \eqref{eq:Smu} and look at the different contributions $(\tau_{l+1}\cdots\tau_{N+1}\circ\hat\mu_n)^\pm_l $.\\
- Case II: For $N+1\geq l\geq n+|\mu|$, the cut at points $l$ and $l+1$ are away from the deformation zone and we have
\begin{eqnarray*}
(\tau_{l+1}\cdots\tau_{N+1}\circ\hat\mu_n)^-_l &=& (l,l+2,\cdots,N+2),\\
(\tau_{l+1}\cdots\tau_{N+1}\circ\hat\mu_n)^+_l &=& (1,\cdots,l-1,l+1,N+3,\cdots).
\end{eqnarray*}
In $(\tau_{l+1}\cdots\tau_{N+1}\circ\hat\mu_n)^-_l$, there are $m=N+2-l$ points, and the derivatives act on all of them.
These terms contribute for $\sum_{m=1}^{N+2-n-|\mu|} C_{m}C^{\mu_n}_{N+k+2-m;k}$ in \eqref{eq:Smu}.\\
- Case III: For $n\leq l \leq n+|\mu|-1$, the cut at points $l$ and $l+1$ is inside the deformation zone. Hence it produces smaller cycles with deformation of size strictly smaller than $|\mu|$. Let $j_*:= \mu_{n}^{-1}(l)$ the pre-image of $l$ by $\mu_{n}$. Since $l$ is left invariant by $\tau_{l+1}\cdots\tau_{N+1}$, it is also the pre-image of $l$ by $\Omega\mu^l_{n}$, i.e. $\Omega\mu^l_{n}(j_*)=l$. The two sub-cycles obtained by cutting at $l$ and $l+1$ are : 
\begin{align*}
(\tau_{l+1}\cdots\tau_{N+1}\circ\hat{\mu}_{n})_{l}^{-} & =(l,\Omega\mu^l_{n}(j_*+1),\Omega\mu^l_{n}(j_*+2),\cdots,\Omega\mu^l_{n}(n+|\mu|-1),n+|\mu|+1, \cdots,N+2)\\
(\tau_{l+1}\cdots\tau_{N+1}\circ\hat{\mu}_{n})_{l}^{+} & =(1,\cdots,\Omega\mu^l_{n}(j_*-1),l+1,N+3,\cdots)
\end{align*}
The string of derivatives $\nabla_{1}\cdots\nabla_{N+2}$ acts on all the points inside the deformation. By construction, as $l$ goes from $n$ to $n+|\mu|-1$, its pre-image $j_*$ takes every value between $n$ and $n+|\mu|-1$ once. Using additionally that $C_{k;0}^{\sigma}$ is independent of $\sigma$, these terms contribute for $\sum_{m=N+3-n-|\mu|}^{N+2-n}C_{m}C_{N+k+2-m;k}$.\\
- Case IV: For $1\leq l\leq n-1$, the cut is before the deformation zone and we have
\begin{eqnarray*}
(\tau_{l+1}\cdots\tau_{N+1}\circ\hat\mu_n)^-_l &=& (l,l+2,\cdots,n,[\cdots\Omega\mu\cdots],n+|\mu|+1,\cdots,N+2),\\
(\tau_{l+1}\cdots\tau_{N+1}\circ\hat\mu_n)^+_l &=& (1,\cdots,l-1,l+1,N+3,\cdots) ,
\end{eqnarray*}
where the notation $[\cdots\Omega\mu\cdots]$ means that we insert the sequence defined by the deformation $\mu$ shifted by the cyclic permutation $\tau_{l+1}\cdots\tau_{N+1}$. The derivatives $\nabla_1\cdots\nabla_{N+2}$ act on all points of $(\tau_{l+1}\cdots\tau_{N+1}\circ\hat\mu_n)^-_l$ which contains the deformation. Again, using that $C^\sigma_{k;0}$ is independent of $\sigma$, these terms contribute  for $\sum_{m=N+3-n}^{N+1} C_{m}C_{N+k+2-m}$ in \eqref{eq:Smu}.
Hence,
\begin{equation*}
S^{\mu_n}_{N+k+2;k}= \sum_{m=1}^{N+2-n-|\mu|} C_{m}C^{\mu_n}_{N+k+2-m;k}
+\sum_{m=N+3-n-|\mu|}^{N+1} C_m C_{N+k+2-m;k},
\end{equation*}
The rest of the proof is by induction on $N$. For $N+2=n+|\mu|$, the first term is absent and hence $S^{\mu_n}_{n+|\mu|+k;k}=S_{n+|\mu|+k;k}$, independently of $\mu$, and thus $C^{\mu_n}_{n+|\mu|+k;k}=C_{n+|\mu|+k;k}$. Next, for $N+2=n+|\mu|$, the first sum is $C_{m}C^{\mu_n}_{n+|\mu|+k;k}$, but we just proved that $C^{\mu_n}_{n+|\mu|+k;k}=C_{n+|\mu|+k;k}$, so that it completes the second sum, and thus $S^{\mu_n}_{n+|\mu|+k+1;k}=S_{n+|\mu|+k+1;k}$. And so on by iteration.
\cqfd

With $N+2=M+n-1$ and $1\leq M \leq |\mu|$, the hole coefficients $S_{N+k+2;k}^{\mu_n}$ involve a deformation starting at $n$, derivatives w.r.t. the $N+1$ first $x$-variables, and the evaluation at zero for the $(N+2)$-th variables. If $M=1$, the series of derivatives are up to the last position before the support of the deformation and the evaluation at zero is at the first position of the deformation, if $|\mu|\leq M\leq 2$, the last derivative and the evaluation are both inside the support of the deformation.

To describe the relations within the deformation we use to the notions of extractions of a deformation introduced in the previous Section \ref{sec:def-extra}. With these definitions we can formulate the following

\begin{lemma}
For $n\leq N+2 \leq n +|\mu|-1$, set $N+2=M+n-1$ with $1\leq M \leq |\mu|$. For $k\geq |\mu|-M$, we let $P=N+k+2$. We have:
 
\begin{eqnarray} \label{eq:Smu-inside}
S_{P;k}^{\mu_n} &=& \sum_{l=1}^{n-1} C^{(\mu_\varpi^{M]})_{n-l+1}}_{n-l+|\mu^{M]}|;k^{M]}}\, C^{(\mu_\varpi^{[M})_{l+1}}_{P-n+l-|\mu^{M]}|; k-k^{M]}}\\
&& + \sum_{ m=1}^{M-1} C^{\mu_\varpi^{[m,M]}}_{|\mu^{[m;M]}|;k^{[m;M]}}\, C^{(\mu_\varpi^{m][M})_{n}}_{P-|\mu^{[m;M]}|; k-k^{[m;M]}} \quad ,\nonumber
\end{eqnarray}
The numbers $k^{[m;M]}$ and $k^{M]}$ are the number of floating variables in the corresponding extractions of the deformation $\mu$:
\begin{eqnarray*}
k^{[m,M]} &=& \# \{j\in[M+1,|\mu|]\ \mathrm{s.t.}\ j\in \mathrm{Im}\, \mu_\varpi^{[m,M]}\}\\
k^{M]} &=& \# \{j\in[M+1,|\mu|]\ \mathrm{s.t.}\ j\in \mathrm{Im}\, \mu_\varpi^{M]}\}
\end{eqnarray*}
 The floating variables in $\mu_\varpi^{M]}$ are those $x_{j+n-1}=y_{k+M-j}$ with $j\in \mathrm{Im}\mu_\varpi^{M]}\cap[M+1,|\mu|]$, and similarly for the other extractions  $\mu_\varpi^{[M}$,  $\mu_\varpi^{[m,M]}$ and  $\mu_\varpi^{m][M}$. The total number of floating variables in the deformation zone is $p:=|\mu|-M$.
\end{lemma}

{\it Proof.} 
The proof follows from a case by case analysis (and part of the difficulty is in the writing). Recall we set $P=N+2+k$ and $N+2=M+n-1$ with $1\leq M\leq |\mu|$. We start from \eqref{eq:Smu}, with $l$ running from $1$ to $N+1$. It involves the action of the product $\tau_{l+1}\cdots\tau_{N+1}$ on the deformation and the breaking of the deformation twisted by this product at the points $l$ and $l+1$. Recall that $\tau_{l+1}\cdots\tau_{N+1}$ is the following cyclic permutation
\[ \tau_{l+1}\cdots\tau_{N+1} = (l+1,l+2,\cdots,N+1,N+2).\]
As in the previous lemma, let $\hat \mu_n:=\mu_n\circ\omega_{N+k+1}$ and $\Omega\mu_{n}^{l}:=\tau_{l+1}\cdots\tau_{N+1}\mu_n$. Let $I_*$ be the pre-image of $M$ by $\mu$, i.e. $I_*=\mu^{-1}(M)$ or equivalently $\mu(I_*)=M$ so that $\mu_n(I_*+n-1)=N+2$ and $\Omega\mu_n^l(I_*+n-1)=l+1$.\\
- Case V: Let us first start with $N+1\geq l \geq n$, and set $l=m+n-1$ with $M-1\geq m \geq 1$ (hence these contributions are not present for $M=1$). All points between $l$ and $N+2$ are inside the deformation. 
Let $i_*$ be the pre-image of $m$ by $\mu$, i.e. $i_*=\mu^{-1}(m)$ or equivalently $\mu(i_*)=m$ so that $\mu_n(i_*+n-1)=\Omega\mu_n^l(i_*+n-1)=l$.
From \eqref{eq:sigma0}, we have to extract the sub-cycles $(\tau_{l+1}\cdots\tau_{N+1}\circ\hat\mu_n)^\pm_l$. We have
\begin{eqnarray*}
&&(\tau_{l+1}\cdots\tau_{N+1}\circ\hat\mu_n)^-_l = (\Omega\mu_n^l(i_*+n-1),\cdots,\Omega\mu_n^l(I_*+n-2)),\\
&&(\tau_{l+1}\cdots\tau_{N+1}\circ\hat\mu_n)^+_l  =\\
&& \hskip -0.8 truecm  (1,\cdots,n-1,\Omega\mu_n^l(n),\cdots,\Omega\mu_n^l(i_*+n-2),\Omega\mu_n^l(I_*+n-1),\cdots,\Omega\mu_n^l(|\mu|+n-1),n+|\mu|,\cdots),
\end{eqnarray*}
if $i_*<I_*$ and $\Omega \mu_n^l(n) \neq l$ (i.e. $i_*\not=1$). For $\Omega \mu_n^l(n)=l$ (i.e. $i_*=1$),
\begin{eqnarray*}
(\tau_{l+1}\cdots\tau_{N+1}\circ\hat\mu_n)^+_l  = (1,\cdots,n-1,\Omega\mu_n^l(I_*+n-1),\cdots,\Omega\mu_n^l(|\mu|+n-1),n+|\mu|,\cdots) .
\end{eqnarray*}
Recall that $\Omega\mu_n^l(i_*+n-1)=l$ and $\Omega\mu_n^l(I_*+n-1)=l+1$. Hence, $(\tau_{l+1}\cdots\tau_{N+1}\circ\hat\mu_n)^-_l $ is equivalent to $\mu_\varpi^{(m,M)}$ and $(\tau_{l+1}\cdots\tau_{N+1}\circ\hat\mu_n)^+_l$ to $ \mu_\varpi^{m)(M}$ but starting at the $n$-th position. The sub-cycles are thus obtained by translating the extractions of $\mu$ obtained by cutting at $m$ and $M$, and we have
\begin{eqnarray*}
(\tau_{l+1}\cdots\tau_{N+1}\circ\hat\mu_n)^-_l &=& \mu_\varpi^{(m,M)},\\
(\tau_{l+1}\cdots\tau_{N+1}\circ\hat\mu_n)^+_l &=& \big( \mu_\varpi^{m)(M}\big)_n,
\end{eqnarray*}
if $m$ and $M$ are naturally ordered along the cycle $\mu$, i.e. if $\mu^{-1}(m)<\mu^{-1}(M)$.
Similarly, 
\begin{eqnarray*}
(\tau_{l+1}\cdots\tau_{N+1}\circ\hat\mu_n)^-_l &=& \big( \mu_\varpi^{M)(m}\big)_n ,\\
(\tau_{l+1}\cdots\tau_{N+1}\circ\hat\mu_n)^+_l &=& \mu_\varpi^{(M,m)},
\end{eqnarray*}
if $m$ and $M$ are anti-ordered along the cycle $\mu$, i.e. if $\mu^{-1}(M)<\mu^{-1}(m)$.\\
- Case VI: Let us now consider $l$ such that $1\leq l\leq n-1$. The action of the permutation $\tau_{l+1}\cdots\tau_{N+1}$ within the deformation is stable, in the sense that it is $l$ independent, except that $N+2\to l+1$. Outside the deformation, it acts as $(1,\cdots,n-1)\to (1,\cdots,l,l+2,\cdots,n)$. Hence, when breaking it by cutting at $l$ and $l+1$, we break it before the deformation zone at point $l$ and inside the deformation zone at the image point of $N+2$ (which is the translation of the point $M$ by $n-1$ step). 
As a consequence, 
\begin{eqnarray*}
(\tau_{l+1}\cdots\tau_{N+1}\circ\hat\mu_n)^-_l &=& (l,l+2,\cdots,n,\Omega\mu^l_n(n),\cdots, \Omega\mu^l_n(I_*+n-2)),\\
(\tau_{l+1}\cdots\tau_{N+1}\circ\hat\mu_n)^+_l &=& (1,\cdots,l-1,l+1,\Omega\mu^l_n(I_*+n)),\cdots,\Omega\mu_n^l(n+|\mu|-1),n+|\mu|,\cdots).
\end{eqnarray*}
Now, in the string of numbers $(\Omega\mu^l_n(n),\cdots, \Omega\mu^l_n(I_*+n-2))$ we recognize $\mu_\varpi^{M]}$ but inserted at the $n-l+1$-th position, and the string of numbers $(\Omega\mu^l_n(I_*+n)),\cdots,\Omega\mu_n^l(n+|\mu|-1))$ coincides with $\mu_\varpi^{[M}$ but inserted at the $l$-th position, thus
\begin{eqnarray*}
(\tau_{l+1}\cdots\tau_{N+1}\circ\hat\mu_n)^-_l &=& \big(\mu_\varpi^{M]}\big)_{n-l+1},\\
(\tau_{l+1}\cdots\tau_{N+1}\circ\hat\mu_n)^+_l &=& \big( \mu_\varpi^{[M}\big)_{l+1} .
\end{eqnarray*}
This proves \eqref{eq:Smu-inside}.\\
- Since \eqref{eq:sigma0} involves the successive derivatives $\nabla_1\cdots\nabla_{N+2}$, with $N+2=M+n-1$, it is clear that the remaining floating variables within the deformation zone are the $x_{j+n-1}$'s with $j=M+1,\cdots,|\mu|$. These variables are attached to the various extractions involved in \eqref{eq:Smu-inside} according to the images of those extractions. Finally, note that $x_{j+n-1}=y_{k+M-j}$ for $P=N+2+k$.
\cqfd

{\bf Remark.} The above formula reproduce the formulas \eqref{eq:Dnsys} for the simple transposition. 
Let the deformation be the transposition $(12)$, ie. $\mu:=(12)=\left(\begin{smallmatrix} 1&2\\2&1\end{smallmatrix}\right)$, with length $|\mu|=2$. \\
For $M=1$, there is only the first term in \eqref{eq:Smu-inside}. We have $\mu^{1]}=\left(\begin{smallmatrix} 1\\2\end{smallmatrix}\right)$ and $\mu^{[1}=\emptyset$, and since $\omega^1_M$ is trivial in this case, $\mu_\varpi^{1]}=\left(\begin{smallmatrix} 1\\2\end{smallmatrix}\right)$ and $\mu_\varpi^{[1}=\emptyset$. As element of $\mathbb{S}_1$, they are both trivial. The number of floating variable in the deformation is $|\mu|-M=1$. It is attached to the extraction $\mu_\varpi^{1]}$ since it is the element of $[M+1,|\mu|]=\{2\}$ in the image of the extractions. Hence, in this case
\[ 
S_{P;k}^{\mu_n} = \sum_{l=1}^{n-1} C_{n-l+1}C_{P+l-n-1;k-1}.
\]
For $M=2$, the two types of terms in \eqref{eq:Smu-inside} contribute. For the first terms, we have $\mu^{2]}=\emptyset$ and $\mu^{[2}=\left(\begin{smallmatrix} 2\\1\end{smallmatrix}\right)$, and since $\omega^1_M=(12)$, we have $\mu_\varpi^{2]}=\emptyset$ and $\mu_\varpi^{[2}=\mathrm{Id}$. We have $|\mu^{2]}|=0$ and $|\mu^{[2}|=1$. For the second terms, with $m=1,M=2$, we have $\mu^{[m,M]}=\mu^{(2,1)}=\left(\begin{smallmatrix} 1\\2\end{smallmatrix}\right)$ and $\mu^{m][M}=\mu^{2)(1}=\left(\begin{smallmatrix} 2\\1\end{smallmatrix}\right)$, and since $\omega^{m+1}_M$ is trivial, $\mu_\varpi^{[1,2]}=\left(\begin{smallmatrix} 1\\2\end{smallmatrix}\right)$ and $\mu_\varpi^{1][2}=\left(\begin{smallmatrix} 2\\1\end{smallmatrix}\right)$. we have $|\mu^{[1,2]}|=|\mu^{1][2}|=1$. As element of $\mathbb{S}_1$, they are both trivial. There are no floating variables. Hence, in this case
\[ 
S_{P;k}^{\mu_n} = \sum_{l=1}^{n-1} C_{n-l}C_{P+l-n;k} + C_{1}C_{P-1;k}.
\]
This coincides with \eqref{eq:Dnsys}.	

	\subsection{Generating functions}
We first introduce the generating functions associated to the relations away from the deformation. For any permutation $\mu\in \mathbb{S}_\mu$  we let, for $q\geq 1$ and $k\geq |\mu|$,
\begin{subequations} \label{eq:def-DSgeneral} 
\begin{align} 
\mathcal{D}_k^{(\mu;q)}(z) &:=\sum_{N\geq 0} C_{N+k+1;k}^{\mu_{q+N}}\, z^N , \label{eq:def-DSgeneral-a} \\
\mathcal{S}_k^{(\mu;q)}(z) &:=\sum_{N\geq 0} {S}_{N+k+2;k}^{\mu_{q+N+1}}\,z^N . \label{eq:def-DSgeneral-b}
\end{align}
\end{subequations} 
As in Section \ref{sec:trans-deformation}, the labels $\mu$ code for the profile of the deformation and $q$ for its location along the chain interval. These functions depend on the $k$ variables $(y_0,\cdots,y_{k-1})$.
We have $\mathcal{D}_k^{(\mu;q)}(0)=C^{\mu_q}_{k+1;k}$, which is the derivative w.r.t. $x_1$ of the deformed loop. Thus
\begin{equation} \label{eq:reconstruction}
 [\mu_q\circ\omega_{k+1}](x_1,\cdots,x_{k+1})= x_1\, \mathcal{D}_k^{(\mu;q)}(0)(y_0,\cdots,y_{k-1}),\ \mathrm{with}\ x_j=y_{k+1-j}.
 \end{equation}
Hence, we have to solve for $\mathcal{D}_k^{(q)}(0)$, for $q\geq 1$. This is done in a series of Propositions.

\begin{proposition} \label{prop:DS-general}
The following recurrence relations hold, for $k\geq|\mu|$:
\begin{subequations} \label{eq:Dmu-away-gene}
\begin{align}
\mathcal{D}_{k+1}^{(\mu;q)}(z) &= \mathcal{S}_k^{(\mu;q-1)}(z) + y_kz^{-1}\,\big(\mathcal{D}_k^{(\mu;q-1)}(z)-\mathcal{D}_k^{(\mu;q-1)}(0)\big) ,\label{eq:Dmu-away-gene-a}\\
\mathcal{S}_k^{(\mu;q)}(z) &= \mathfrak{c}(z)\, \mathcal{D}_k^{(\mu;q)}(z), \quad \mathrm{for}\ q\geq 2 . \label{eq:Dmu-away-gene-b}
\end{align}
\end{subequations}
These equations are equivalent to \eqref{eq:Dgene-recur-intro}.
\end{proposition}

{\it Proof.}
(i) Eq.\eqref{eq:Dmu-away-gene-a} is a direct consequence of $C^{\mu_n}_{P;k+1}=S^{\mu_n}_{P;k}+y_kC^{\mu_n}_{P;k}$.\\
(ii) Eq.\eqref{eq:Dmu-away-gene-b} follows by checking that it is equivalent to the first equation in \eqref{eq:Dmu-sys1}. 
\cqfd

We now introduce generating functions appropriate for the recurrence relations within the bulk of the deformation, with $k'\geq 0$ ($k=k'+p'$) and $0\leq p'\leq |\mu|-1$ ($M=|\mu|-p'$),
\begin{eqnarray} \label{eq:def-gene-bis}
\widehat{\mathcal{D}}_{k';p'}^{\mu}(z) &:=& \sum_{N'\geq 0} C^{\mu_{N'+1}}_{N'+k'+|\mu|;k'+p'}\, z^{N'} ,\\
\widehat{\mathcal{S}}_{k';p'}^{\mu}(z) &:=& \sum_{N'\geq 0} S_{N'+k'+|\mu|;k'+p'}^{\mu_{N'+1}}\,z^{N'} ,
\end{eqnarray}
The first term in $\widehat{\mathcal{S}}_{k';p'}^{\mu}(z)$ for $p'=|\mu|-1$ is ill defined, we set it to $0$ by convention.
The sums are over the locations of the deformation, fixing the number $k'$ of floating variables after the deformation and the number $p'$ of variables inside the deformation. The total number of variables is $k=k'+p'$, the total number of points is $N'+k'+|\mu|$ and the position of the deformation is $n=N'+1$. 

We have
\begin{subequations} \label{eq:DS-initial}
\begin{align}
 \mathcal{D}^{(\mu;1)}_{k+|\mu|}(z) &= \widehat{\mathcal{D}}^{\mu}_{k+1;|\mu|-1}(z) , \label{eq:DS-initial-a}\\
  \mathcal{S}^{(\mu;1)}_{k+|\mu|}(z) &= z^{-1}\big(\widehat{\mathcal{S}}^{\mu}_{k+1;|\mu|-1}(z)-\widehat{\mathcal{S}}^{\mu}_{k+1;|\mu|-1}(0)\big) .\label{eq:DS-initial-b}
  \end{align}
  \end{subequations}

We can then recast the recurrence relation \eqref{eq:Smu-inside}.

\begin{proposition} \label{prop:DS-general-through}
We have:\\
(i) For all $0\leq p'\leq |\mu|-1$, and $M=|\mu|-p'$, we have:
\begin{subequations} \label{eq:Dmu-inside-gene}
\begin{align}
\widehat{\mathcal{D}}_{k';p'+1}^{\mu}(z) &= \widehat{\mathcal{S}}_{k';p'}^{\mu}(z) + y_{k'+p'}\, \widehat{\mathcal{D}}_{k';p'}^{\mu}(z) ,\label{eq:Dmu-inside-gene-a}\\
\widehat{\mathcal{S}}^{\mu}_{k';p'}(z) &=
z^{-1}\Big(\widehat{\mathcal{D}}^{\mu_\varpi^{M]}}_{0;k^{M]}}(z) - \widehat{\mathcal{D}}^{\mu_\varpi^{M]}}_{0;k^{M]}}(0)\Big)\,\Big(\widehat{\mathcal{D}}^{\mu_\varpi^{[M}}_{k';k^{[M}}(z)-\widehat{\mathcal{D}}^{\mu_\varpi^{[M}}_{k';k^{[M}}(0)\Big) \label{eq:Dmu-inside-gene-b} \\
& ~~~~~~ + \sum_{m=1}^{M-1} \widehat{\mathcal{D}}^{\mu_\varpi^{[m,M]}}_{0;k^{[m,M]}}(0) \widehat{\mathcal{D}}^{\mu_\varpi^{m][M}}_{k';k^{m][M}}(z) , \nonumber 
\end{align}
\end{subequations}
with $k^{[M}+k^{M]}=p'$ and $k^{m][M}+k^{[m,M]}=p'$ (recall that $p'$ is the number of variables in the deformation zone). 
By convention, the last sum in \eqref{eq:Dmu-inside-gene-b} is absent for $p'=|\mu|-1$ ($M=1$), and we choose the normalization $\widehat{\mathcal{S}}^{\mu}_{k';|\mu|-1}(0)=0$.\\
 These floating variables in $\mu_\varpi^{M]}$ are those $x_{j+n-1}=y_{k+M-j}$ with $j\in \mathrm{Im}\mu_\varpi^{M]}\cap[M+1,|\mu|]$, and similarly for the other extractions  $\mu_\varpi^{[M}$,  $\mu_\varpi^{[m,M]}$ and  $\mu_\varpi^{m][M}$. The total number of floating variables in the deformation zone is $p:=|\mu|-M$, the other variables $y_{k'-1},\cdots,y_0$ are away from the deformation zone.\\
(ii) Moreover,
\begin{equation} \label{eq:Dmu-initial}
\widehat{\mathcal{D}}_{k';0}^{\mu}(z) = \Big[z^{1-|\mu|}\,\mathcal{C}_{k'}(z) \Big]_+,
\end{equation}
 where  $\big[\cdots\big]_+$ means the part with positive degrees of the Laurent series. 
\end{proposition}

{\it Proof.}
(i) Eq.\eqref{eq:Dmu-inside-gene-a} is a direct consequence of $C^{\mu_n}_{P;k+1}=S^{\mu_n}_{P;k}+y_kC^{\mu_n}_{P;k}$.\\
(ii) To verify that \eqref{eq:Dmu-inside-gene-b} is equivalent to \eqref{eq:Smu-inside} we rewrite the later. In eq.\eqref{eq:Smu-inside}, we set $n=N'+1$, $k=k'+p'$ with $M+p'=|\mu|$ so that $P=N+2+k=N'+k'+|\mu|$. Then \eqref{eq:Smu-inside} can be written as
\begin{eqnarray*} 
S_{N'+k'+|\mu|;k'+p'}^{\mu_{N'+1}} &=& \sum_{l=1}^{N'} C^{(\mu_\varpi^{M]})_{N'-l+2}}_{N'+|\mu^{M]}|-l+1;k^{M]}}\, C^{(\mu_\varpi^{[M})_{l+1}}_{k'+|\mu^{[M}|+l; k'+k^{[M}}\\
&& + \sum_{ m=1}^{M-1} C^{\mu_\varpi^{[m,M]}}_{|\mu^{[m;M]}|;k^{[m;M]}}\, C^{(\mu_\varpi^{m][M})_{N'+1}}_{N'+k'+\mu^{m][M}|; k'+k^{m][M}} \quad ,\nonumber
\end{eqnarray*}
where $|\mu^{M]}|+|\mu^{[M}|=|\mu|-1$, $|\mu^{[m,M]}|+|\mu^{m][M}|=|\mu|$ are the respective sizes of the extractions, and $k^{M]}+k^{[M}=p'$, $k^{[m,M]}+k^{m][M}=p'$ the respecting number of floating variables on those extractions.\\
(ii) Eq.\eqref{eq:Dmu-initial} is equivalent to $C^{\mu_n}_{N+k+1;k}=C_{N+k+1;k}$ for $N\geq n+|\mu|-2$, independently of the deformation $\mu$.
\cqfd

Moreover, the stabilisation phenomena observed in the case of the regular loop hold in general. For any deformation, there exists a generating function for the loop expectation values with an arbitrary number of points.

\begin{theorem} \label{th:defom-Dseries}
For all deformations $\mu$, of finite size support, and all locations $q\geq2$, there exists a generating function $\overline{D}^{\mu;q}$ depending an infinite number of variables $y_0,y_1,\cdots$ such that
\begin{equation} \label{eq:Dmu-infini}
[\mu_q\circ\omega_{k+1}]({\bm x}) = x_1\, \overline{D}^{\mu;q}(y_0,\cdots,y_{k-1},0,0,\cdots), 
\end{equation}
with $k\geq|\mu|$ and $y_l=x_{k+1-l}$.\\
That is: the expectation values with $k+1$ points are obtained by evaluating the generating function $\overline{D}^{\mu;q}$ with all variables $y_k,y_{k+1},\cdots$ set to zero. This function $\bar D^{\mu;q}$ is recursively constructed using (\ref{eq:Dmu-away-gene},\ref{eq:Dmu-inside-gene}) with initial condition \eqref{eq:Dmu-initial}.
\end{theorem}

{\it Proof.}
The relation \eqref{eq:Dmu-away-gene-b}, $\mathcal{S}_k^{(\mu;q)}(z) = \mathfrak{c}(z)\, \mathcal{D}_k^{(\mu;q)}(z)$, for $q\geq 2$ and $k\geq|\mu|$, evaluated at $z=0$ implies that $\mathcal{S}_k^{(\mu;q)}(0) = \mathcal{D}_k^{(\mu;q)}(0)$, or equivalently, $S^{\mu;q}_{k+2;k}=C^{\mu;q}_{k+1;k}$. By the definition of the hole decomposition $S^{\mu;q}_{k+2;k}=C^{\mu;q}_{k+2;k+1}\vert_{x_2=0}$. Hence, 
\[ \mathcal{D}_k^{(\mu;q)}(0)(y_0,\cdots,y_{k-1})= \mathcal{D}_{k+1}^{(\mu;q)}(0)(y_0,\cdots,y_{k-1},y_k=0) ,\]
since $\mathcal{D}_k^{(\mu;q)}(0)=C^{\mu;q}_{k+1;k}$. By iterating at infinitum, we infer that there exists a function $\bar D^{\mu;q}$ depending an infinite number of variables $y_0,y_1,\cdots$ such that
\[ \mathcal{D}_k^{(\mu;q)}(0)(y_0,\cdots,y_{k-1}) = \overline{D}^{\mu;q}(y_0,\cdots,y_{k-1},0,0,\cdots).\]
Since $\mathcal{D}_k^{(\mu;q)}(0)=\nabla_1[\mu_q\circ\omega_{k+1}]=x_1^{-1}[\mu_q\circ\omega_{k+1}]$ this proves \eqref{eq:Dmu-infini}.
\cqfd

Eqs.\eqref{eq:Dmu-inside-gene} with the initial condition \eqref{eq:Dmu-initial} allow to determine $\widehat{\mathcal{D}}_{k';p'}^{\mu}(z)$ and $\widehat{\mathcal{S}}_{k';p'}^{\mu}(z)$ recursively, starting from $p'=0$ and increasing $p'$ up to $|\mu|-1$. Indeed, starting from the initial condition \eqref{eq:Dmu-initial} for $\widehat{\mathcal{D}}_{k';0}^{\mu}(z)$ and $\widehat{\mathcal{S}}_{k';0}^{\mu}(z)$ given by \eqref{eq:Dmu-inside-gene-b} we compute $\widehat{\mathcal{D}}_{k';1}^{\mu}(z)$ using \eqref{eq:Dmu-inside-gene-a}. Then by iteration using the just determined $\widehat{\mathcal{D}}_{k';1}^{\mu}(z)$ and $\widehat{\mathcal{S}}_{k';0}^{\mu}(z)$ from \eqref{eq:Dmu-inside-gene-b}, we get $\widehat{\mathcal{D}}_{k';2}^{\mu}(z)$ using \eqref{eq:Dmu-inside-gene-a}. This continues up to determining $\widehat{\mathcal{D}}_{k';|\mu|-1}^{\mu}(z)$. Eq.\eqref{eq:Dmu-inside-gene-b} for $p'=|\mu|-1$ gives $\widehat{\mathcal{S}}_{k';|\mu|-1}^{\mu}(z)$. This yields $ \mathcal{D}^{(\mu;1)}_{k+|\mu|}(z)$ and $ \mathcal{S}^{(\mu;1)}_{k+|\mu|}(z)$ since,
\begin{eqnarray*}
 \mathcal{D}^{(\mu;1)}_{k+|\mu|}(z) &=& \widehat{\mathcal{D}}^{\mu}_{k+1;|\mu|-1}(z) ,\\
  \mathcal{S}^{(\mu;1)}_{k+|\mu|}(z) &=& z^{-1}\big(\widehat{\mathcal{S}}^{\mu}_{k+1;|\mu|-1}(z)-\widehat{\mathcal{S}}^{\mu}_{k+1;|\mu|-1}(0)\big)
\end{eqnarray*}
Once we know $\mathcal{D}^{(\mu;1)}_{k+|\mu|}(z)$ and $\mathcal{S}^{(\mu;1)}_{k+|\mu|}(z)$, we can iterate the recurrence relations \eqref{eq:Dmu-away-gene} and determine $\mathcal{D}^{(\mu;q)}_{k+|\mu|}(z)$ for all $q\geq 1$, and thus all expectation values $[\mu_q\circ\omega_{k+1}]({\bm x})$ via the reconstruction formula \eqref{eq:reconstruction}.

{\bf Remark.}	
Let us show how to recover the results \eqref{eq:Dmu-inside-gene} for a simple permutation from the above formulation. We have to match the different definitions of generating functions.

For $M=1$ ($p'=1$), the extraction $\mu_\varpi^{1]}$ and $\mu_\varpi^{[1}$ are trivial with $|\mu^{1]}|=1$, $k^{1]}=1$ and $|\mu^{[1}|=0$, $k^{[1}=0$, and we have
\begin{eqnarray*}
\widehat{\mathcal{D}}^{\mu_\varpi^{1]}}_{0;1}(z) - \widehat{\mathcal{D}}^{\mu_\varpi^{1]}}_{0;1}(0) = z\,\mathcal{C}_1(z),\
\widehat{\mathcal{D}}^{\mu_\varpi^{[1}}_{k';0}(z) - \widehat{\mathcal{D}}^{\mu_\varpi^{[1}}_{k';0}(0) = z\,\mathcal{C}_{k'}(z).
\end{eqnarray*}
Eq.\eqref{eq:Dmu-inside-gene-b} becomes (with the convention $\widehat{\mathcal{S}}^{\mu}_{k';1}(0)=0$)
\[
\widehat{\mathcal{S}}^{\mu}_{k';1}(z) = z\,\mathcal{C}_1(z)\, \mathcal{C}_{k'}(z)
\]
For $M=2$ ($p'=0$), we have $\mu_\varpi^{2]}=\emptyset$ and $\mu_\varpi^{[2}=\mathrm{Id}$ with $|\mu_\varpi^{2]}|=0$ and $|\mu_\varpi^{[2}|=1$, so that
\begin{eqnarray*}
\widehat{\mathcal{D}}^{\mu_\varpi^{2]}}_{0;0}(z) - \widehat{\mathcal{D}}^{\mu_\varpi^{2]}}_{0;0}(0) = z\mathfrak{c}(z),\
\widehat{\mathcal{D}}^{\mu_\varpi^{[2}}_{k';0}(z) = \mathcal{C}_{k'}(z).
\end{eqnarray*}
We also have that $\mu_\varpi^{[1,2]}$ and $\mu_\varpi^{1][2}$ are both trivial with $|\mu^{[1,2]}|=|\mu^{1][2}|=1$ so that
\begin{eqnarray*}
\widehat{\mathcal{D}}^{\mu_\varpi^{[1,2]}}_{0;0}(0) = \mathfrak{c}(0)=1,\
\widehat{\mathcal{D}}^{\mu_\varpi^{1][2}}_{k';0}(z) = \mathcal{C}_{k'}(z).
\end{eqnarray*}
Eq.\eqref{eq:Dmu-inside-gene-b} becomes 
\[
\widehat{\mathcal{S}}^{\mu}_{k';0}(z) = \mathfrak{c}(z)\big(\mathcal{C}_{k'}(z)-\mathcal{C}_{k'}(0)\big)+\mathfrak{c}(0)\, \mathcal{C}_{k'}(z).
\]
From the definition of the generating functions for a transposition deformation, we have
\begin{eqnarray*}
\mathcal{D}^{(1)}_{k+1}(z) = \widehat{\mathcal{D}}_{k;1}(z) &,& 
\mathcal{D}^{(0)}_{k}(z) -\mathcal{D}^{(0)}_{k}(0) = z\, \widehat{\mathcal{D}}_{k;0}(z) ,\\
z\, \mathcal{S}^{(1)}_{k+1}(z) = \widehat{\mathcal{S}}_{k;1}(z) -\widehat{\mathcal{S}}_{k;1}(0) &,&
\mathcal{S}^{(0)}_{k}(z) =\widehat{\mathcal{S}}_{k;0}(z) .
\end{eqnarray*}
The initial condition is $\widehat{\mathcal{D}}_{k;0}(z)= \widehat{\mathcal{D}}_{k;0}^{\mu}(z) = \big[z^{-1}\,\mathcal{C}_{k'}(z) \big]_+= z^{-1}\big(\mathcal{C}_k(z)-\mathcal{C}_k(0)\big)$.
Making the correspondence between those quantities then shows that the system (\ref{eq:Dmu-away-gene},\ref{eq:Dmu-inside-gene}) coincides with \eqref{eq:Dn-gener}.

\section{Appendix} \label{sec:appendix}

\subsection{The exchange relations for  $[\omega_P]$ and $[\tau_n\circ\omega_P]$} \label{sec:exchange-Tn}

Here we give the proof that the pair $[\omega_P]$ and $[\tau_n\circ\omega_P]$ satisfy the exchange relation \eqref{eq:moves}, as claimed in Section \ref{sec:omega-Tn}.

{\it Proof.}
  The proof will be by induction on $n$ for all $P$. Let us recall the decomposition $[\sigma]=A_n(\sigma) +B_n(\sigma)x_n +C_n(\sigma)x_{n+1} +D_n(\sigma)x_nx_{n+1}$. \\
- Let us first prove \eqref{eq:moves-d} for the pair $[\omega_P]$ and $[\tau_n\circ\omega_P]$. It amounts to show that $\nabla_n\nabla_{n+1}[\tau_n\circ\omega_P]=\nabla_n\nabla_{n+1}[\omega_P]$. For any $\sigma$ we have
 \[ D_n(\sigma)=\sum_{j=1}^{n-2}x_1\cdots x_j \nabla_n\nabla_{n+1}[\sigma]_{j+1}^o + \sum_{j=n+1}^P\nabla_n\nabla_{n+1}x_1\cdots x_{j}\,[\sigma]_{j+1}^o.\]
 The last terms with $j\geq n+1$ coincide for $\sigma=\omega_P$ and $\sigma=\tau_n\circ\omega_P$, thanks to \eqref{eq:Dnsys-a}. The terms with $j\leq n-2$ for $\sigma=\tau_n\circ\omega_P$ reads
 \[ \nabla_n\nabla_{n+1}[\tau_n\circ\omega_P]_{j+1}^o = \sum_{k=1}^{j}D_{j+1-k|0}\nabla_n\nabla_{n+1}D^{(n+k-j-1)}_{k|P-j-1}({\bm x}_{\geq j+2}) .\]
 But $D^{(n+k-j-1)}_{k|P-j-1}({\bm x}_{\geq j+2})= \nabla_1\cdots\nabla_k[\tau_{n+k-j-1}\circ\mathfrak{c}_{P-j-1}]$ by definition. Hence, 
 \[ \nabla_n\nabla_{n+1}D^{(n+k-j-1)}_{k|P-j-1}({\bm x}_{\geq j+2})=\nabla_n\nabla_{n+1}D_{k|P-j-1}({\bm x}_{\geq j+2}) \]
  and $\nabla_n\nabla_{n+1}[\tau_n\circ\omega_P]_{j+1}^o=\nabla_n\nabla_{n+1}[\omega_P]_{j+1}^o$ for $j\leq n-2$.\\
- Let us now prove (\ref{eq:moves-c},\ref{eq:moves-b}) for the pair $[\omega_P]$ and $[\tau_n\circ\omega_P]$. For any $\sigma$ we have,
\begin{eqnarray*}
B_n(\sigma)=\sum_{j=1}^{n-2}x_1\cdots x_j \nabla_n[\sigma]_{j+1}^o\vert_{x_{n+1}=0} + x_1\cdots x_{n-1}[\sigma]_{n+1}^o,\\
C_n(\sigma)=\sum_{j=1}^{n-2}x_1\cdots x_j \nabla_{n+1}[\sigma]_{j+1}^o\vert_{x_{n}=0} + x_1\cdots x_{n-1}\nabla_{n+1}[\sigma]_{n}^o.
\end{eqnarray*}
Thanks to the expression \eqref{eq:Dnsys-d} of $[\tau_n\circ\omega_P]_{j+1}^o$ in terms of $D^{(n+k-j-1)}_{k|P-j-1}({\bm x}_{\geq j+2})$, for $j\leq n-2$, the contributions of the terms $x_1\cdots,x_j$ in this decomposition fulfil the moves (\ref{eq:moves-c},\ref{eq:moves-c})  by the induction hypothesis.\\
Proving (\ref{eq:moves-c},\ref{eq:moves-b}) for the terms $x_1\cdots x_{n_1}$ in this decomposition amounts to verify that
\begin{subequations} \label{eq:moveBnCn}
\begin{align}
~[\tau_n\circ\omega_P]_{n+1}^o= \nabla_{n+1} [\omega_P]_{n}^o + D_{1|0}D_{n|P-n-1}({\bm x}_{\geq n+2}), \label{eq:moveBnCn-a}\\
\nabla_{n+1} [\tau_n\circ\omega_P]_{n}^o = [\omega_P]_{n+1}^o -D_{1|0}D_{n|P-n-1}({\bm x}_{\geq n+2}). \label{eq:moveBnCn-b}
\end{align}
\end{subequations}
To prove \eqref{eq:moveBnCn-a}, recall that $[\omega_P]_{j+1}^o = \sum_{k=1}^{j}  D_{j+1-k|0} D_{k|P-j-1}({\bm x}_{\geq j+2})$ so that 
\begin{eqnarray*}
\nabla_{n+1} [\omega_P]_{n}^o &=& \sum_{k=1}^{n-1}  D_{n-k|0} \nabla_{n+1} D_{k|P-n}({\bm x}_{\geq n+1})\\
&=& \sum_{k=1}^{n-1}  D_{n-k|0} D_{k+1|P-n-1}({\bm x}_{\geq n+2})\\
&=& [\tau_n\circ\omega_P]_{n+1}^o - D_{1|0}D_{n|P-n-1}({\bm x}_{\geq n+2}),
\end{eqnarray*}
where in the last line we use formula \eqref{eq:Dnsys-b} expressing $[\tau_n\circ\omega_P]_{n}^o$ in terms of the $D_{k|P-k}$'s.\\ 
To prove \eqref{eq:moveBnCn-b}, we notice that, from \eqref{eq:Dnsys-c} and $\nabla_{n+1}D_{n-k|1}(x_{n+1})=D_{n+1-k|0}$, we have
\begin{eqnarray*}
\nabla_{n+1}[\tau_n\circ\omega_P]_{n}^o &=& \sum_{k=1}^{n-1}D_{n+1-k|0}D_{k|P-n-1}({\bm x}_{\geq n+2}) \\
&=& [\omega_P]_{n+1}^o -D_{1|0}D_{n|P-n-1}({\bm x}_{\geq n+2}).
\end{eqnarray*}
- Finally, we have to prove (\ref{eq:moves-a}) for the pair $[\omega_P]$ and $[\tau_n\circ\omega_P]$. For any $\sigma$ we have,
\[
A_n(\sigma)= \sum_{j=1}^{n-1} x_1\cdots x_j\,[\sigma]_{j+1}^o\vert_{x_n=0=x_{n+1}} .
\]
Again the contribution of the terms $x_1\cdots x_j$ for $1\leq j\leq n-2$ in this decomposition fulfil (\ref{eq:moves-a}) thanks to the induction hypothesis and the formula \eqref{eq:Dnsys-d} for $[\tau_n\circ\omega_P]_{j+1}^o$ for $1\leq j\leq n-2$. Let thus look at the term $x_1\cdots x_{n-1}$. Using $D_{n-k|1}(x_{n+1})= -D_{n+1-k|0}(1-x_{n+1})$, we have
\begin{eqnarray*}
~[\tau_n\circ\omega_P]_{n}^o\vert_{x_{n+1}=0} &=&  \sum_{k=1}^{n-1}D_{n-k|1}(x_{n+1})\vert_{x_{n+1}=0}D_{k|P-n-1}({\bm x}_{\geq n+2}) ,\\
&=&  - \sum_{k=1}^{n-1}D_{n+1-k|0}D_{k|P-n-1}({\bm x}_{\geq n+2}) .
\end{eqnarray*}
On the other hand, since $ [\omega_P]_{j+1}^o = \sum_{k=1}^{j}  D_{j+1-k|0} D_{k|P-j-1}({\bm x}_{\geq j+2}) $ and $D_{k|P-n}(x_{n+1},{\bm x}_{\geq n+2})= [\omega_{P+k-n}]_{k+1}^o$, we have
\begin{eqnarray*}
[\omega_P]_{n}^o\vert_{x_{n+1}=0} &=& \sum_{k=1}^{n-1}  D_{n-k|0} D_{k|P-n}({\bm x}_{\geq j+2})\vert_{x_{n+1}=0} \\
&=& \sum_{k=1}^{n-1}  D_{n-k|0} \sum_{l=1}^k D_{k+1-l|0}D_{l|P-n-1}({\bm x}_{\geq j+2}) \\
&=& - \sum_{k=1}^{n-1}D_{n+1-k|0}D_{k|P-n-1}({\bm x}_{\geq n+2}) = [\tau_n\circ\omega_P]_{n}^o\vert_{x_{n+1}=0} ,
\end{eqnarray*}
where we used $D_{m|0}= - \sum_{n=1}^{m-1}D_{n|0}D_{m-n|0}$ in the last line.
Thus we proved the exchange relations (\ref{eq:moves}) for the pair $[\omega_P]$ and $[\tau_n\circ\omega_P]$.
\cqfd

\subsection{Examples of extractions} \label{app:exemple}

In order to make the notations from Section \ref{sec:general-deformation} clear, we illustrate the different cases pictured in Figure \ref{fig:differentcases} on specific examples. For a deformation $\mu_n$, let $\hat \mu_n=\mu_n\circ\omega_P$ with $\omega_P$ the regular with $P$, as in the main text.

\noindent
\textbf{I.} For $N\leq n-3$, $l\leq N+1$, \\ let $P=10,n=7,l=2,N=4,|\mu|=2$. 
\begin{align*}
\mu & =\begin{pmatrix}\boldsymbol{1} & \boldsymbol{2}\\
\boldsymbol{2} & \boldsymbol{1}
\end{pmatrix},\\
\mu_{n} & =\begin{pmatrix}1 & 2 & 3 & 4 & 5 & 6 & \boldsymbol{7} & \boldsymbol{8} & 9 & 10\\
1 & \underset{l}{2} & \underset{l+1}{3} & 4 & 5 & \underset{N+2}{6} & \boldsymbol{8} & \boldsymbol{7} & 9 & 10
\end{pmatrix},\\
\tau_{l+1}\cdots\tau_{N+1}\mu_{n} & =\begin{pmatrix}1 & 2 & 3 & 4 & 5 & 6 & \boldsymbol{7} & \boldsymbol{8} & 9 & 10\\
1 & \underset{l}{2} & 4 & 5 & \underset{N+2}{6} & \underset{l+1}{3} & \boldsymbol{8} & \boldsymbol{7} & 9 & 10
\end{pmatrix},\\
(\tau_{l+1}\cdots\tau_{N+1}\circ\hat \mu_{n})_{l}^{-} & =\begin{pmatrix}\underset{l}{2} & 4 & 5 & \underset{N+2}{6}\end{pmatrix},\\
(\tau_{l+1}\cdots\tau_{N+1}\circ\hat \mu_{n})_{l}^{+} & =\begin{pmatrix}1 & \underset{l+1}{3} & \boldsymbol{8} & \boldsymbol{7} & 9 & 10\end{pmatrix}.
\end{align*}

\noindent
\textbf{II.} For $N\geq n+|\mu|-2$, $N+1\geq l\geq n+|\mu|$, \\
let $P=10,n=3,l=6,N=7,|\mu|=2.$
\begin{align*}
\mu & =\begin{pmatrix}\boldsymbol{1} & \boldsymbol{2}\\
\boldsymbol{2} & \boldsymbol{1}
\end{pmatrix},\\
\mu_{n} & =\begin{pmatrix}1 & 2 & \boldsymbol{3} & \boldsymbol{4} & 5 & 6 & 7 & 8 & 9 & 10\\
1 & 2 & \boldsymbol{4} & \boldsymbol{3} & 5 & \underset{l}{6} & \underset{l+1}{7} & 8 & \underset{N+2}{9} & 10
\end{pmatrix},\\
\tau_{l+1}\cdots\tau_{N+1}\mu_{n} & =\begin{pmatrix}1 & 2 & \boldsymbol{3} & \boldsymbol{4} & 5 & 6 & 7 & 8 & 9 & 10\\
1 & 2 & \boldsymbol{4} & \boldsymbol{3} & 5 & \underset{l}{6} & 8 & \underset{N+2}{9} & \underset{l+1}{7} & 10
\end{pmatrix},\\
(\tau_{l+1}\cdots\tau_{N+1}\circ\hat \mu_{n})_{l}^{-} & =\begin{pmatrix}\underset{l}{6} & 8 & \underset{N+2}{9}\end{pmatrix},\\
(\tau_{l+1}\cdots\tau_{N+1}\circ\hat \mu_{n})_{l}^{+} & =\begin{pmatrix}1 & 2 & \boldsymbol{4} & \boldsymbol{3} & 5 & \underset{l+1}{7} & 10\end{pmatrix}.
\end{align*}

\noindent
\textbf{III.} For $N\geq n+|\mu|-2$, $n\leq l\leq n+|\mu|-1$, 
\\ let $P=10,n=3,l=4,N=7,|\mu|=4$,
\begin{align*}
\mu & =\begin{pmatrix}\boldsymbol{1} & \boldsymbol{2} & \underset{j_{*}}{\boldsymbol{3}} & \boldsymbol{4}\\
\boldsymbol{3} & \boldsymbol{4} & \boldsymbol{2} & \boldsymbol{1}
\end{pmatrix},\\
\mu_{n} & =\begin{pmatrix}1 & 2 & \boldsymbol{3} & \boldsymbol{4} & \underset{j_{*}}{\boldsymbol{5}} & \boldsymbol{6} & 7 & 8 & 9 & 10\\
1 & 2 & \underset{l+1}{\boldsymbol{5}} & \boldsymbol{6} & \underset{l}{\boldsymbol{4}} & \boldsymbol{3} & 7 & 8 & \underset{N+2}{9} & 10
\end{pmatrix},\\
\tau_{l+1}\cdots\tau_{N+1}\mu_{n} & =\begin{pmatrix}1 & 2 & \boldsymbol{3} & \boldsymbol{4} & \boldsymbol{5} & \boldsymbol{6} & 7 & 8 & 9 & 10\\
1 & 2 & \boldsymbol{6} & \boldsymbol{7} & \underset{l}{\boldsymbol{4}} & \boldsymbol{3} & 8 & \underset{N+2}{9} & \underset{l+1}{\boldsymbol{5}} & 10
\end{pmatrix},\\
(\tau_{l+1}\cdots\tau_{N+1}\circ\hat \mu_{n})_{l}^{-} & =\begin{pmatrix}\underset{l}{\boldsymbol{4}} & \boldsymbol{3} & 8 & \underset{N+2}{9}\end{pmatrix},\\
(\tau_{l+1}\cdots\tau_{N+1}\circ\hat \mu_{n})_{l}^{+} & =\begin{pmatrix}1 & 2 & \boldsymbol{6} & \boldsymbol{7} & \underset{l+1}{\boldsymbol{5}} & 10\end{pmatrix}.
\end{align*}

\noindent
\textbf{IV. }For $N\geq n+|\mu|-2$, and $1\leq l\leq n-1$, \\
let $P=10,l=2,n=4,N=8,|\mu|=2$,
\begin{align*}
\mu & =\begin{pmatrix}\boldsymbol{1} & \boldsymbol{2}\\
\boldsymbol{2} & \boldsymbol{1}
\end{pmatrix},\\
\mu_{n} & =\begin{pmatrix}1 & 2 & 3 & \boldsymbol{4} & \boldsymbol{5} & 6 & 7 & 8 & 9 & 10\\
1 & \underset{l}{2} & \underset{l+1}{3} & \boldsymbol{5} & \boldsymbol{4} & 6 & 7 & 8 & 9 & \underset{N+2}{10}
\end{pmatrix},\\
\tau_{l+1}\cdots\tau_{N+1}\mu_{n} & =\begin{pmatrix}1 & 2 & 3 & \boldsymbol{4} & \boldsymbol{5} & 6 & 7 & 8 & 9 & 10\\
1 & \underset{l}{2} & 4 & \boldsymbol{6} & \boldsymbol{5} & 7 & 8 & 9 & \underset{N+2}{10} & \underset{l+1}{3}
\end{pmatrix},\\
(\tau_{l+1}\cdots\tau_{N+1}\circ\hat \mu_{n})_{l}^{-} & =\begin{pmatrix}\underset{l}{2} & 4 & \boldsymbol{6} & \boldsymbol{5} & 7 & 8 & 9 & \underset{N+2}{10}\end{pmatrix},\\
(\tau_{l+1}\cdots\tau_{N+1}\circ\hat \mu_{n})_{l}^{+} & =\begin{pmatrix}1 & \underset{l+1}{10}\end{pmatrix}.
\end{align*}

\noindent
\textbf{V.} For $N+1\geq l\geq n$, $M-1\geq m\geq1$, \\
let $P=10,n=3,l=4,N=5,M=N+2-(n-1)=5,m=l-(n-1)=2,|\mu|=7$.
\begin{align*}
\mu & =\begin{pmatrix}\underset{i_{*}}{\boldsymbol{1}} & \boldsymbol{2} & \boldsymbol{3} & \boldsymbol{4} & \boldsymbol{5} & \boldsymbol{6} & \underset{I_{*}}{\boldsymbol{7}}\\
\boldsymbol{2} & \boldsymbol{6} & \boldsymbol{4} & \boldsymbol{7} & \boldsymbol{3} & \boldsymbol{1} & \boldsymbol{5}
\end{pmatrix},\\
\tau_{m+1}\cdots\tau_{M-1}\mu=\omega_{M}^{m+1}\mu & =\begin{pmatrix}\boldsymbol{1} & \boldsymbol{2} & \boldsymbol{3} & \boldsymbol{4} & \boldsymbol{5} & \boldsymbol{6} & \boldsymbol{7}\\
\boldsymbol{2} & \boldsymbol{6} & \boldsymbol{5} & \boldsymbol{7} & \boldsymbol{4} & \boldsymbol{1} & \boldsymbol{3}
\end{pmatrix},\\
\mu_{n} & =\begin{pmatrix}1 & 2 & \boldsymbol{3} & \boldsymbol{4} & \boldsymbol{5} & \boldsymbol{6} & \boldsymbol{7} & \boldsymbol{8} & \boldsymbol{9} & 10\\
1 & 2 & \underset{l}{\boldsymbol{4}} & \boldsymbol{8} & \boldsymbol{6} & \boldsymbol{9} & \underset{l+1}{\boldsymbol{5}} & \boldsymbol{3} & \underset{N+2}{\boldsymbol{7}} & 10
\end{pmatrix},\\
\tau_{l+1}\cdots\tau_{N+1}\mu_{n} & =\begin{pmatrix}1 & 2 & \boldsymbol{3} & \boldsymbol{4} & \boldsymbol{5} & \boldsymbol{6} & \boldsymbol{7} & \boldsymbol{8} & \boldsymbol{9} & 10\\
1 & 2 & \underset{l}{\boldsymbol{4}} & \boldsymbol{8} & \underset{N+2}{\boldsymbol{7}} & \boldsymbol{9} & \boldsymbol{6} & \boldsymbol{3} & \underset{l+1}{\boldsymbol{5}} & 10
\end{pmatrix},\\
(\tau_{l+1}\cdots\tau_{N+1}\circ\hat \mu_{n})_{l}^{-} & =\begin{pmatrix}\underset{l}{\boldsymbol{4}} & \boldsymbol{8} & \underset{N+2}{\boldsymbol{7}} & \boldsymbol{9} & \boldsymbol{6} & \boldsymbol{3}\end{pmatrix},\\
(\tau_{l+1}\cdots\tau_{N+1}\circ\hat \mu_{n})_{l}^{+} & =\begin{pmatrix}1 & 2 & \underset{l+1}{\boldsymbol{5}} & 10\end{pmatrix}.
\end{align*}
\begin{align*}
\mu_{\varpi}^{(m,M)} & =\omega_{M}^{m+1}\begin{pmatrix}\mu{}^{-1}(m) & \mu{}^{-1}(m)+1 & \cdots & \mu{}^{-1}(M)-1\\
m & \mu(\mu{}^{-1}(m)+1) & \cdots\mu( & \mu{}^{-1}(M)-1)
\end{pmatrix}\\
 & =\begin{pmatrix}\boldsymbol{1} & \boldsymbol{2} & \boldsymbol{3} & \boldsymbol{4} & \boldsymbol{5} & \boldsymbol{6}\\
\boldsymbol{2} & \boldsymbol{6} & \boldsymbol{5} & \boldsymbol{7} & \boldsymbol{4} & \boldsymbol{1}
\end{pmatrix}\equiv\begin{pmatrix}\boldsymbol{3} & \boldsymbol{4} & \boldsymbol{5} & \boldsymbol{6} & \boldsymbol{7} & \boldsymbol{8}\\
\underset{l}{\boldsymbol{4}} & \boldsymbol{8} & \underset{N+2}{\boldsymbol{7}} & \boldsymbol{9} & \boldsymbol{6} & \boldsymbol{3}
\end{pmatrix} \\
& \implies(\tau_{l+1}\cdots\tau_{N+1}\circ\hat \mu_{n})_{l}^{-}\\
(\mu_{\varpi}^{m)(M})_{n} & =\Big(\omega_{M}^{m+1}\begin{pmatrix}1 & \cdots & \mu{}^{-1}(m)-1 & \mu^{(-1)}(M) & \cdots & |\mu|\\
\mu(1) & \cdots & \mu(\mu^{-1}(m)-1) & M & \cdots & \mu(|\mu|)
\end{pmatrix}\Big)_{n}\\
 & =\Big(\begin{pmatrix}\boldsymbol{7}\\ \boldsymbol{3} \end{pmatrix}\Big)_{n}
 =\begin{pmatrix}1 & 2 & \boldsymbol{9} & 10\\
1 & 2 & \boldsymbol{5}_{l+1} & 10
\end{pmatrix} \\
& \implies(\tau_{l+1}\cdots\tau_{N+1}\circ\hat \mu_{n})_{l}^{+}
\end{align*}

\noindent
\textbf{VI.} For $1\leq l\leq n-1$,\\
 let $P=14,n=6,l=3,N=7,M=N+2-(n-1)=4,|\mu|=7$ : 
\begin{align*}
\mu & =\begin{pmatrix}\boldsymbol{1} & \boldsymbol{2} & \boldsymbol{3} & \boldsymbol{4} & \underset{I_{*}}{\boldsymbol{5}} & \boldsymbol{6} & \boldsymbol{7}\\
\boldsymbol{2} & \boldsymbol{5} & \boldsymbol{7} & \boldsymbol{6} & \boldsymbol{4} & \boldsymbol{1} & \boldsymbol{3}
\end{pmatrix}\\
\tau_{1}\cdots\tau_{M-1}\mu=\omega_{M}^{1}\mu & =\begin{pmatrix}\boldsymbol{1} & \boldsymbol{2} & \boldsymbol{3} & \boldsymbol{4} & \boldsymbol{5} & \boldsymbol{6} & \boldsymbol{7}\\
\boldsymbol{3} & \boldsymbol{5} & \boldsymbol{7} & \boldsymbol{6} & \boldsymbol{1} & \boldsymbol{2} & \boldsymbol{4}
\end{pmatrix}\\
\mu_{n} & =\begin{pmatrix}1 & 2 & 3 & 4 & 5 & \boldsymbol{6} & \boldsymbol{7} & \boldsymbol{8} & \boldsymbol{9} & \boldsymbol{10} & \boldsymbol{11} & \boldsymbol{12} & 13 & 14\\
1 & 2 & \underset{l}{3} & \underset{l+1}{4} & 5 & \boldsymbol{7} & \boldsymbol{10} & \boldsymbol{12} & \boldsymbol{11} & \underset{N+2}{\boldsymbol{9}} & \boldsymbol{6} & \boldsymbol{8} & 13 & 14
\end{pmatrix}\\
\tau_{l+1}\cdots\tau_{N+1}\mu_{n} & =\begin{pmatrix}1 & 2 & 3 & 4 & 5 & \boldsymbol{6} & \boldsymbol{7} & \boldsymbol{8} & \boldsymbol{9} & \boldsymbol{10} & \boldsymbol{11} & \boldsymbol{12} & 13 & 14\\
1 & 2 & \underset{l}{3} & 5 & \boldsymbol{6} & \boldsymbol{8} & \boldsymbol{10} & \boldsymbol{12} & \boldsymbol{11} & \underset{l+1}{4} & \boldsymbol{7} & \underset{N+2}{\boldsymbol{9}} & 13 & 14
\end{pmatrix}\\
(\tau_{l+1}\cdots\tau_{N+1}\circ\hat \mu_{n})_{l}^{-} & =\begin{pmatrix}3_{l} & 5 & \boldsymbol{6} & \boldsymbol{8} & \boldsymbol{10} & \boldsymbol{12} & \boldsymbol{11}\end{pmatrix}\equiv\begin{pmatrix}1 & 3 & \boldsymbol{4} & \boldsymbol{6} & \boldsymbol{8} & \boldsymbol{10} & \boldsymbol{9}\end{pmatrix}\\
(\tau_{l+1}\cdots\tau_{N+1}\circ\hat \mu_{n})_{l}^{+} & =\begin{pmatrix}1 & 2 & \underset{l+1}{4} & \boldsymbol{7} & \underset{N+2}{\boldsymbol{9}} & 13 & 14\end{pmatrix}
\end{align*}
\begin{align*}
\mu^{M]} & =\begin{pmatrix}1 & \cdots & \mu^{-1}(M)-1\\
\mu(1) & \cdots & \mu(\mu^{-1}(M)-1)
\end{pmatrix}\\
 & =\begin{pmatrix}\boldsymbol{1} & \boldsymbol{2} & \boldsymbol{3} & \boldsymbol{4}\\
\boldsymbol{2} & \boldsymbol{5} & \boldsymbol{7} & \boldsymbol{6}
\end{pmatrix}\\
\omega_{M}^{1}\mu^{M]} & =\begin{pmatrix}\boldsymbol{1} & \boldsymbol{2} & \boldsymbol{3} & \boldsymbol{4}\\
\boldsymbol{3} & \boldsymbol{5} & \boldsymbol{7} & \boldsymbol{6}
\end{pmatrix}\\
(\mu_{\varpi}^{M]})_{n-l+1} & =\begin{pmatrix}1 & 2 & 3 & \boldsymbol{4} & \boldsymbol{5} & \boldsymbol{6} & \boldsymbol{7}\\
1 & 2 & 3 & \boldsymbol{6} & \boldsymbol{8} & \boldsymbol{10} & \boldsymbol{9}
\end{pmatrix}\equiv\begin{pmatrix}3 & 4 & 5 & \boldsymbol{6} & \boldsymbol{7} & \boldsymbol{8} & \boldsymbol{9}\\
3_{l} & 5 & \boldsymbol{6} & \boldsymbol{8} & \boldsymbol{10} & \boldsymbol{12} & \boldsymbol{11}
\end{pmatrix}\\
 & \implies(\tau_{l+1}\cdots\tau_{N+1}\circ\hat \mu_{n})_{l}^{-}\\
\mu^{[M} & =\begin{pmatrix}\mu^{-1}(M)+1 & \cdots & |\mu|\\
\mu(\mu^{-1}(M)+1) & \cdots & \mu(|\mu|)
\end{pmatrix}\\
 & =\begin{pmatrix}\boldsymbol{6} & \boldsymbol{7}\\
\boldsymbol{1} & \boldsymbol{3}
\end{pmatrix}\\
\omega_{M}^{1}\mu^{[M} & =\begin{pmatrix}\boldsymbol{6} & \boldsymbol{7}\\
\boldsymbol{2} & \boldsymbol{4}
\end{pmatrix}\\
(\mu_{\varpi}^{[M})_{l+1}=(\omega_{M}^{1}\mu^{[M})_{l+1} & =\begin{pmatrix}1 & 2 & 3 & \boldsymbol{9} & \boldsymbol{10} & 13 & 14\\
1 & 2 & 3 & \boldsymbol{5} & \boldsymbol{7} & 13 & 14
\end{pmatrix}\equiv\begin{pmatrix}1 & 2 & \boldsymbol{10} & \boldsymbol{11} & \boldsymbol{12} & 13 & 14\\
1 & 2 & \underset{l+1}{4} & \boldsymbol{7} & \underset{N+2}{\boldsymbol{9}} & 13 & 14
\end{pmatrix}\\
 & \implies(\tau_{l+1}\cdots\tau_{N+1}\circ\hat \mu_{n})_{l}^{+}
\end{align*}


\newpage

\begin{thebibliography}{99}

\bibitem{Kipnis99} C. Kipnis and C. Landim, {\it Scaling limits of interacting particle systems}, Springer, Berlin, (1999).
\bibitem{Liggett99} T. Liggett, {\it Stochastic interacting systems: contact, voter and exclusion processes}, Fund. Principles Math. Sciences 324, Springer, Berlin (1999).
\bibitem{Spohn91} H. Spohn, {\it Large scale dynamics of interacting particles},  Springer, Berlin, (1991).

\bibitem{SSEP} C. Kipnis, S. Olla and S. Varadhan, {\it Hydrodynamics and large deviation for simple exclusion processes}, Commun. Pure Appl. Math. 42, 115-137 (1989).
\bibitem{Derrida_Review} B. Derrida, {\it Non-equilibrium steady states: fluctuations and large deviations of the density and of the current}, J. Stat. Mech., P07023, (2007).\\
B. Derrida, {\it Microscopic versus macroscopic approaches to non-equilibrium systems}, J. Stat. Mech. 2011, P01030 (2011).
\bibitem{Mallick_Review}K. Mallick, {\it The Exclusion Process: A paradigm for non-equilibrium behaviour}, Physica A: Stat. Mech. and Appl., 418, 1-188 (2015).

\bibitem{Fluctu95} G. Gallavotti and E.G.D. Cohen, {\it Dynamical ensembles in non-equilibrium statistical mechanics}, Phys. Rev. Lett. 74, 2694 (1995).
\bibitem{Jar97} C. Jarzynski, {\it Nonequilibrium equality for free energy differences}, Phys. Rev. Lett. 78, 2690-2693 (1997).
\bibitem{Crooks99} G.E. Crooks, {\it Entropy production fluctuation theorem and the nonequilibrium work relation for free energy differences}, Phys. Rev. E 60, 2721-2726 (1999).

\bibitem{Maes99} C. Maes, {\it The Fluctuation Theorem as a Gibbs Property.}, J. Stat. Phys. 95, 367-392, (1999).
\bibitem{Maes_bis} C. Maes and  K. Natocny, {\it Time reversal and entropy}, J. Stat. Phys. 110, 269-310, (2003).

\bibitem{MFT} L. Bertini, A. De Sole, D. Gabrielli, G. Jona-Lasinio, and C. Landim, {\it Current fluctuations in stochastic lattice gases}, Phys. Rev. Lett. 94, 030601 (2005).\\
L. Bertini, A. De Sole, D. Gabrielli, G. Jona-Lasinio and C. Landim, {\it Macroscopic fluctuation theory}, Rev. Mod. Phys. 87(2), 593 (2015).

\bibitem{GHD1} O. A. Castro-Alvaredo, B. Doyon and T. Yoshimura, {\it Emergent Hydrodynamics in Integrable Quantum Systems Out of Equilibrium}, Phys. Rev. X 6, 041065 (2016).
\bibitem{GHD2} B. Bertini, M. Collura, J. De Nardis and M. Fagotti, {\it Transport in Out-of-Equilibrium XXZ Chains: Exact Profiles of Charges and Currents},  Phys. Rev. Lett. 117, 207201 (2016).

\bibitem{Membrane1} M. Mezei, {\it Membrane theory of entanglement dynamics from holography}, Phys. Rev. D 98, 106025, (2018).
\bibitem{Membrane2} T. Zhou and A. Nahum. {\it Emergent statistical mechanics of entanglement in random unitary circuits}. Physical Review B, 99(17):174205, (2019).
\bibitem{Membrane3} M.J. Gullans and D. A. Huse, {\it Entanglement Structure of Current-Driven Diffusive Fermion Systems}, Phys. Rev. X 9, 021007 (2019).
\bibitem{Membrane4} C. A. Agon and M. Mezei, {\it Bit threads and the membrane theory of entanglement dynamics}, arXiv:1910.12909, (2019).
\bibitem{Membrane5} T. Zhou and A. Nahum. {\it The entanglement membrane in chaotic many-body systems}, arXiv:1912.12311, (2019).

\bibitem{BBJ1} M. Bauer, D. Bernard and T. Jin, {\it Equilibrium Fluctuations in Maximally Noisy Extended Quantum Systems}. SciPost Phys. 6:45, (2019).

\bibitem{BJ2019} D. Bernard and T. Jin, {\it Open quantum symmetric simple exclusion process}, Phys. Rev. Lett. 123, 080601 (2019).

\bibitem{JKB2020} T. Jin, A. Krajenbrink and D. Bernard, {\it From stochastic spin chains to quantum Kardar-Parisi-Zhang dynamics}, arXiv:2001.04278. 

\bibitem{Profile} 
M. Znidaric, {\it Exact solution for a diffusive non-equilibrium steady state of an open quantum chain}, J. Stat. Mech. 2010, L05002 (2010).\\
V. Eisler, {\it Crossover between ballistic and diffusive transport: The Quantum Exclusion Process}, J. Stat. Mech. P06007 (2011).\\
M.V. Medvedyeva,  F.H.L. Essler, and T. Prosen, {\it Exact Bethe Ansatz Spectrum of a Tight-Binding Chain with Dephasing Noise}, Phys. Rev. Lett. 117, 137202 (2016).


\bibitem{discret-geom} S. Devadoss and J. O'Rourke, {\it Discrete and Computational Geometry}, Princeton University Press, (2011).

\bibitem{arkani_et_al} N. Arkani-Hamed, Y. Bai, S. He and G. Yan, {\it Scattering Forms and the Positive Geometry of Kinematics, Color and the Worldsheet}, JHEP 05, 096 (2018).

\bibitem{integrable} See e.g.
B.A. Dubrovin, I.M. Krichever, S.P. Novikov, {\it Integrable systems. I. }, in: Arnold V.I., Novikov S.P. (eds), Dynamical Systems IV. Encyclopaedia of Mathematical Sciences, vol 4. Springer,  (2001).\\
O. Babelon, D. Bernard, M. Talon, {\it Introduction to classical integrable systems}, Cambridge University Press (2003).\\
A.R. Chowdhury, A.G. Choudhury, {\it  Quantum integrable systems}, Research Notes in Mathematics Series 435, Chapman $\&$ Hall, (2004).

\bibitem{fomin} S. Fomin and A. Zelevinsky, {\it Cluster algebras. I: Foundations}, J. Am. Math. Soc. 15, 497-529 (2002).\\
S. Fomin and A. Zelevinsky, {\it Cluster algebras II: Finite type classification}, Invent. Math.154, 63-121 (2003).

\bibitem{cluster-asso}
S. Fomin and N. Reading, {\it Root systems and generalized associahedra}, Geometric combinatorics, 63-131, IAS/Park City Math. Ser., 13, Amer. Math. Soc., (2007).

\bibitem{Brenier} Y. Brenier, {\it Permutations and PDE}, Oberwolfach Seminar October 2018, http://www.math.ens.fr/-brenier/

\bibitem{BD2004} T. Bodineau and B. Derrida, {\it Current Fluctuations in Non-equilibrium Diffusive Systems: An Additivity Principle}, Phys. Rev. Lett. 92, 180601 (2004).

\bibitem{michel} We thank Michel Bauer for pointing out this connexion.


\end{thebibliography}
\end{document}